\documentclass[twocolumn]{aastex701}

\begin{document}

\title{A Detailed Model Atmosphere Analysis of Cool White Dwarfs in DESI DR1}

\author[orcid=0000-0001-6098-2235,sname='Kilic']{Mukremin Kilic} 
\affiliation{Homer L. Dodge Department of Physics and Astronomy, University of Oklahoma, 440 W. Brooks St., Norman, OK, 73019 USA}
\email[show]{kilic@ou.edu}

\author[orcid=0000-0003-2368-345X,sname='Bergeron']{Pierre Bergeron}
\affiliation{D\'epartement de Physique, Universit\'e de Montr\'eal, C.P. 6128, Succ. Centre-Ville, Montr\'eal, QC H3C 3J7, Canada}
\email[show]{bergeron@astro.umontreal.ca}

\author[orcid=0000-0002-9632-1436,sname='Blouin']{Simon Blouin}
\affiliation{Department of Physics and Astronomy, University of Victoria, Victoria BC V8W 2Y2, Canada}
\email{sblouin@uvic.ca}

\author[0000-0001-7143-0890]{Adam Moss} 
\affiliation{Department of Astronomy, University of Florida, Bryant Space Science Center, Stadium Road, Gainesville, FL 32611, USA}
\email{Adam.G.Moss-1@ou.edu}

\author[0000-0002-0948-4801]{Matthew J. Green}
\affiliation{Homer L. Dodge Department of Physics and Astronomy, University of Oklahoma, 440 W. Brooks St., Norman, OK, 73019 USA}
\email{matthew.j.green-2@ou.edu}

\author[orcid=0009-0009-9105-7865,sname='Jewett']{Gracyn Jewett}
\affiliation{Homer L. Dodge Department of Physics and Astronomy, University of Oklahoma, 440 W. Brooks St., Norman, OK, 73019 USA}
\email{gjewett@ou.edu}

\author[0000-0002-6153-9304,sname='Barrientos']{Manuel Barrientos} 
\affiliation{Homer L. Dodge Department of Physics and Astronomy, University of Oklahoma, 440 W. Brooks St., Norman, OK, 73019 USA}
\email{mbarrientos@ou.edu }

\author[0009-0000-7416-5228]{Alexander L. Albright}
\affiliation{Homer L. Dodge Department of Physics and Astronomy, University of Oklahoma, 440 W. Brooks St., Norman, OK, 73019 USA}
\email{Alexander.L.Albright-1@ou.edu}

\author[0009-0003-3096-7618]{Alexander R. Gleason}
\affiliation{Homer L. Dodge Department of Physics and Astronomy, University of Oklahoma, 440 W. Brooks St., Norman, OK, 73019 USA}
\email{Alexander.R.Gleason-1@ou.edu }

\author[0000-0002-4462-2341,sname='Brown']{Warren R.\ Brown}
\affiliation{Center for Astrophysics, Harvard \& Smithsonian, 60 Garden Street, Cambridge, MA 02138, USA}
\email{wbrown@cfa.harvard.edu}

\author[0000-0003-2947-5889]{Nagaraj Vernekar}
\affiliation{Homer L. Dodge Department of Physics and Astronomy, University of Oklahoma, 440 W. Brooks St., Norman, OK, 73019 USA}
\email{nv101@ou.edu}

\author[0000-0001-7294-9766]{Michael R. Hayden}
\affiliation{Homer L. Dodge Department of Physics and Astronomy, University of Oklahoma, 440 W. Brooks St., Norman, OK, 73019 USA}
\email{mrhayden@ou.edu}

\begin{abstract}

We present a detailed model atmosphere analysis of cool white dwarfs in the Dark Energy Spectroscopic Instrument Data Release 1 (DESI DR1). Our sample includes 25,642 unique targets with $G_{\rm BP}-G_{\rm RP}>0$. Unlike the hot DA white dwarf sample in DESI DR1, we do not find a significant discrepancy between the photometric and spectroscopic masses for cool DAs. Hence, DESI’s calibration problems for broad lines have a negligible effect for cooler DAs with narrower lines. Magnetic DAs are found everywhere, and not just on the crystallization sequence, indicating that crystallization induced dynamos cannot solely explain the origin of magnetism in white dwarfs. A detailed analysis of cool DC and DZ white dwarfs indicates that the H/He abundance ratio in He-atmosphere white dwarfs increases at lower temperatures. Based on the currently available models, this is the only way to keep the DC masses consistent with the average white dwarf mass of $0.6~M_\odot$. Combined with the analysis of the hot white dwarfs presented previously, this paper completes the analysis of 44,963 white dwarf candidates with DESI DR1 spectra. We use this sample to constrain the fraction of He-atmosphere white dwarfs as a function of temperature, and demonstrate that the He-fraction increases significantly below 10,000 K due to convective mixing. We also highlight rare systems, including new extremely low-mass, DA+DB, and DA+DQ binaries. 

\end{abstract}

\keywords{\uat{White dwarf stars}{1799} --- \uat{DA stars}{348} --- \uat{DB stars}{358} --- \uat{DC stars}{0} --- \uat{DQ stars}{1849} --- \uat{DZ stars}{1848}}

\section{Introduction}

The number of spectroscopically confirmed white dwarfs increased from $\sim$$10^3$ in the late eighties \citep{mccook87} to $\sim$$10^4$ by the late 2000s.
This dramatic increase was due to the technological advances in fiber-fed spectroscopy and specifically through the Sloan Digital Sky Survey \citep[SDSS,][]{eisenstein06,kleinman13,kepler16}. We are on the verge of increasing the sample size by yet another order of magnitude, thanks to Gaia astrometry \citep{gaia18,gaia21} and the ongoing and planned wide-field multiplexed spectroscopic surveys like the Dark Energy Spectroscopic Instrument \citep[DESI,][]{desi26}, SDSS-V \citep{kollmeier19}, 4-metre Multi-Object Spectroscopic Telescope \citep[4MOST,][]{dejong19}, and the William Herschel Telescope Enhanced Area Velocity Explorer \citep[WEAVE,][]{jin24}. 

There are $\approx$359,000 high-probability white dwarf candidates in the Gaia sample \citep{gentile21}. Many of these candidates will be observed by the aforementioned surveys. The large sample size will help us better characterize the white dwarf population in the solar neighborhood, including the various spectral types and their temperature and mass distributions.
Stellar evolution leaves its mark on the white dwarf luminosity and mass functions as the resulting distributions depend on the star formation history, the initial mass function, the initial-final mass relation, white dwarf cooling physics, and the impacts of close binary evolution. Furthermore, element transport mechanisms like radiative levitation, winds, gravitational settling, diffusion, convection, and external accretion lead to spectral evolution as white dwarfs cool
\citep[see][for a review]{bedard24b}. Hence, the results from these multiplexed surveys will significantly improve our understanding of the white dwarf evolution. Furthermore, when the sample size grows by an order of magnitude, this usually leads to unexpected discoveries, including rare objects. DESI, SDSS-V, 4MOST, and WEAVE are destined to be great discovery machines.

DESI DR1 \citep{desi26} recently provided a sneak peak into the outcomes from these large scale spectroscopic surveys by releasing spectroscopy of $\sim$45,000 Gaia white dwarf candidates. Given the large sample size, and the sheer task of visually inspecting each model fit, we split the DESI DR1 sample into hot (blue) and cool (red) white dwarfs. In Paper I of this series \citep{kilic26}, we presented  a detailed model atmosphere analysis of 19,321 hot white dwarfs with $G_{\rm BP}-G_{\rm RP}\leq0$, which corresponds to $T_{\rm eff}\gtrsim10,000$ K. Here, we extend our analysis to
the Gaia white dwarf candidates with $G_{\rm BP}-G_{\rm RP}>0$, which make up the majority of the DESI DR1 white dwarf sample.

\begin{figure}
\center
\includegraphics[width=3.2in]{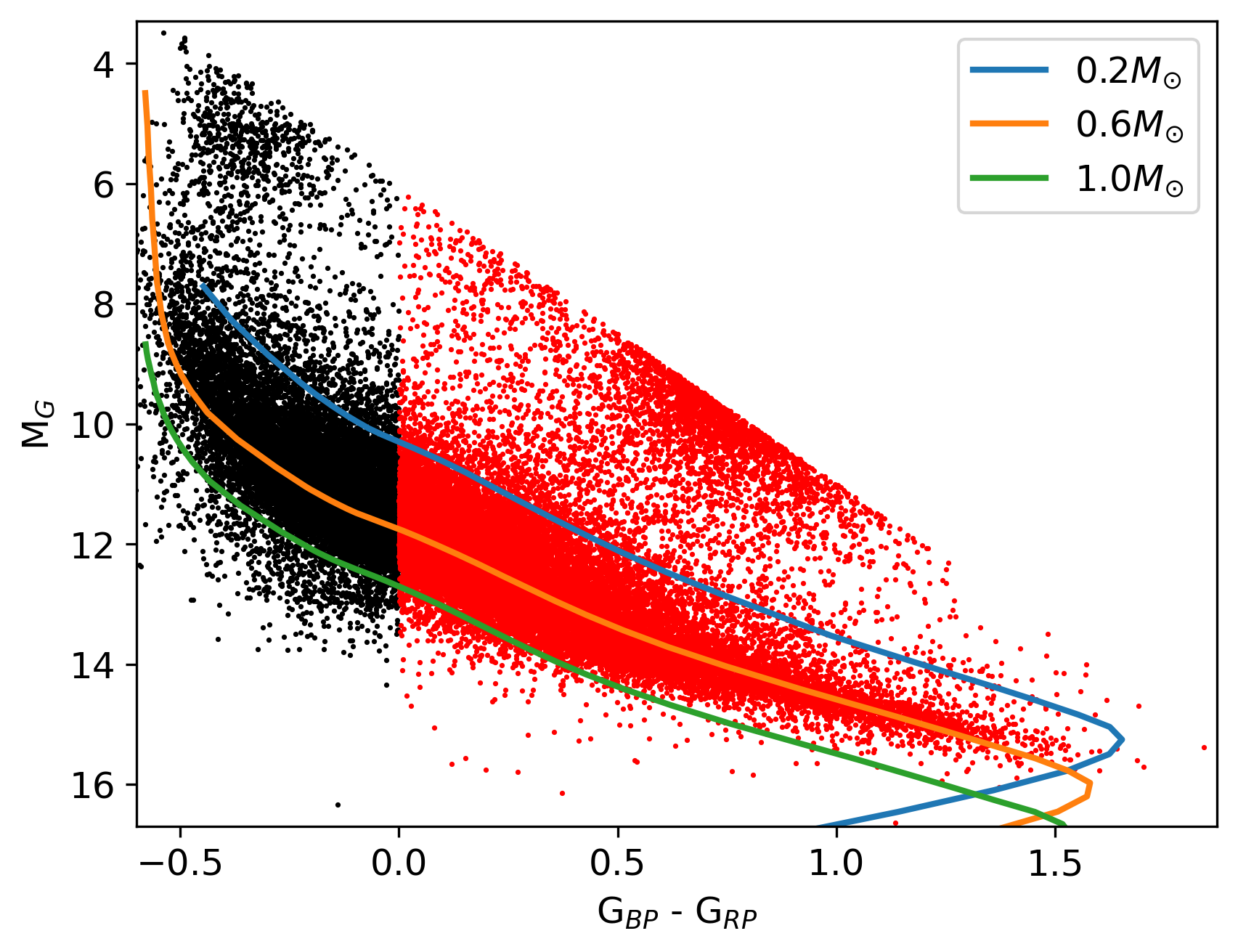}
\caption{DESI DR1 white dwarf sample in the H-R diagram. Black points mark the hot white dwarf sample analyzed in Paper I. Red points mark the cool white dwarf sample analyzed in this paper. Evolutionary tracks for pure H atmosphere white dwarfs with $M=$ 0.2, 0.6, and $1.0~M_\odot$ are shown for comparison.}
\label{fighr} 
\end{figure}

Figure \ref{fighr} shows the color-magnitude diagram of the DESI DR1 white dwarf sample. Black points mark the hot/blue white dwarf sample
presented in Paper I, whereas the red points mark the cool/red white dwarfs candidates analyzed here. The colored lines show the improved evolutionary
tracks (see section \ref{secmodel}) for pure hydrogen atmosphere white dwarfs with $M=$ 0.2, 0.6, and $1.0~M_\odot$. These tracks overlap with the majority of the white dwarfs in our sample. However, both the hot (black points) and cool (red points) samples show a significant concentration of over-luminous objects above the main white dwarf sequence. The hot white dwarf sample is contaminated by subdwarf stars with $M_G \sim5$, whereas the over-luminous objects in the cooler sample are dominated by white dwarf + main-sequence binaries.

For spectral classification, we first compared the DESI spectra with the predictions from the best-fitting pure H and pure He atmosphere white dwarf
models. We visually inspected each model fit to classify each object. We then performed a tailored analysis of each spectral type,
and re-inspected all of the model fits to verify the assigned spectral type and identify additional subtypes. 

We follow the spectral classification system of \citet{sion83}, with an uppercase D for degenerate, followed by an uppercase letter for the primary
spectroscopic type, with additional letters added for secondary or tertiary spectroscopic features. Just like for O, B, and A type main-sequence stars, we denote white dwarfs with ionized helium, neutral helium, and hydrogen features with the spectral types DO, DB, and DA (here a `D' is added at the beginning to indicate a degenerate star). We use DC for white dwarfs with featureless optical spectra (continuum emission), DQ for carbon features, and DZ for other metal features. We add `H' to indicate magnetism detected through Zeeman split and/or shifted spectral features, where DAH would refer to a magnetic DA white dwarf with a pure H atmosphere. 

The DESI DR1 sample includes a number of DA white dwarfs where the hydrogen lines are heavily broadened through van der Waals interactions in a helium-dominated atmosphere. 
We label these objects as He-DA white dwarfs. Even though they have helium-dominated atmospheres, the effective temperature is too cool for
He $\lambda5876$ to be visible. Hence, only H lines are visible in these stars.
We refer the reader to \citet{wesemael93} for a detailed review of the optical spectra for all major white dwarf spectral types and subtypes.
About half of the DESI DR1 spectra have
a signal-to-noise ratio (S/N) below 10 (see Figure 2 in \citealt{kilic26}), hence some of these classifications could be improved with higher S/N data, especially for cooler objects with relatively weaker lines. We add `:' for uncertain spectral types, and `?' for questionable classifications. 

We present our model fits for various spectral types in Section \ref{secmodel}, and discuss the overall properties of the DESI cool white dwarf sample, including the DC, DQ, DZ, and magnetic  white dwarf populations in Section \ref{secdis}. This section also includes the results on the spectral evolution of white dwarfs covering the entire effective temperature range from 100,000 down to 5000 K, and a brief discussion of the expected outcomes from future DESI data releases and the other current and planned multi-plexed spectorscopic surveys. We conclude in Section \ref{seccon}.

\section{Model Atmosphere Analysis}
\label{secmodel}

\subsection{Color-Magnitude Diagrams}
\label{seccmd}

\begin{figure}
\includegraphics[width=3.4in, clip=true, trim=0.2in 0.4in 0.2in 1.1in]{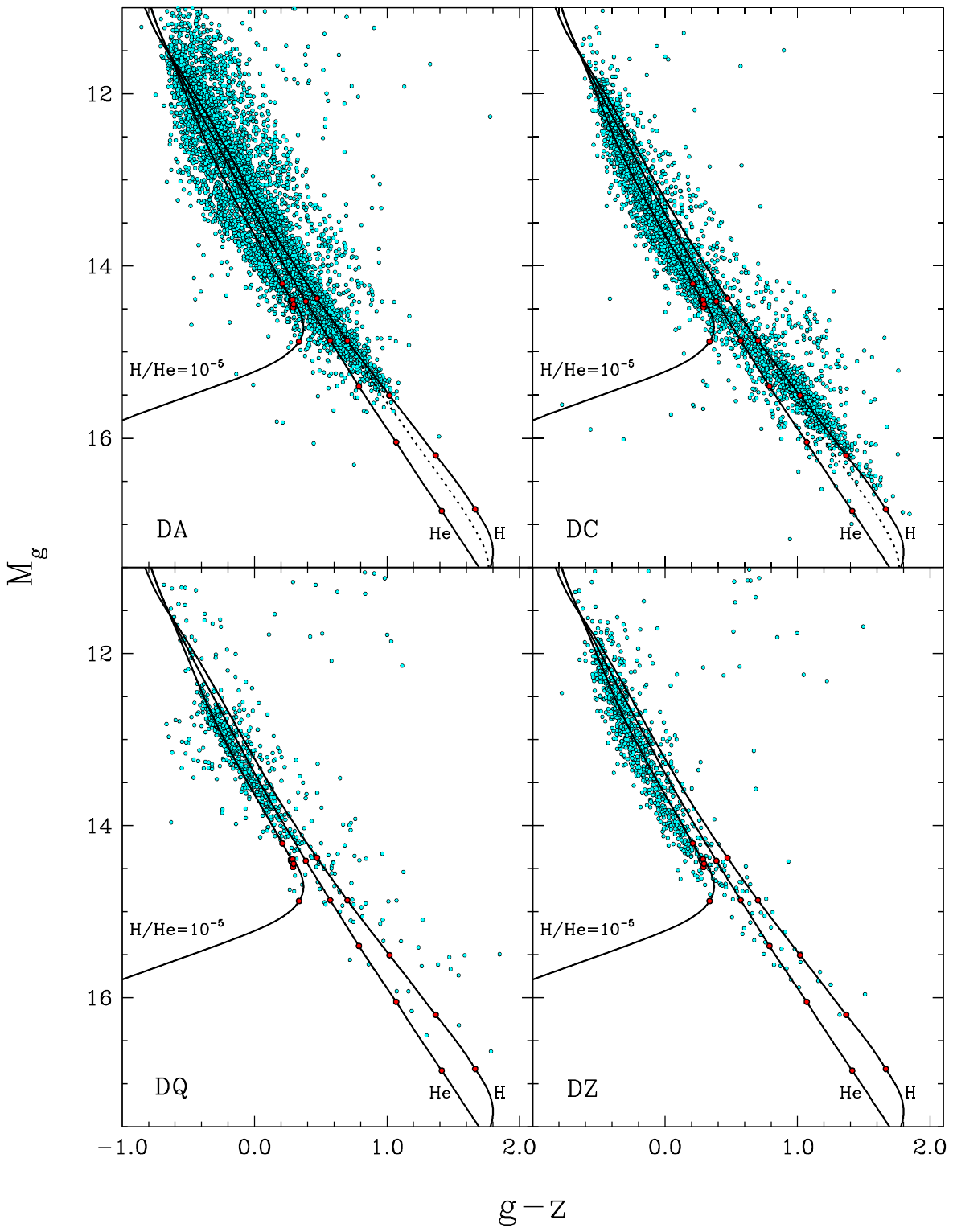}
\caption{Color-magnitude diagrams of DA, DC, DQ, and DZ white dwarfs in our sample. Solid lines show the evolutionary tracks for 
$0.6~M_\odot$ pure H, pure He, and mixed H/He atmosphere white dwarfs with $\log$ H/He = $-5$. The dotted lines in the top panels show
the pure H models without the correction to the H$_3^+$ partition function (see text). The red dots on each sequence correspond to effective temperatures ranging from 6000 K down to 4000 K (from top to bottom) in 500 K steps.} 
\label{figcmd} 
\end{figure}

Before presenting the detailed model atmosphere analysis of our sample, we first explore the global properties of the various spectral types using color-magnitude diagrams. The top panels in Figure \ref{figcmd} show color-magnitude diagrams for DA and DC white dwarfs. We restrict this figure to DAs and DCs with $<5$\% and $<10$\% distance errors, respectively. Solid lines show the evolutionary tracks for $0.6~M_\odot$ pure H, pure He, and mixed H/He atmosphere white dwarfs with $\log$ H/He = $-5$; details of these models are provided in the next subsections for each spectral type. 

\citet{Kowalski2026} recently speculated that an overabundance of H$_3^+$ in cool H-atmosphere models could explain the low-mass problem reported at the cool end of the white dwarf sequence \citep{caron23,sahu25b,kilic25a}. However, the origin of this potential overabundance remained unclear. Motivated by this work, we identified an issue in the calculation of the H$_3^+$ partition function used in all of our model atmosphere codes, which relies on the fit provided by \citet{neale95}. Unlike the standard practice in astrophysics, \citet{neale95} include the degeneracy associated with nuclear spins, leading to an inconsistency in the chemical equilibrium calculations implemented in our atmosphere codes. Correcting this issue—specifically by reducing the partition function by a factor of $2^3 = 8$—results in lower H$_3^+$ abundances and, consequently, reduced H$^-$ opacity, thereby affecting the entire spectral energy distribution of cool H-rich white dwarfs. The dotted lines in the top panels show the pure-H models computed without this correction and should be compared to the solid tracks labeled H. With the correction included (also adopted in Figure \ref{fighr}), cool DA and DC white dwarfs lie much closer to the $0.6~M_\odot$ sequences in color--magnitude diagrams, which should therefore help correct the low-mass problem highlighted in previous studies.
 
\begin{figure*}
\hspace{-0.4in}
\includegraphics[width=2.55in, clip=true, trim=0.4in 0.8in 0.2in 1.1in]{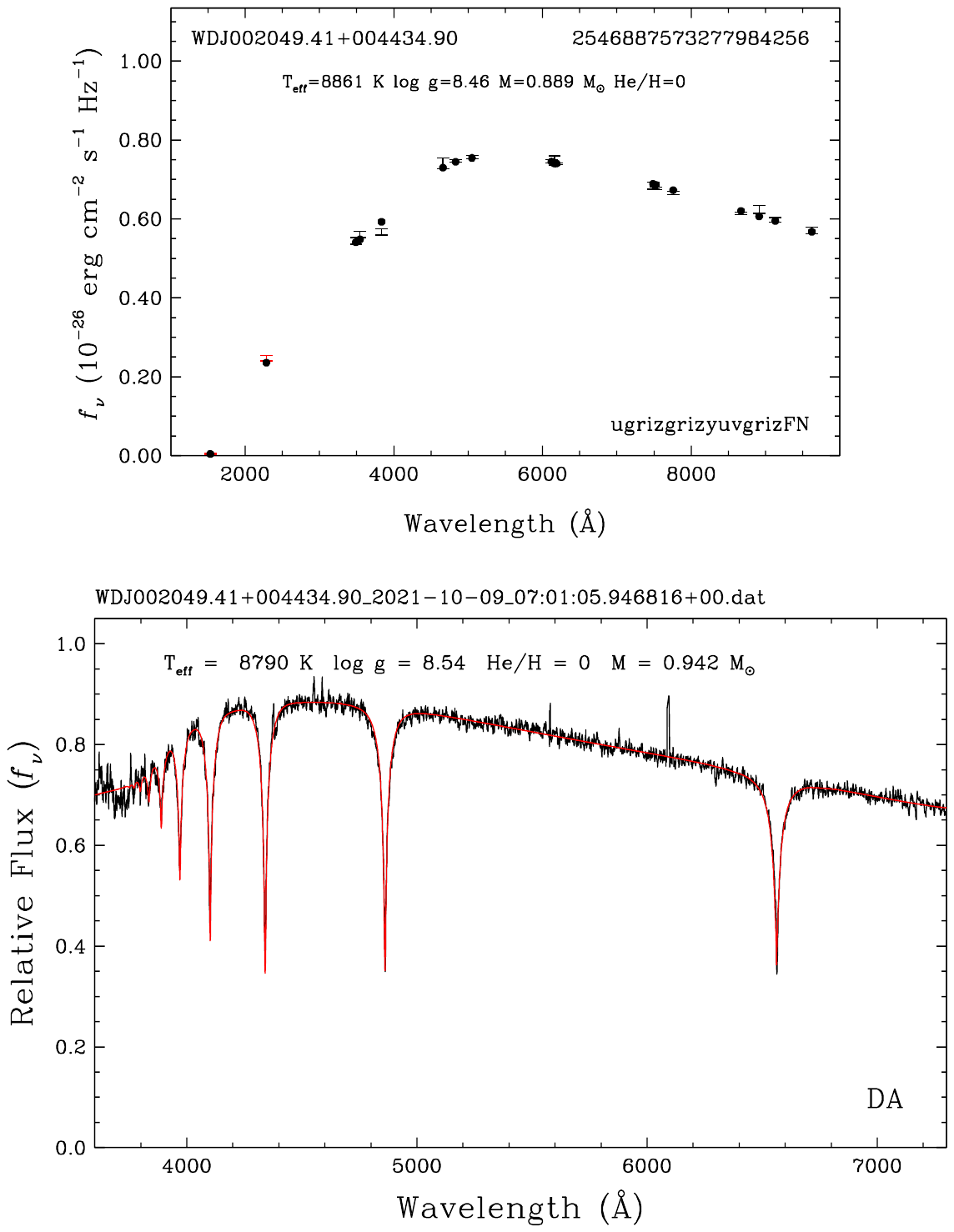}
\includegraphics[width=2.55in, clip=true, trim=0.4in 0.8in 0.2in 1.1in]{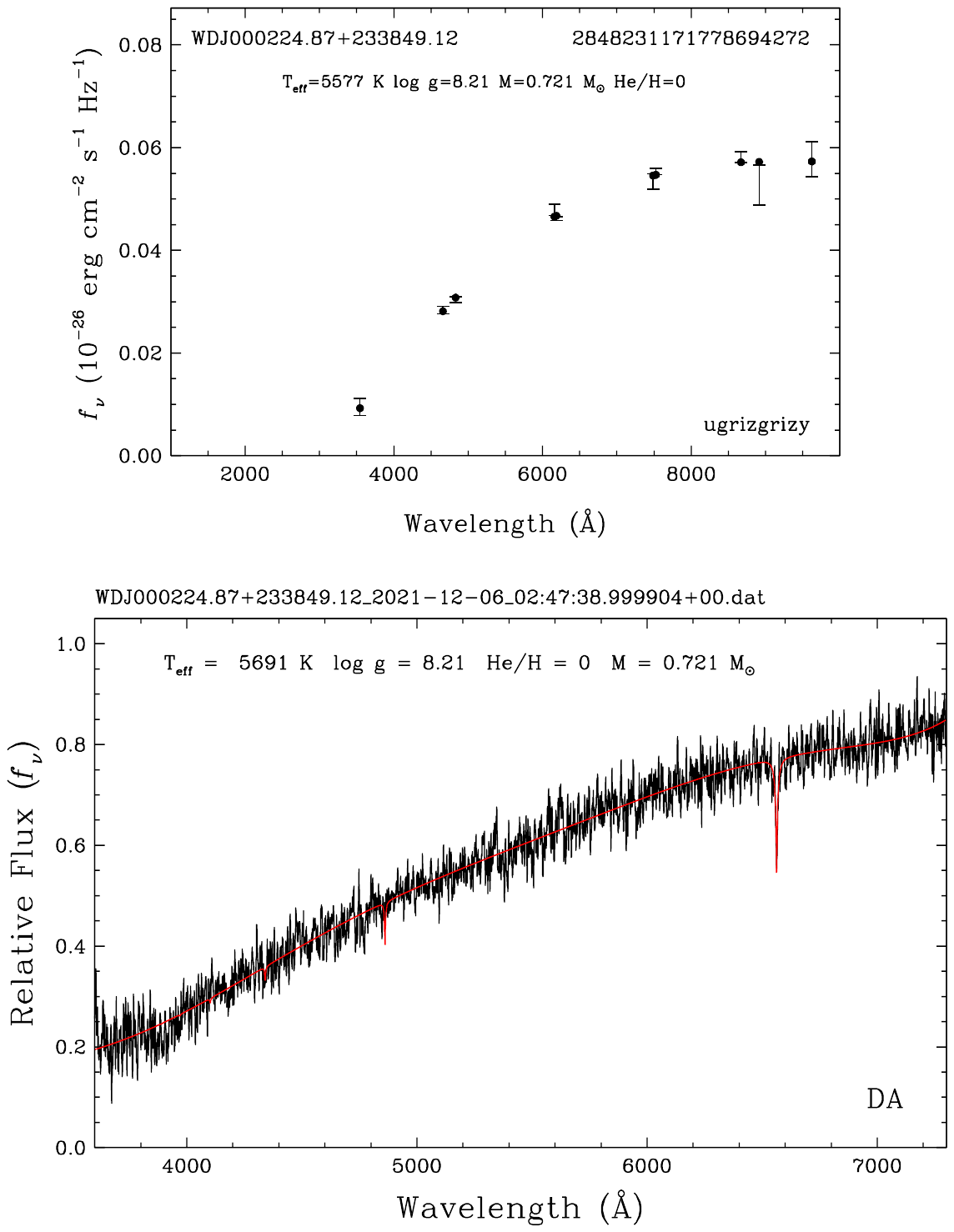}
\includegraphics[width=2.55in, clip=true, trim=0.4in 0.8in 0.2in 1.1in]{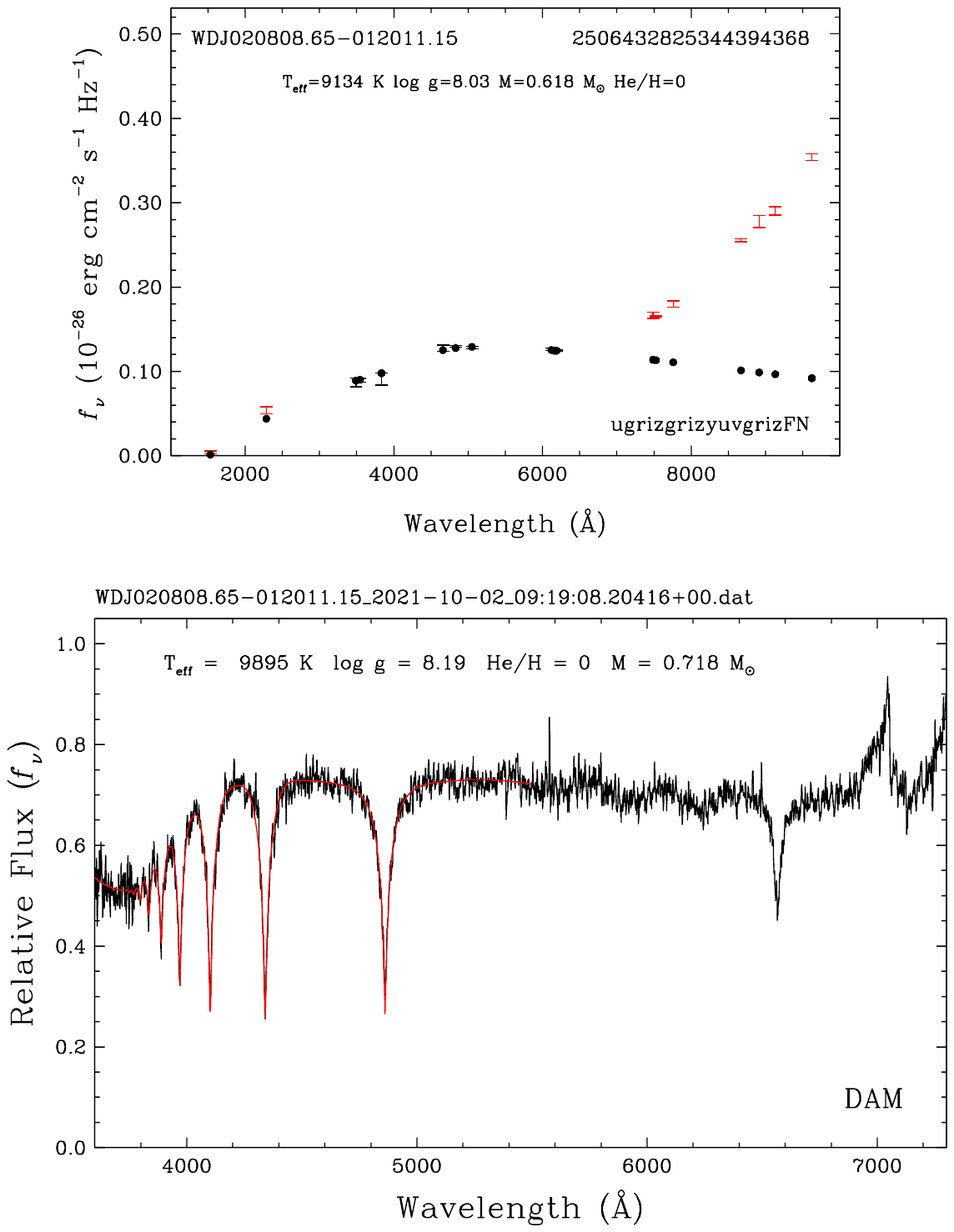}
\caption{Model fits to three cool DA white dwarfs. The top panels show the best-fitting pure hydrogen (filled dots) atmosphere models to the photometry (error bars). These panels also include the white dwarf name, Gaia Source ID, the file name of the spectrum (including the DESI observation date), and the photometry used in the fitting. The bottom panels show the fits to the DESI spectra. Since the spectroscopic method becomes unreliable for DA white dwarfs cooler than 6500 K with relatively weak H lines (like the object in the middle), we force the photometric $\log{g}$ on their spectroscopic fits. 
Red-colored photometric errors bars indicate data that are excluded from the model fits, for instance the GALEX data in the UV, or the red photometric bands for the objects contaminated by an M dwarf companion.
Here and in the following figures, we smooth the DESI spectra by 5 points (with 0.8 \AA\ per point) for display purposes only.}
\label{figda}
\end{figure*}
 
A comparison between the observed sequences and the evolutionary tracks shown in Figure \ref{figcmd} for various compositions is informative.
The DA sequence follows the predictions from the pure H atmosphere models, and extends down to $M_g = 15.5$ and $g-z=1.0$, which correspond
to $T_{\rm eff}\approx5000$ K. H$\alpha$ disappears below this temperature, hence those stars become DC white dwarfs beyond those limits.
A significant number of low-mass white dwarfs are also visible above the main DA and DC sequences shown in these diagrams. 

The DC sequence has to be treated carefully. Since H$\alpha$ is visible above 5000 K, DC stars hotter than that temperature must have helium-rich atmospheres. The DC sequence indeed follows the He-rich tracks closely down to about $M_g=14.5$. However, the majority of the coolest DC white dwarfs with $g-z\geq0.8$ instead follow the pure H model sequence, suggesting that they must have atmospheres dominated by hydrogen.

The bottom panels in Figure \ref{figcmd} show the color-magnitude diagrams for all DQ and DZ white dwarfs in the sample. These sequences are also informative: even though both DQ and DZ sequences extend down to
$M_g\sim16$, both samples are dominated by objects with $M_g\leq15.0$. The relative scarcity of DQ and DZ white dwarfs below this limit suggests that we are seeing the impacts of spectral
evolution directly in color-magnitude diagrams, where there is a scarcity of helium-rich atmospheres among the faintest and coolest white dwarfs.
We revisit the discussion of spectral evolution in Section \ref{secdis} below. 

\subsection{DA White Dwarfs}

We use both photometric and spectroscopic methods \citep{bergeron92a,bergeron19} to obtain independent constraints on the physical parameters of the DA white dwarfs, with the exclusion of DAZ stars and magnetic DAs, which are discussed in Section \ref{sec:DZ} and \ref{secmag}, respectively. We rely on the photometry from the SDSS, Pan-STARRS, and SkyMapper, and astrometry from Gaia DR3, with distance measurements from \citet{bailer21}. We include Gaia photometry if no other photometric data is available. We also include GALEX photometry in the figures, but not in the fits themselves. We omit problematic photometry (shown in red in the figures), where contamination from nearby objects is clearly visible. We correct for reddening based on the mean $A_V$ for each object from \citet{gentile21}. 

Spectroscopic model fits for DA white dwarfs usually suffer from a degeneracy between a hot and a cold solution. To break this degeneracy, we use the photometric temperature as the initial seed for the spectroscopic fits, but for some objects we have to force the hotter solution to better fit the spectrum. For objects with multiple DESI spectra, we always use the highest S/N spectrum in our fits. Additional details of our fitting procedures are described in \citet{kilic26}. For brevity, here we discuss the special considerations that are required for the analysis of cool DAs (and the other spectral types below). We provide the model fits for all stars in our sample on Zenodo, which can be accessed via the DOI \href{https://doi.org/10.5281/zenodo.20775938}{10.5281/zenodo.20775938}.

For DA white dwarfs, we rely on the LTE models described in \citet{tremblay09} with improved Stark broadening profiles. These pure H models are similar to those used in our analysis of the hot white dwarf sample in DESI DR1, and we refer the reader to section 3 in \citet{kilic26} for additional details. The main difference here is that our DA model grid covers $T_{\rm eff}$ ranging from 1500 to 150,000 K and $\log{g}$ from 6.0 to 9.5. Convection is included using the ML2/$\alpha=0.7$ parameterization of the mixing-length theory, but we include the 3D hydrodynamical corrections from \citet{tremblay13,tremblay15} for all of the spectroscopic parameters presented in our figures and tables. Note that non-ideal effects in the equation-of-state are also included in the cooler models following \citet{blouin18a}, but these are not expected to play a major role in the context of DA stars, which are all hotter than $\approx 5000$ K.

To calculate masses and cooling ages, we use the \citet{bedard20} evolutionary models with C/O cores, $q({\rm He})\equiv M_{\rm He}/M_{\star}=10^{-2}$, and $q({\rm H})=10^{-4}$ and $10^{-10}$, which are representative of H- and He-atmosphere white dwarfs, respectively. In Paper I, we relied on the \citet{althaus13} evolutionary models for He-core white dwarfs if the mass derived from C/O core models is below $0.35~M_\odot$. Adopting the same limit for the cool white dwarf sample produces an artificial over-density of objects around $0.35~M_\odot$. Here, we instead utilize the He-core models below $M=0.20~M_\odot$. The resulting parameters for low-mass DAs should not be trusted in any case, since they are based on the assumption of a single star in the system, while many are found in binary systems \citep{marsh95}.

\begin{deluxetable*}{lrcccccccc}
\tablecolumns{10} \tablewidth{0pt}
\tablefontsize{\tiny}
\tablecaption{Physical Parameters of the Cool White Dwarfs in DESI DR1.\label{tabpar}}
\tablehead{\colhead{Object} & \colhead{Gaia SourceID} & \colhead{Type} & \colhead{Comp} & \colhead{$T_{\rm eff,phot}$} & \colhead{$\log{g, {\rm phot}}$} & \colhead{Mass,phot} & \colhead{$T_{\rm eff,spec}$} & \colhead{$\log{g,{\rm spec}}$} & \colhead{Mass,spec}\\
 & & & & (K) & (cm s$^{-2}$) & ($M_\odot$) & (K) & (cm s$^{-2}$) & ($M_\odot$)}
\startdata
WDJ000006.67+144122.20   & 2771972786192217600 & DC & logH/He=$-2.99$ &  7854 $\pm$ 129 & 8.276 $\pm$ 0.057 & 0.752 $\pm$ 0.052 &\nodata&\nodata&\nodata\\
WDJ000006.76$-$004653.87 & 2449845669146456576 & DA & H                 & 10918 $\pm$ 166 & 8.020 $\pm$ 0.068 & 0.614 $\pm$ 0.056 & 11584 & 8.144 & 0.692\\
WDJ000007.82+304606.35   & 2873402974373195008 & DA & H                 &  7649 $\pm$ 116 & 7.473 $\pm$ 0.087 & 0.336 $\pm$ 0.047 &  7798 & 7.710 & 0.438\\
WDJ000011.43$-$040315.24 & 2447815253423324544 & DC & logH/He=$-0.10$ &  5577 $\pm$  12 & 8.152 $\pm$ 0.010 & 0.666 $\pm$ 0.009 &\nodata&\nodata&\nodata\\
WDJ000011.56$-$085008.38 & 2441060055844546816 & DQ & logC/He=$-5.74$ &  7649 $\pm$  58 & 7.620 $\pm$ 0.058 & 0.377 $\pm$ 0.037 &  7649 & 7.620 & 0.377\\
WDJ000012.01$-$030831.10 & 2448158399834236928 & DA & H                 &  8037 $\pm$ 164 & 8.247 $\pm$ 0.131 & 0.750 $\pm$ 0.116 &  8230 & 8.049 & 0.625\\
WDJ000014.89$-$120216.48 & 2421622820569194240 & DC & logH/He=$-4.48$ & 10173 $\pm$ 443 & 7.838 $\pm$ 0.163 & 0.490 $\pm$ 0.118 &\nodata&\nodata&\nodata\\
WDJ000022.54$-$105142.20 & 2422606780396753024 & DA & H                 &  8345 $\pm$  78 & 8.002 $\pm$ 0.044 & 0.597 $\pm$ 0.036 &  8333 & 7.994 & 0.592\\
WDJ000023.46+021434.32   & 2738745819678741504 & DA & H                 & 10159 $\pm$ 170 & 7.108 $\pm$ 0.266 & 0.240 $\pm$ 0.065 & 10380 & 7.673 & 0.430\\
WDJ000024.51+214108.98   & 2846816370896369280 & DA & H                 &  7506 $\pm$ 106 & 7.968 $\pm$ 0.062 & 0.574 $\pm$ 0.051 &  7232 & 7.694 & 0.428\\
\enddata
\tablecomments{This table is available in its entirety in machine-readable format in the online journal. A portion is shown here for guidance regarding its form and content.}
\end{deluxetable*}

Figure \ref{figda} shows our photometric (top) and spectroscopic (bottom panels) model fits to three cool DAs, including a DA+M dwarf.
The top left panel shows the SDSS $ugriz$, Pan-STARRS $grizy$, and SkyMapper $uvgriz$ photometry (error bars)
along with the predicted fluxes (filled dots) from the best-fitting pure hydrogen atmosphere model for WDJ002049.41+004434.90.
The labels in the same panel give the white dwarf name, Gaia Source ID, and the photometry used in the fitting. 

Remarkably, this object has photometry measurements available in 18 different filters, including GALEX FUV and NUV. The best-fitting photometric model has
$T_{\rm eff}=8861\pm49$ K and $M=0.889\pm0.007~M_\odot$. Even though the GALEX photometry is not used in the model fits, the best-fitting model provides an excellent match in the UV. The bottom left panel shows the spectroscopic fit to the DESI spectrum of the same star. A major difference between the spectroscopic fits presented here and in the hot white dwarf sample in Paper I is that we now use the entire spectrum for all spectral types, and
not just the spectral ranges shown in the figures presented in Paper I. 
The spectroscopic model fit here provides an excellent match to the DESI spectrum, though the best-fitting model has a slightly larger surface gravity and mass.

The middle panels in Figure \ref{figda} show our model fits to a cooler DA white dwarf, WDJ000224.87+233849.12, where only a weak H$\alpha$ feature is visible in the DESI spectrum. The spectroscopic method becomes unreliable for cooler DA white dwarfs, since it is difficult to constrain
both the effective temperature and surface gravity based on a few weak lines observed in their spectra. 
We use the photometric method to constrain $\log{g}$, and force this solution in the spectroscopic fit if the spectroscopic $T_{\rm eff}<6500$ K, or $T_{\rm eff}<7000$ K and H$\alpha$ equivalent width $<20$ \AA. Even though surface gravities and masses are forced to be identical, this provides an independent spectroscopic temperature estimate. The resulting spectroscopic fit with a slightly hotter temperature provides an excellent match to the DESI spectrum of this target. 

Given our color selection, our sample includes a large number of white dwarf + M dwarfs. The right panels in Figure \ref{figda}
show our model fits to WDJ020808.65$-$012011.15. Here the contamination from the M dwarf is clearly visible in the photometry beyond the $r$--band and in the DESI spectrum beyond 6000 \AA. To avoid contamination problems in our model fits, we exclude
all photometry beyond the $r$--band and restrict our spectroscopic model fits to $\lambda<5500$ \AA\ for all white dwarf + M dwarf systems including DA+M, DB+M, and DO+M. Our photometric
and spectroscopic model fits provide excellent matches to the blue portion of the spectral energy distribution of this DA white dwarf. Note that some of these white dwarf + M dwarf systems may
be chance alignments. Our main interest in this paper is constraining the parameters of the white dwarfs in each system, and figuring out which
objects are chance alignments is beyond the scope of this paper.

\citet{rolland18} found 28 He-DA candidates in the SDSS Data Release 7 white dwarf catalog \citep{kleinman13}. Adding the 9 He-DA found in Paper I, we identify a total of 74 He-DA white dwarfs in DESI DR1, all of which are found in the temperature range of $\sim$8000 to 11,500 K. We rely on a model grid of He-atmosphere white dwarfs (see Section \ref{secdc}) with $\log$ H/He = $-5$ to $-2$ by steps of 1 dex. Given the relatively broad and shallow H lines in these stars, we force the photometric temperature and surface gravity, but fit the H/He ratio using the DESI spectrum. 

Table \ref{tabpar} presents the physical parameters for all white dwarfs in our sample, including the results from the photometric and spectroscopic fits. For some spectral types, only one set of solutions is presented. For example, only the photometric parameters are given for DC white dwarfs. For DAs and DBs where the photometric $\log{g}$ is forced, we provide both sets of solutions so that an independent spectroscopic temperature
estimate is also available.

\begin{figure}
\center
\includegraphics[width=2.6in, clip=true, trim=1.4in 0.7in 1.75in 1.1in]{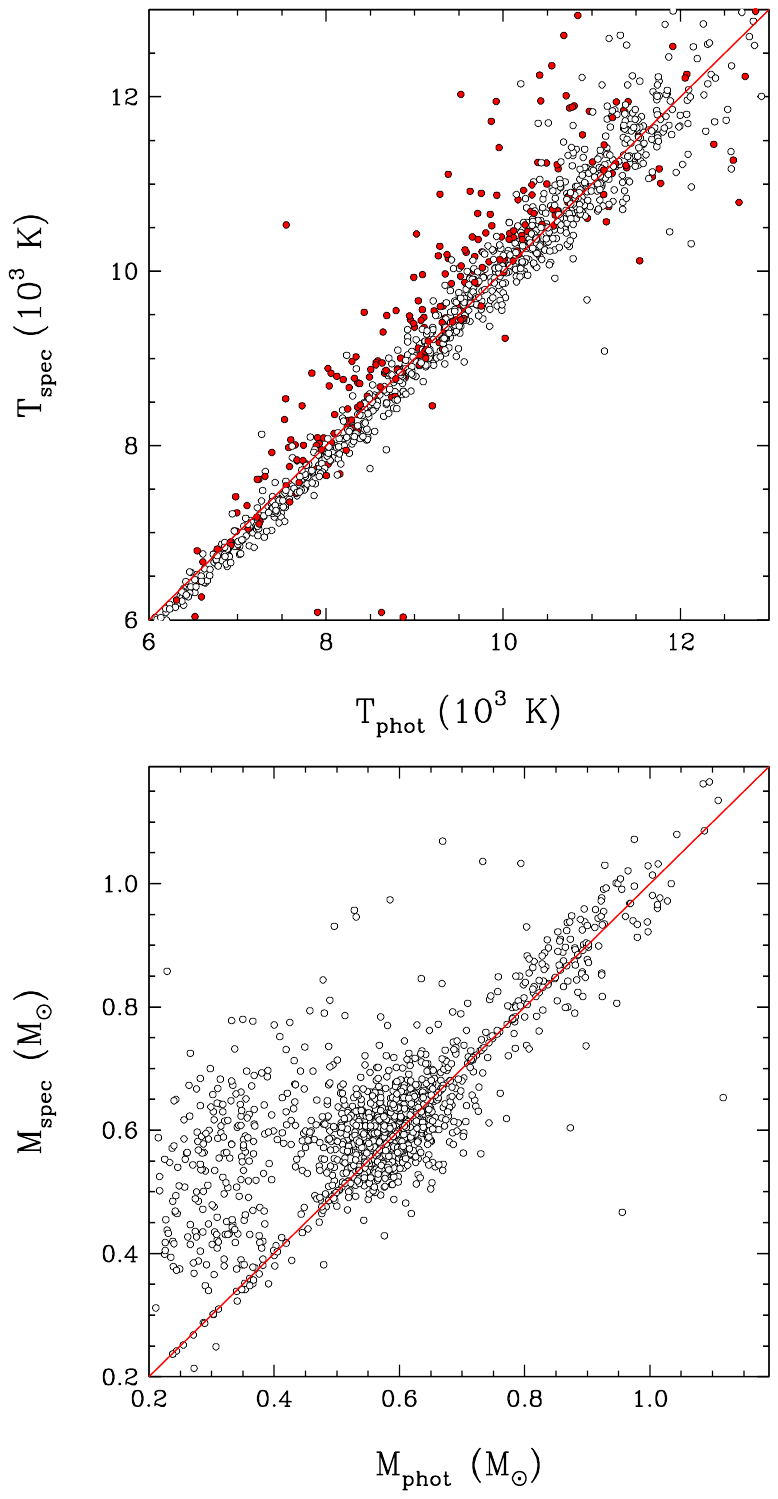}
\caption{Photometric and spectroscopic temperatures and masses for cool DA white dwarfs in DESI DR1. The line of equality in each panel is shown in red. This figure is restricted to objects with distance accuracy better than 10\% and S/N $>20$ spectra in DESI. The red points in the top panel mark objects with photometric masses below $0.5~M_\odot$.}
\label{figcompda} 
\end{figure}

In Paper I, we found a systematic offset between the photometric and spectroscopic solutions for the hot DA white dwarfs in DESI DR1, where
the spectroscopic masses are systematically higher by 0.05-0.06 $M_\odot$. Figure \ref{figcompda} shows a comparison between the photometric
and spectroscopic temperatures and masses for the cool DA white dwarf sample in DESI DR1. Here we restrict the comparison to objects with $<10$\% distance accuracy and S/N $>20$ DESI spectra. Excluding the double degenerate candidates with $M_{\rm phot}<0.5~M_\odot$,
there is a fairly good agreement between the photometric and spectroscopic parameters of this sample. 

Many of the double degenerate candidates appear over-luminous in photometry due to the extra flux from a companion \citep{marsh95}.
For example, for an unresolved binary system of two identical DAs with the same temperature and mass, $M=0.6~M_\odot$, the measured spectroscopic mass will also be $0.6~M_\odot$, but the object would be twice as luminous. This would lead to a radius estimate inflated by $\sqrt{2}$ based on the photometric fit assuming a single star. Hence, the resulting photometric mass would be significantly lower than $M=0.6~M_\odot$. 
If there is a large discrepancy between the temperatures of the binary components, the photometric and spectroscopic methods may provide wildly different parameters for those stars under the assumption of a single star \citep{bedard17}. The red points in the top panel mark objects with photometric masses below $0.5~M_\odot$. Indeed, most of the scatter in the top panel is due to these double degenerate candidates. 

Restricting our sample of DAs to stars with distance errors smaller than 10\%, S/N $>20$ DESI spectra, $T_{\rm eff}= 7000$-11,000 K (the lower limit is set by the fact that the spectroscopic method becomes unreliable below that temperature for DAs), and
$M\geq0.5~M_\odot$ (to remove the majority of the double degenerate candidates), the average differences between the spectroscopic and
photometric temperature and mass are 26 K and $0.006~M_\odot$, respectively.  Hence, unlike
the hot DA white dwarf sample, we do not find significant systematic differences between the photometric and spectroscopic parameters for the cool DA sample in DESI (excluding the binary candidates). In Paper I, we attributed the systematic offsets to problems with the broad hydrogen line profiles
in DESI spectroscopy data. The hydrogen lines get significantly weaker in cool DAs; compare the example DA spectra shown in
Figure \ref{figda} here and the hot DA shown in Figure 4 of \citet{kilic26}. Hence, DESI's calibration problems for broad lines appear to have
a negligible effect for cooler DAs with narrower and weaker lines. 

\subsection{DB White Dwarfs}

For DB white dwarfs, we follow a procedure similar to that used for DA stars, except that, if present, hydrogen lines are used to constrain the H/He ratio, and this composition is enforced in the photometric fits, rather than adopting the nearest 0.5 dex value as in Paper I. We rely on the LTE model atmospheres described in \citet{bergeron11}, with improvements to neutral broadening discussed in \citet{genest19} and the improved Stark-broadened \ion{He}{1} profiles from \citet[][the so-called B25 semi-analytical profiles including line dissolution]{tremblay26}. Our previous model grid covered the temperature range from 11,000 to
50,000 K and $\log{g}$ from 7 to 9. However, \ion{He}{1} $\lambda5876$ remains visible below 11,000 K in DESI spectra. In addition, some of the DB spectral energy distributions are likely contaminated. To account for these issues, we extended our model grid all the way down to 6000 K for He-rich atmosphere white dwarfs with $\log$ H/He ranging from $-6.5$ to $-1.5$ in steps of 0.5 dex; we use models with H/He = 0 in cases where no H lines are detected. As explained in Paper I, we also refrain from applying the 3D hydrodynamical corrections \citep{cuka21} to the spectroscopic parameters of DB stars.

\begin{figure}
\center
\includegraphics[width=3in, clip=true, trim=0.4in 0.8in 0.3in 1.1in]{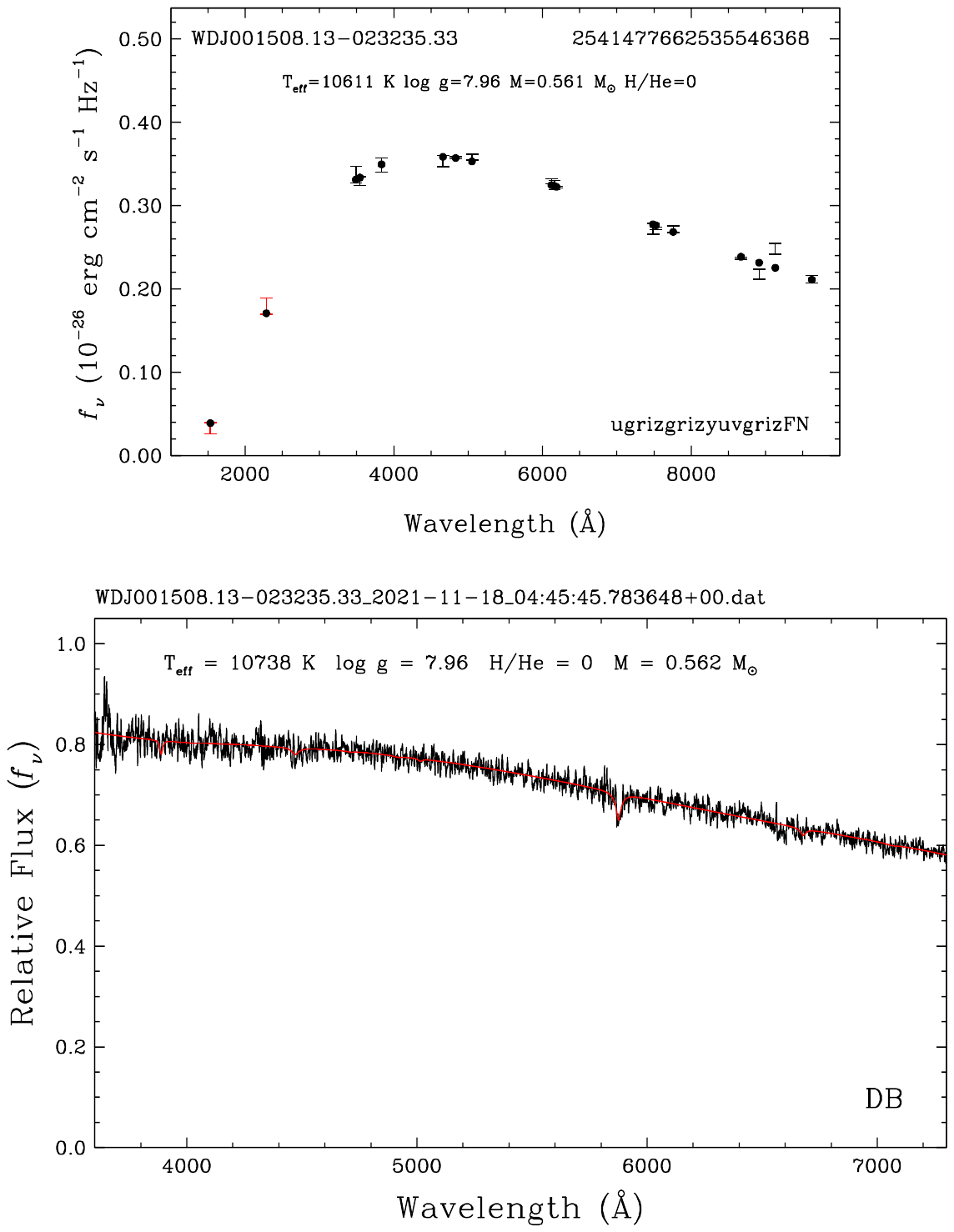}
\caption{Example model fits to a cool DB white dwarf, where the only visible He line at 5876 \AA\ is too weak for meaningful constraints. Here, we force the photometric $\log{g}$ in the spectroscopic fit.}
\label{figdb}
\end{figure}

\begin{figure*}
\hspace{-0.3in}
\includegraphics[width=2.55in, clip=true, trim=0.4in 0.8in 0.1in 1.1in]{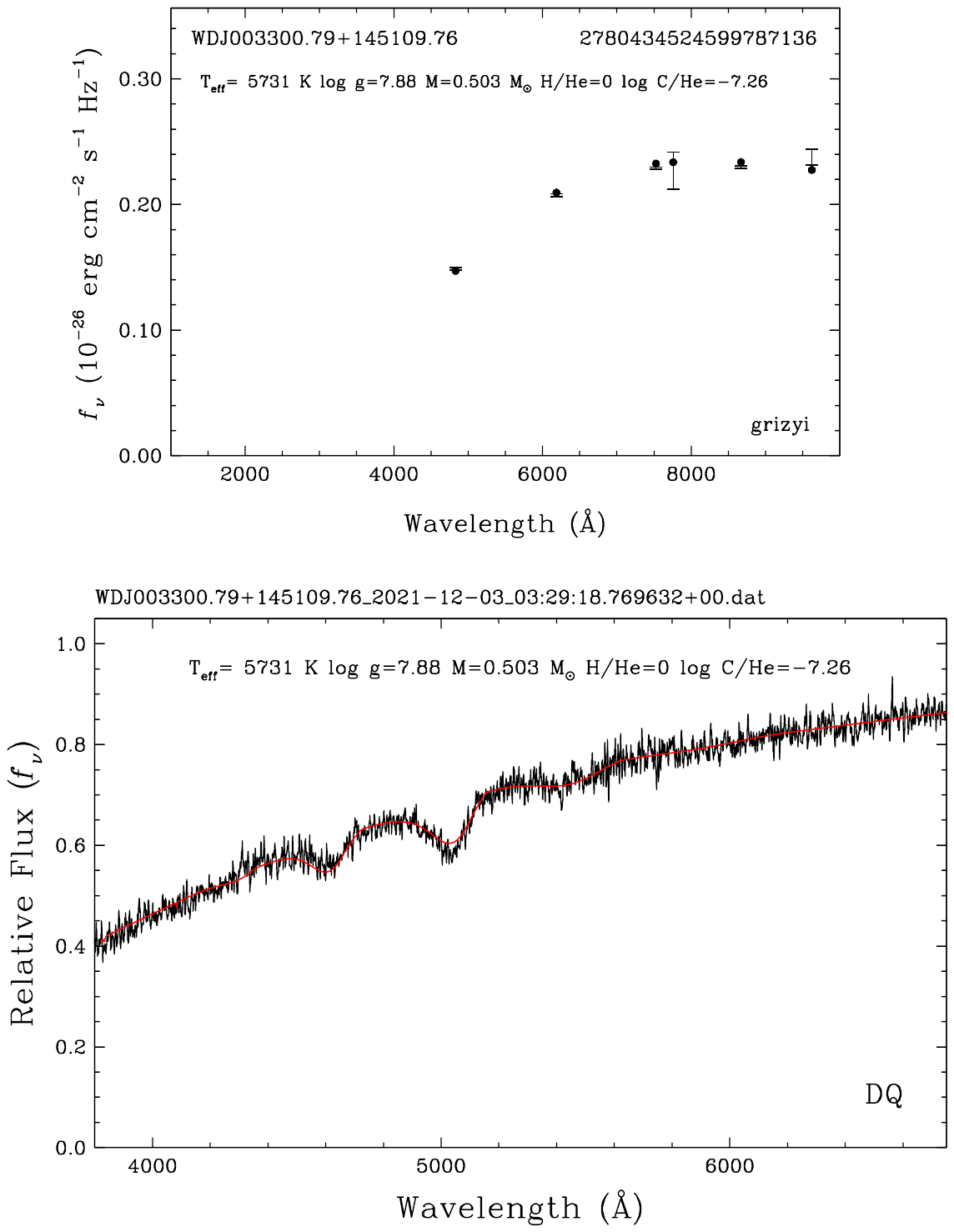}
\includegraphics[width=2.55in, clip=true, trim=0.4in 0.8in 0.1in 1.1in]{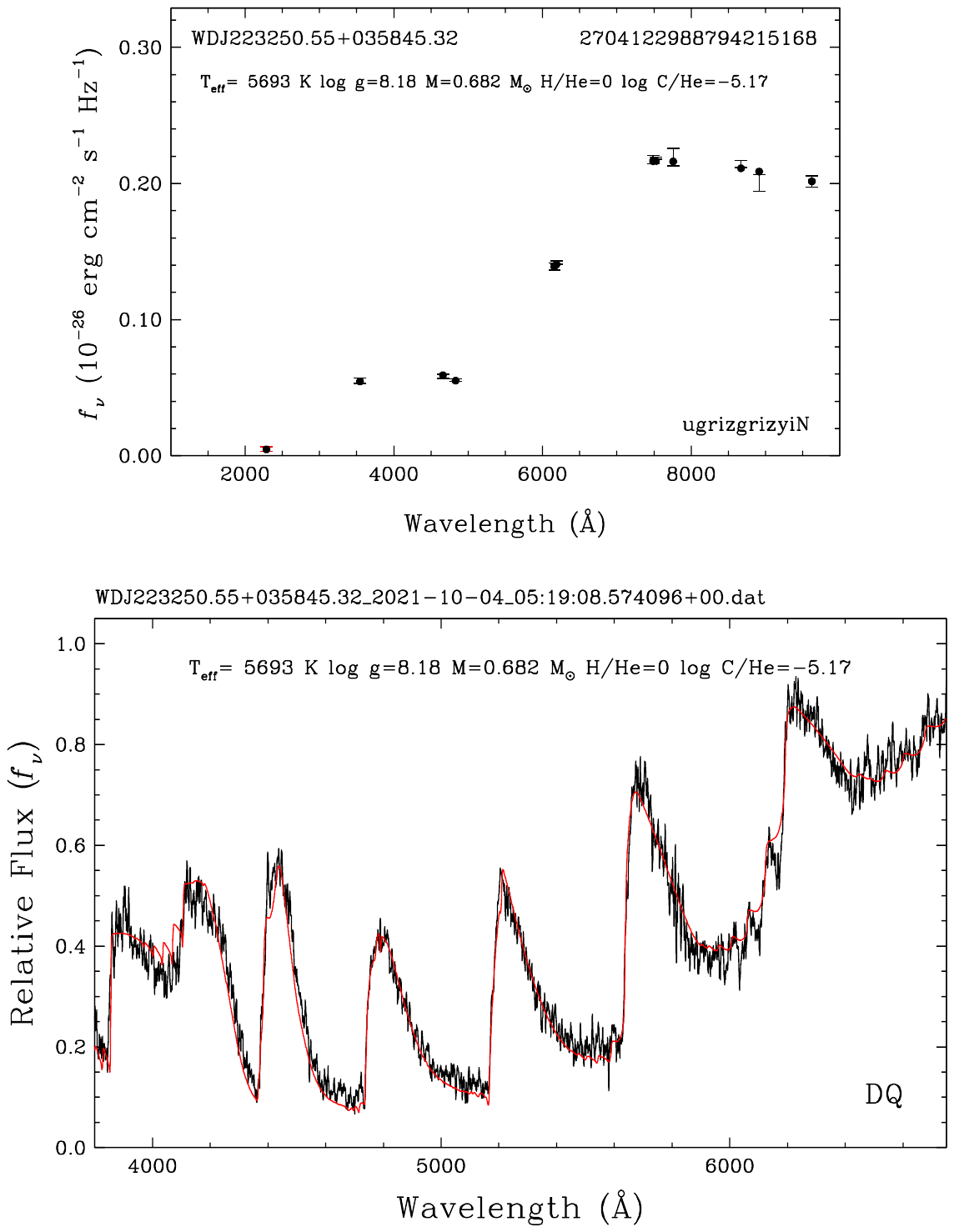}
\includegraphics[width=2.55in, clip=true, trim=0.4in 0.8in 0.1in 1.1in]{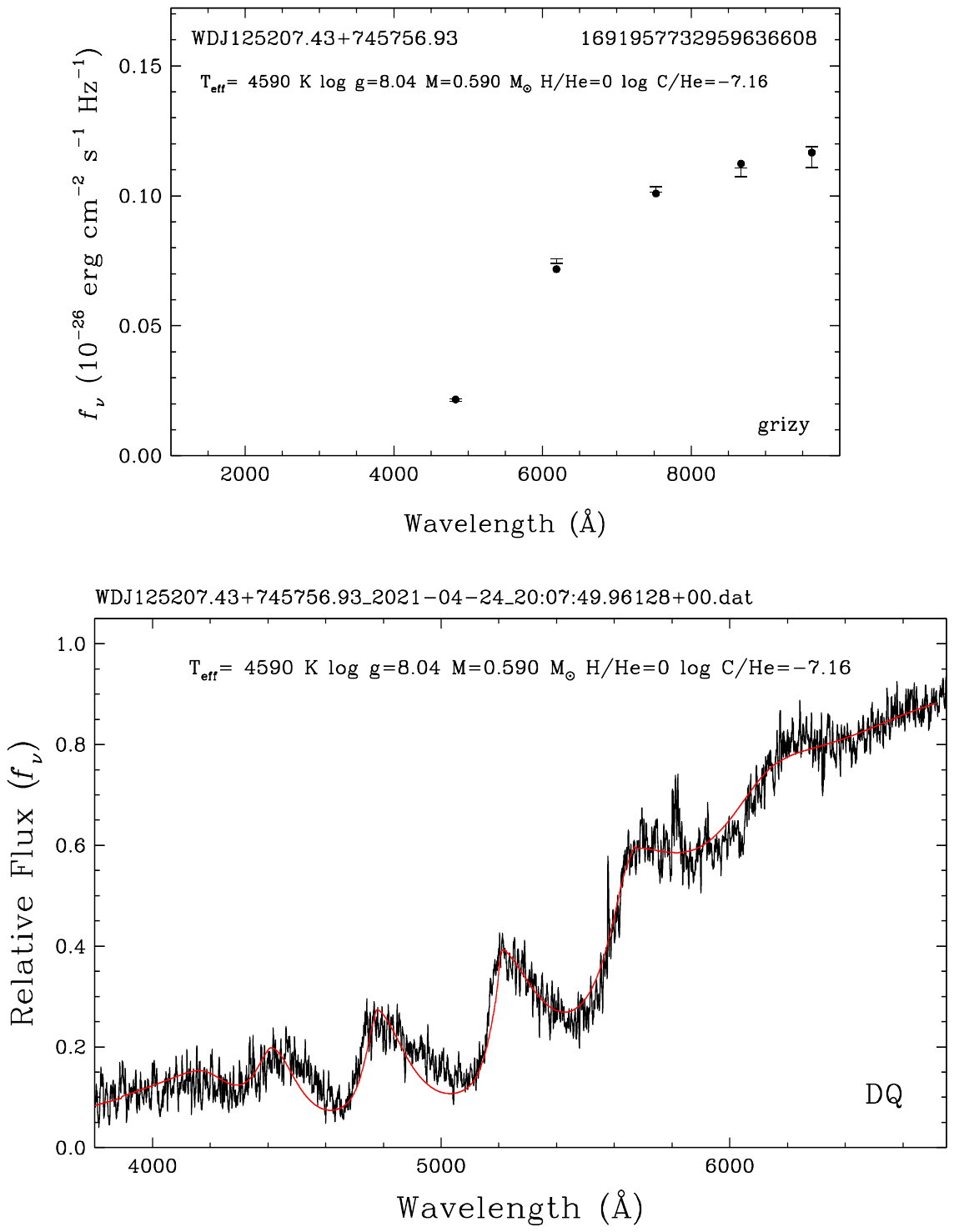}
\caption{Model fits to three cool DQ white dwarfs. The atmospheric models provide an excellent match to both photometry and spectroscopy for these stars.}
\label{figdq}
\end{figure*}

DB white dwarfs hotter than about 13,000 K show relatively strong He lines, which enable independent spectroscopic and photometric fits on their
parameters. However, cooler DBs show diminishingly weaker lines. Hence, just like the cool DA stars, the spectroscopic method becomes unreliable for
the cooler DBs. \citet[][see also \citealt{genest19}]{tremblay26} showed that this limit corresponds to an equivalent width threshold of about 5 \AA\ for the \ion{He}{1} $\lambda4471$
line. Even though the spectroscopic method does not provide reliable constraints on the surface gravities of such stars with weak lines, the spectrum
still contains information on the temperature. By forcing the photometric $\log{g}$ in the spectroscopic fits, we are able to obtain independent spectroscopic constraints on the effective temperatures of these stars.

Fits to hot DB stars have already been presented in Paper I, while Figure \ref{figdb} shows example model fits to the DB white dwarf WDJ001508.13$-$023235.33, where the He lines are too weak for meaningful
constraints on surface gravity. The photometric spectral energy distribution of this star is best-fit by a model with $T_{\rm eff}= 10,612\pm71$ K
and $\log{g}=7.965\pm0.015$. Forcing this surface gravity in the spectroscopic solution, the DESI spectrum of this object is best explained
by a model with a slightly higher $T_{\rm eff}=10,738$ K. This slight difference in temperature also leads to a slightly different mass based on the evolutionary models. 

We identify 14 objects with H and He lines for which our model fits under the assumption of single stars fail to provide acceptable fits. These are unresolved DA+DB binaries. We include our model fits assuming single stars in the \href{https://doi.org/10.5281/zenodo.20775938}{Zenodo archive}, but they are also fitted as binary systems and discussed further in Section \ref{secdqda}. We do not discuss the DB stars further because they are mostly covered in Paper I (the hot white dwarf sample in DESI).

\subsection{DQ White Dwarfs} 

There are two types of DQ white dwarfs. The first class consists of warm DQs with $T_{\rm eff}\gtrsim11,000$ K and $M\gtrsim1~M_\odot$. 
The DQ sample in Paper I is dominated by warm DQs with atomic C lines. Given their unusual compositions, masses, and kinematics, warm DQs are likely merger remnants \citep{dunlap15,cheng19,shen23,bedard24,kilic25b,ould26}. The second class consists of classical DQs that are an order of magnitude more prevalent than warm DQs in the local white dwarf samples. These more numerous DQs are a natural result of convective dredge up of carbon in He-atmosphere white dwarfs \citep{pelletier86,bedard22b} and they are found below $T_{\rm eff}\sim10,000$ K \citep{coutu19,koester19}.

For warm DQs, we rely on the He-free model atmosphere grid presented in \citet{kilic25b}. We adopt a similar approach in which the spectrum is fitted to determine the C/H ratio, along with an independent estimate of the effective temperature, while assuming the photometric $\log{g}$. This choice is motivated by our view that current DQ models are not sufficiently reliable to treat the spectroscopic $\log{g}$ as a free parameter. The photometric fit provides independent measurements of $T_{\rm eff}$ and $\log{g}$ (i.e., the radius), with $\log$ C/H constrained by spectroscopy. We iterate between the spectroscopic and photometric fits until a consistent solution is achieved. As for the DB stars, we do not discuss these objects further as they are already covered in Paper I.

We adopt a similar fitting procedure for the cool DQs, except that we begin with the photometric technique to estimate $T_{\rm eff}$ and $\log{g}$, and then force these parameters to fit the C$_2$ Swan bands to constrain the C/He ratio. Given the abundances derived from the spectroscopic fit, we iterate between the photometric and spectroscopic analyses until a consistent solution is obtained. Our analysis relies on an extended version of the model grids described in \citet{blouin19}. Because cool DQs span a wide range of C/He values—decreasing with decreasing temperature—we employ multiple model grids covering the full parameter space, with $T_{\rm eff}$ ranging from 4000 to 16,000 K, $\log{g}$ from 7.0 to 9.0, and $\log$ C/He from $-8.0$ to $-1.0$. We also assume H-free models. While undetectable traces of hydrogen can influence the atmospheric parameters of helium-atmosphere white dwarfs, the free electrons contributed by carbon are generally far more significant in DQ stars. Indeed, \citet{coutu19} investigated the effects of trace hydrogen in DQ model fits and found that its presence has a negligible impact on the derived parameters. Our sample also includes 6 DQZ white dwarfs, which are fit using metal-free DQ models. 

Figure \ref{figdq} shows our model fits to three classical DQs in our sample. The left and middle panels illustrate two DQ stars with very similar temperatures, but significantly different C abundances. The object shown in the middle panels is $\sim$100 times more C-rich than the one in the left panels. This increased C abundance leads to  numerous Swan bands seen in the DESI spectrum. Regardless of the C abundances, the DQ models presented in \citet{blouin19} provide an  excellent match to both the photometry and spectroscopy data for these stars.
 
The right panels in Figure \ref{figdq} show the model fits to one of the coolest DQ white dwarfs ever found, WDJ125207.43+745756.93, with  shifted and rounded features. \citet{kowalski10} demonstrated that the shifted bands in cool DQs are likely the pressure-shifted bands of C$_2$  in the fluid-like atmospheres of these stars. \citet{blouin19b} included a density-driven shift of the electronic transition energy of the Swan bands based on these results and found very good fits to their sample of DQ stars. Here we rely on the same models to fit the shifted Swan bands. Our best-fitting model for this star has $T_{\rm eff}=4590\pm19$ K, $M=0.588\pm0.018~M_\odot$ and $\log$ C/He = $-7.16$, and it provides a good match to the DESI spectrum.

Our sample includes a number of low-mass DQs with $M\sim0.2~M_\odot$. These are probably in unresolved double-degenerate systems, and the resulting fits (under the assumption of a single star) and the C/He abundances should be used with caution. Our sample also includes 11 unresolved DA+DQ systems that display both C$_2$ Swan bands and Balmer lines. We include our DQ model fits under the assumption of single stars in the \href{https://doi.org/10.5281/zenodo.20775938}{Zenodo archive}, but they are also fitted as a binary system and discussed further in Section \ref{secdqda}. 
  
\subsection{Metal-Line White Dwarfs}\label{sec:DZ}

\begin{figure*}
\hspace{-0.3in}
\includegraphics[width=2.5in, clip=true, trim=0.4in 0.8in 0.1in 1.1in]{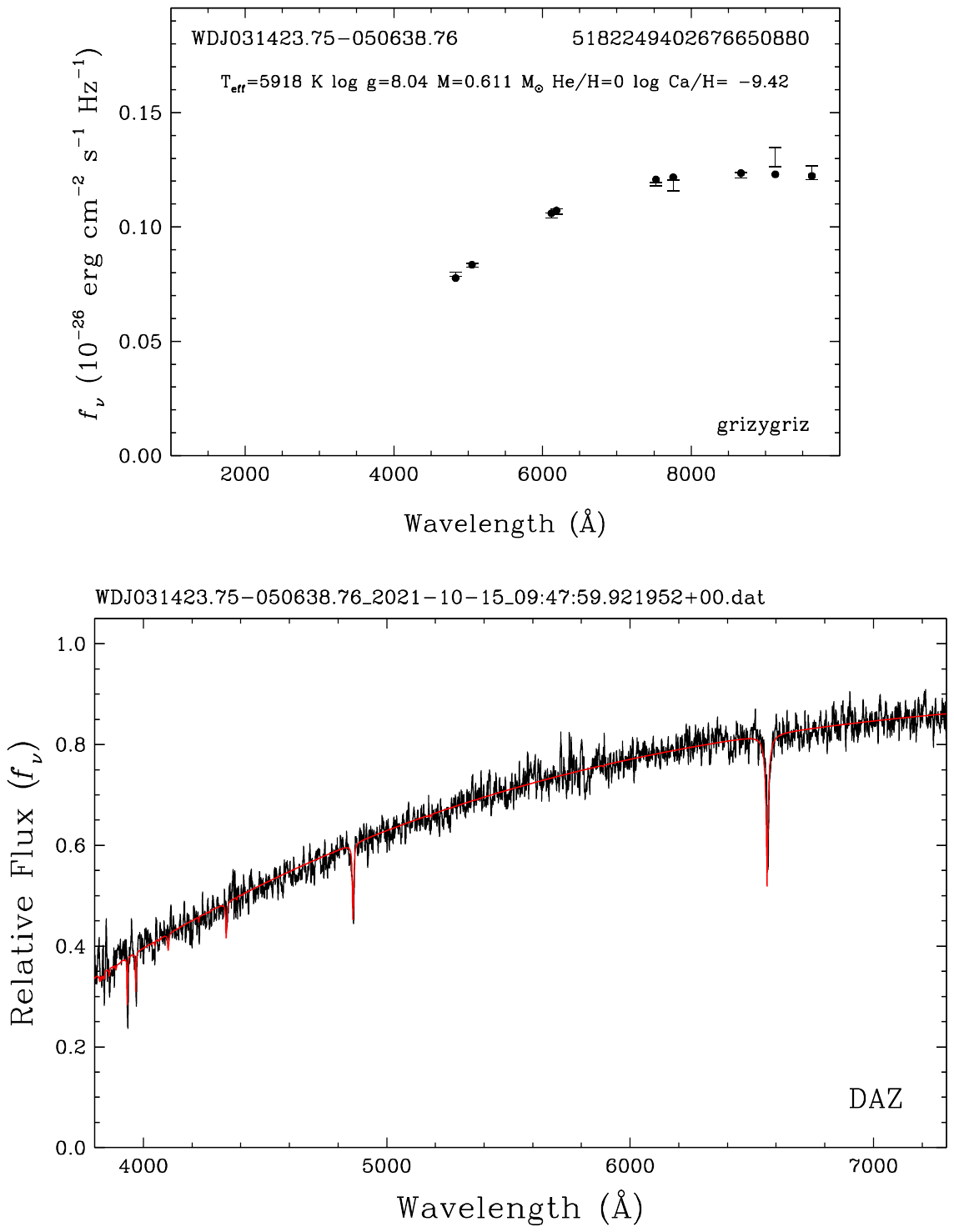}
\includegraphics[width=2.5in, clip=true, trim=0.4in 0.8in 0.1in 1.1in]{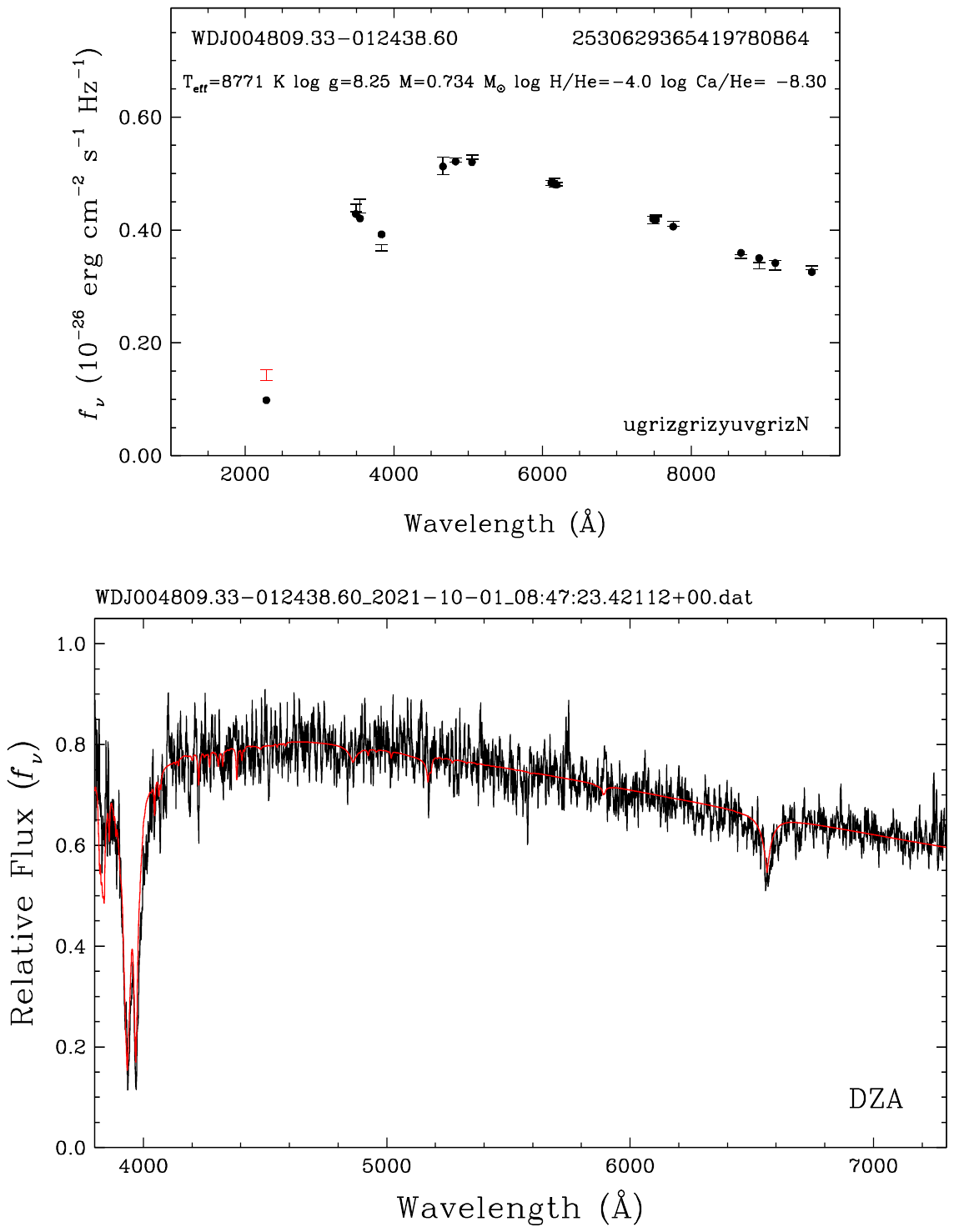}
\includegraphics[width=2.5in, clip=true, trim=0.4in 0.8in 0.1in 1.1in]{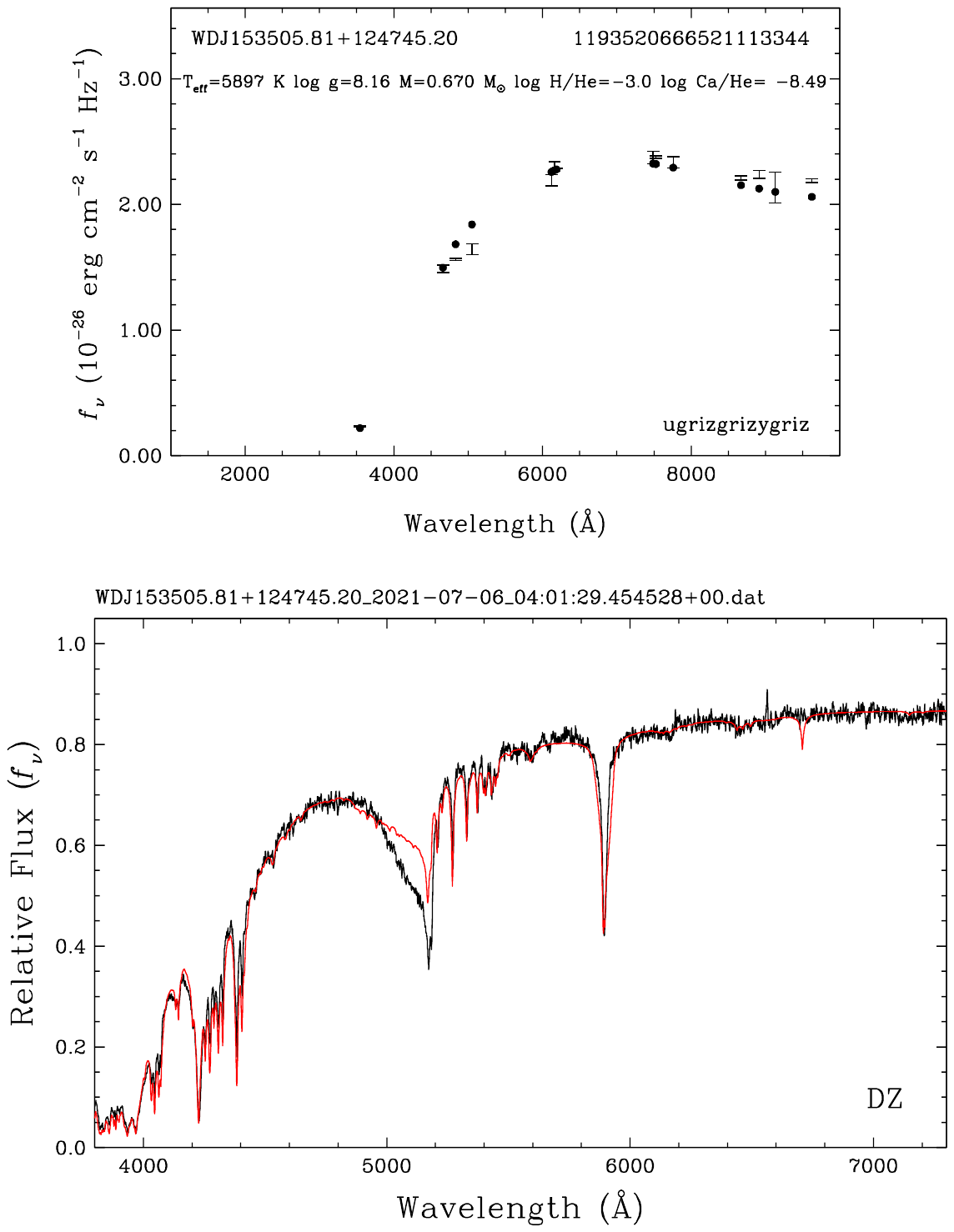}
\caption{Example model fits to DAZ, DZA, and DZ white dwarfs under the assumption of chondritic metal abundance ratios.}
\label{figdz}
\end{figure*}

For all white dwarfs in our sample showing metal lines (DBZ, DBAZ, DAZ, DZA, and DZ), we take a similar approach to our analysis of the DQ white dwarfs and start with the photometric technique to determine the temperature and surface gravity. We then fit the DESI spectrum to constrain Ca/He (or Ca/H) and assume that the abundance ratios of the other heavy elements match the CI chondrites. We then use the models with that abundance to revise the photometric fits, and repeat the photometric and spectroscopic fits until a consistent solution is found.

For the He-atmosphere white dwarfs, we rely on extended versions of the model grids by \citet{blouin18a,blouin18b}, but we include the HeH$^+$ ion as well as the correction to the H$_3^+$ partition function discussed in Section \ref{seccmd}. We use multiple grids covering $T_{\rm eff}$ ranging from 4000 to 19,500 K, $\log{g}$ from 7.0 to 9.0, $\log$ Ca/He from $-12$ to $-6$, and $\log$ H/He from $-6$ to 0, as well as H-free models. We use a separate and specially tailored model grid for the IR-faint DZ WDJ080440.63+223949.68 \citep[see][]{blouin18b}.  
For the H-atmosphere white dwarfs, we rely on He-free models with $T_{\rm eff}$ ranging from 4000 to 9500 K, $\log{g}$ from 7.0 to 9.0, and $\log$ Ca/H from $-10.5$ to $-5.5$. 

One of the major uncertainties in the mass measurements of metal-line white dwarfs is the H/He ratio in the atmosphere. \citet{bergeron19} and \citet{coutu19}
showed that undetectable traces of hydrogen can have a significant effect on the derived masses of He-atmosphere white dwarfs, including DZ stars with no detectable H$\alpha$ feature. For instance, fitting the DESI spectra of DZ white dwarfs with H-free atmosphere models, we find that the average mass is systematically higher than $0.6~M_\odot$. Repeating these fits with atmosphere models with $\log$ H/He = $-5$ or $-6$, the effective temperatures and masses decrease on average by $\sim$200 K and $\sim$$0.04~M_\odot$, respectively. These values are similar to the offsets found by \citet[][see their Figure 5 and 7]{coutu19}. Hence, it is impossible to precisely constrain the mass of an individual DZ star due to the unknown H/He ratio in the atmosphere.

We adopt the following procedure for all metal-line white dwarfs in our sample, which is similar to that used by \citet{coutu19}. For each object, we explore various H/He abundance ratios from our model grid (with a 1 dex resolution in H/He) and pick the best fit solution based on a $\chi^2$ analysis. Luckily, even for DZ stars that do not show H features, the metallic line profiles are still sensitive to the H abundance and they provide an excellent proxy to estimate the H/He ratio. However, in some cases, and in particular for warmer objects, we had to assume H/He ratios based on the visibility limit of H$\alpha$. This visibility limit ranges from $\log$ H/He = $-6$ above $T_{\rm eff}$ = 10,000 K to $-3$ near 6000 K. Hence, large amounts of hydrogen can be hidden in cooler DZs. On the other hand,
if H$\alpha$ is visible, we visually inspect the model fits and tweak the H/He ratio to better match the observed H$\alpha$ line. 

Depending on the dominant element in the atmosphere, metal-line white dwarfs come in various flavors. Figure \ref{figdz} shows example model fits to DAZ, DZA, and DZ white dwarfs to demonstrate the differences between what we can learn from these spectra; examples of DBAZ fits are displayed in Figure 12 of Paper I. 

\begin{figure*}
\hspace{-0.3in}
\includegraphics[width=2.5in, clip=true, trim=0.4in 0.8in 0.1in 1.1in]{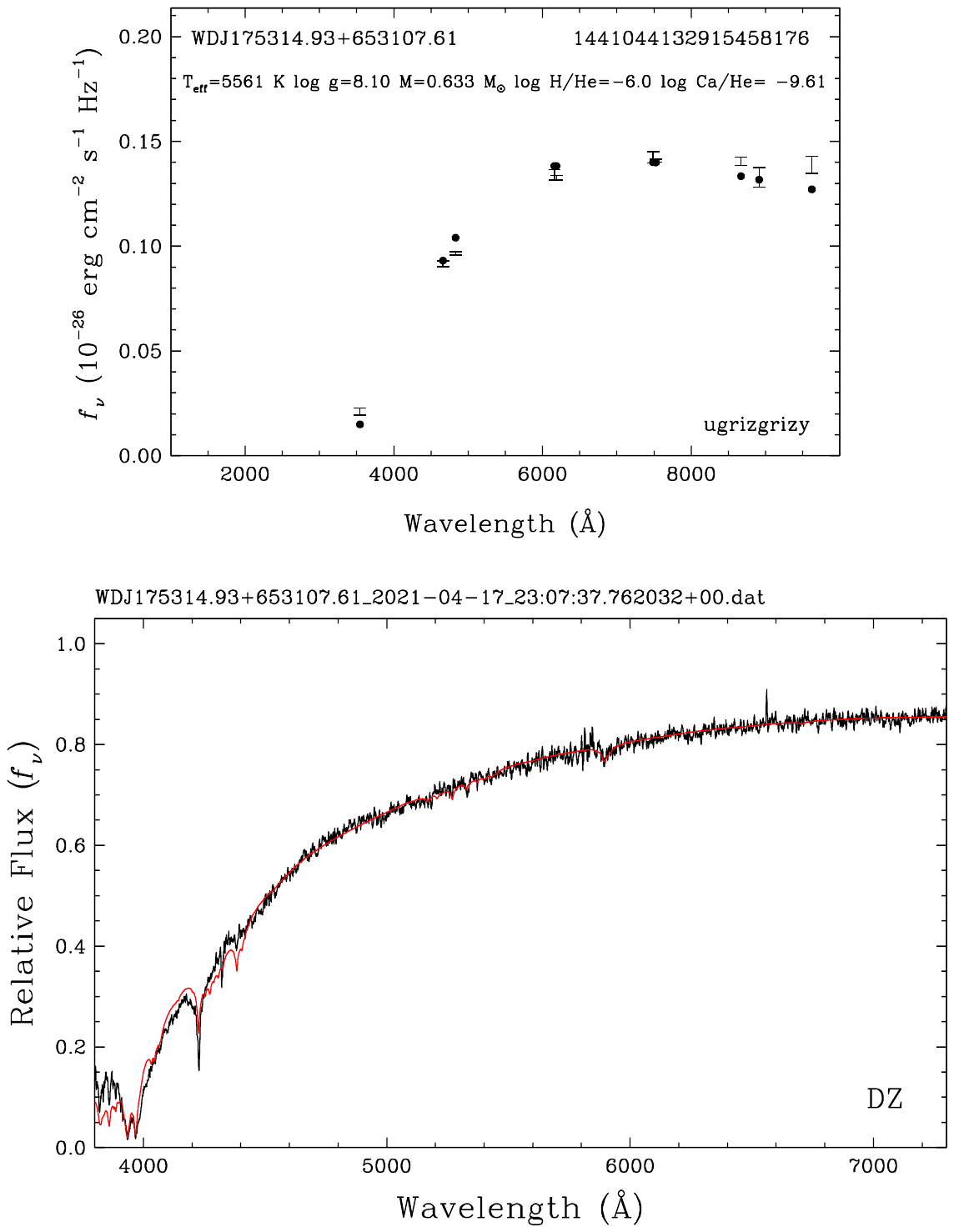}
\includegraphics[width=2.5in, clip=true, trim=0.4in 0.8in 0.1in 1.1in]{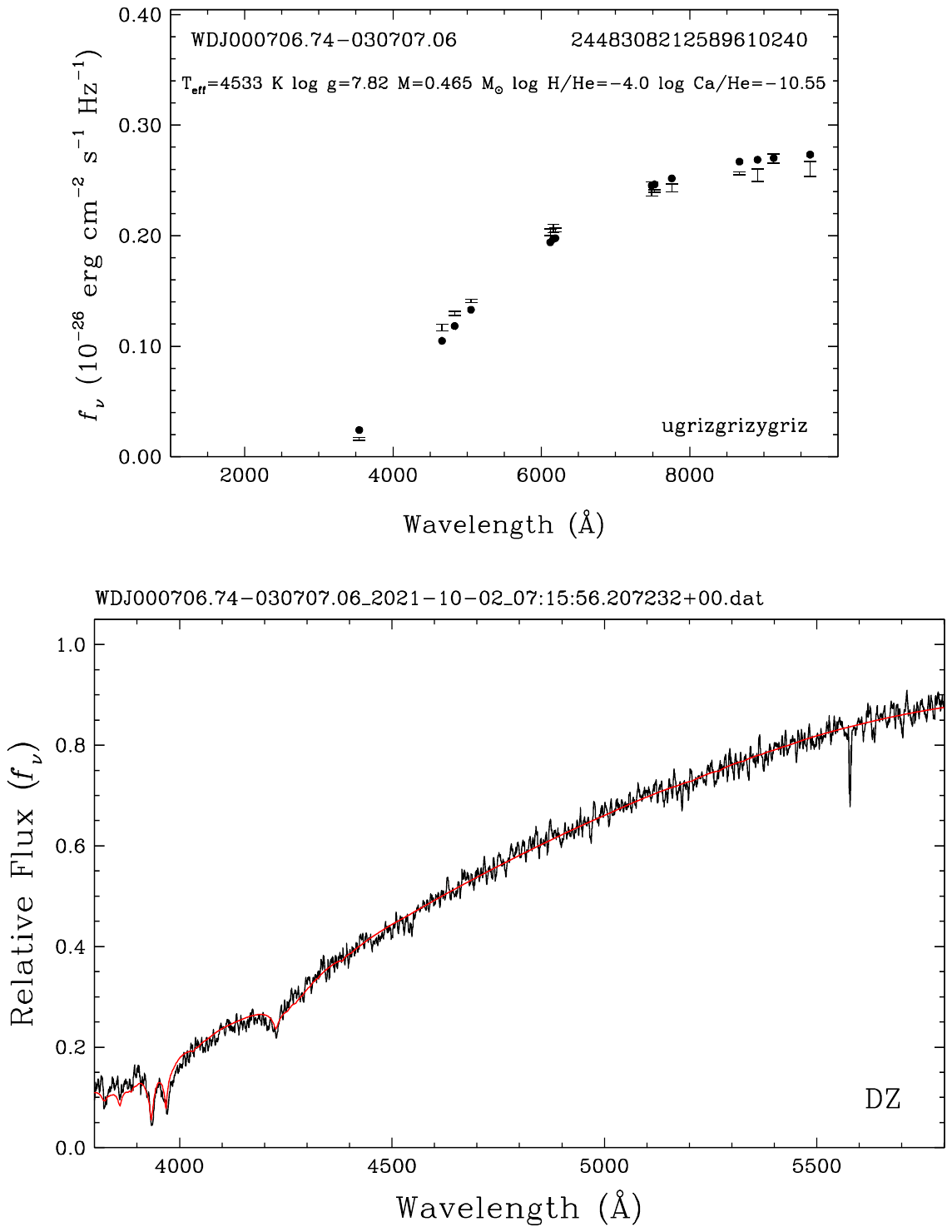}
\includegraphics[width=2.5in, clip=true, trim=0.4in 0.8in 0.1in 1.1in]{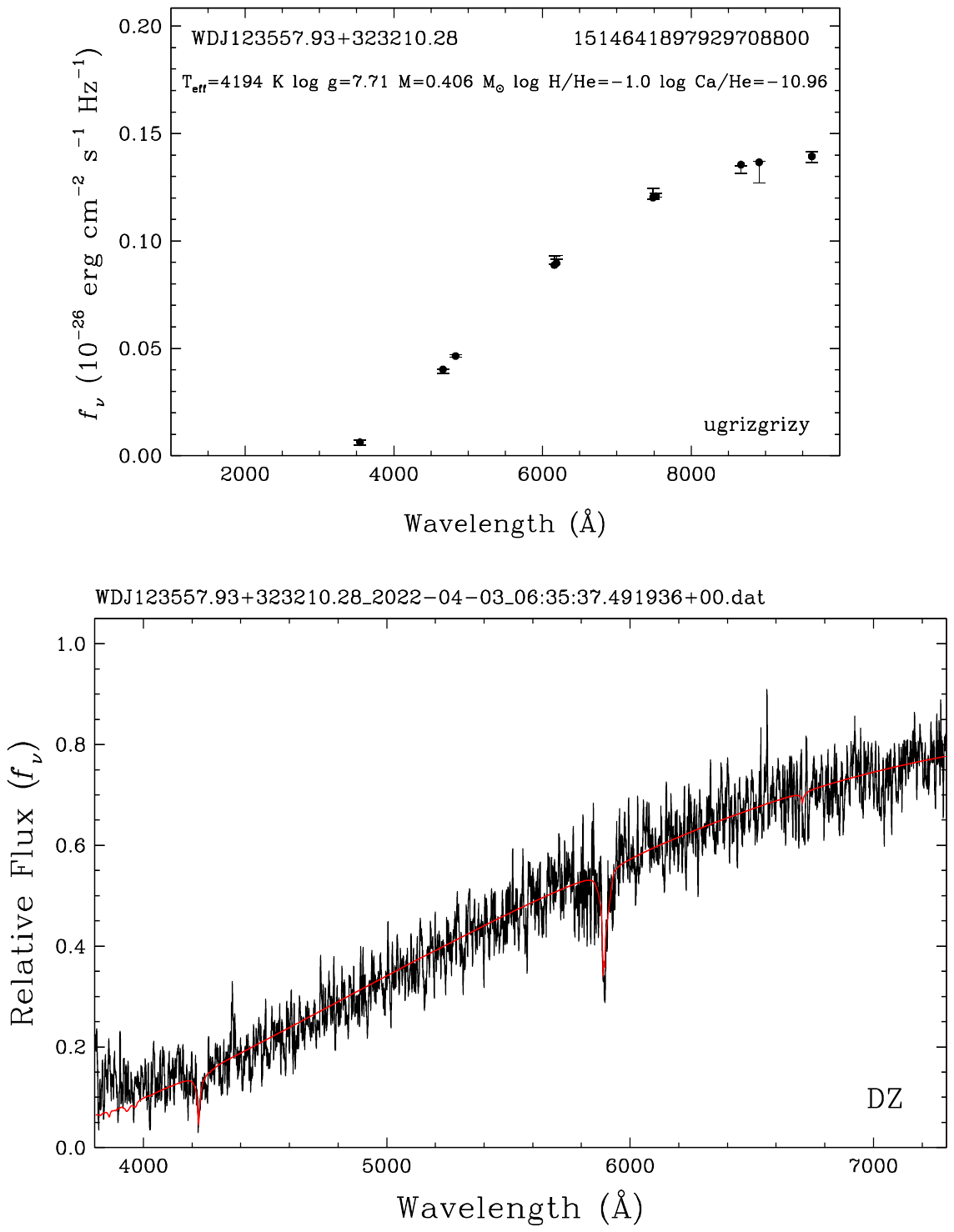}
\caption{Example model fits to three cool DZs with $T_{\rm eff}$ decreasing from left to right.}
\label{figcooldz}
\end{figure*}

The left panels in Figure \ref{figdz} show our model fits to the DAZ WDJ031423.75$-$050638.76. Given the relatively sharp Balmer lines observed in
the DESI spectrum, this object clearly has a H-dominated atmosphere, and a model with $\log {\rm Ca/H}= -9.42$ (and no helium) provides an excellent
match to the DESI spectrum including the relatively weak \ion{Ca}{2} H+K lines.  For comparison, the middle panels show our fits to a DZA white dwarf with relatively strong \ion{Ca}{2} absorption and a broad H$\alpha$ feature, which is heavily broadened through van der Waals interactions in a
He-dominated atmosphere.  Using the DESI spectrum, we constrain the atmospheric composition of WDJ004809.33$-$012438.60 to be $\log {\rm Ca/He}= -8.3$ and $\log {\rm H/He}= -4.0$. Interestingly, the \ion{Ca}{2} H+K features are strong enough to cause a significant dip in flux in the SkyMapper $v$-band photometry (compared to the $u$ and $g$-band data) shown in the top panel. Hence, SkyMapper $v$-band data could be useful for identifying similar DZ white dwarfs with strong Ca absorption features.

The right panels in Figure \ref{figdz} show our model fits to the cool DZ WDJ153505.81+124745.20, where we see additional absorption features from Mg and Na. Our model fits under the assumption of chondritic abundance ratios provide an excellent fit to the overall DESI spectrum, with
the exception of the Mg feature near $\lambda 5175$. The problem with the Mg feature is not unique to this object, cool DZ white dwarfs are known to have higher
Mg/Ca ratios on average than the chondritic ratio \citep{blouin20}. Mg diffuses more slowly than Ca in white dwarf atmospheres. Hence at diffusive
equilibrium, the average DZ accreting chondritic material has a higher Mg/Ca ratio than the chondritic Mg/Ca ratio. In other words, in a large sample of DZs that accrete planetesimals with various Mg/Ca ratios, we expect to find more DZs where the observed Mg \textsc{i} b triplet at $\lambda5175$ is stronger than predicted based on the chondritic ratios. Even though these model fits could be improved by fitting for the abundance ratios of each element, the relatively large number of DZs in the DESI sample prohibits us from performing a tailored analysis of each system.

Figure \ref{figcooldz} shows additional model fits to three cool DZs with
$T_{\rm eff}<6000$ K with temperature decreasing from left to right. The left panels show the model fits to WDJ175314.93+653107.61, which shows
\ion{Ca}{2} H+K lines and \ion{Ca}{1} $\lambda4226$. Based on the strength of these lines and the overall spectral energy distribution, we constrain
the H/He ratio in this $T_{\rm eff}=5561$ K star to $\log$ H/He = $-6$. The middle panels show the model fits to a cooler object, WDJ000706.74$-$030707.06 with $T_{\rm eff}=4533$ K. We see the same \ion{Ca}{1} and \ion{Ca}{2} lines here, and the best-fitting model with $\log$ H/He = $-4$ provides an excellent match to the spectral energy distribution of this star. Finally, the right panels show the model fits to an even cooler DZ, WDJ123557.93+323210.28 with $T_{\rm eff}=4194$ K, which shows \ion{Ca}{1} and \ion{Na}{1} lines. Our model fits require a relatively H-rich atmosphere with $\log$ H/He = $-1$. 

\begin{figure}
\center
\includegraphics[width=3in, clip=true, trim=0.2in 2.7in 0.7in 4.2in]{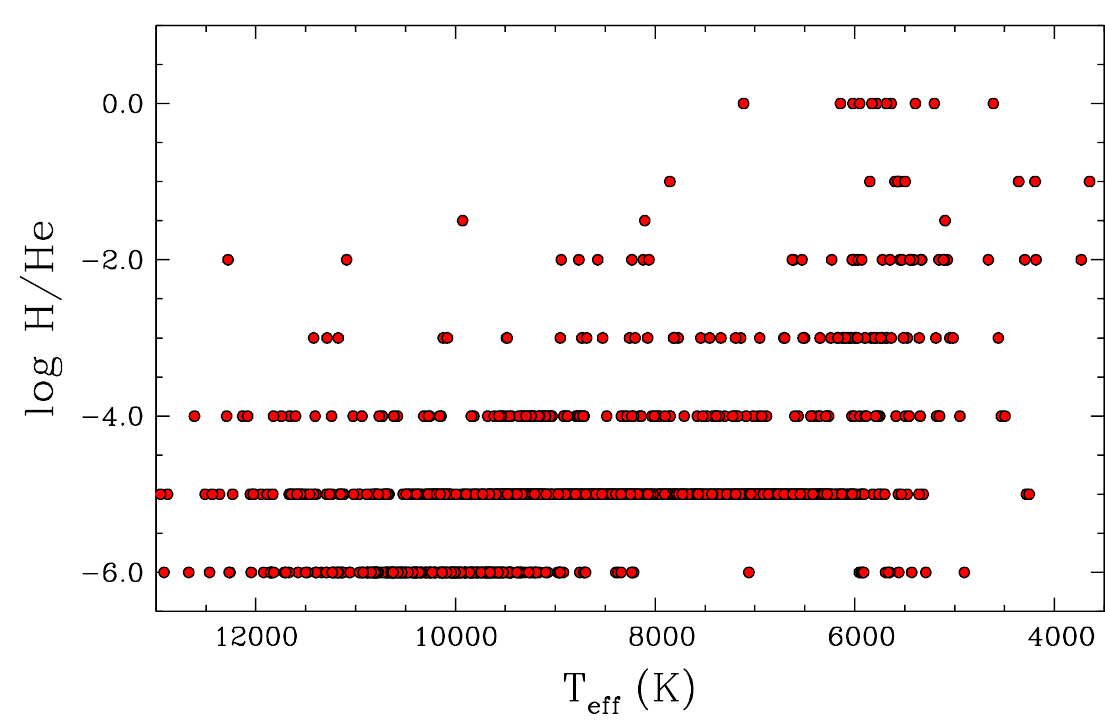}
\caption{H/He ratios as a function of $T_{\rm eff}$ for all metal-line white dwarfs in our sample. These have been measured either by fitting the H$\alpha$ profile, or constrained by the metal line profiles, or set by the visibility limit of H$\alpha$.}
\label{figdzh} 
\end{figure}

\begin{figure*}
\hspace{-0.4in}
\includegraphics[width=2.55in, clip=true, trim=0.4in 0.8in 0.2in 1.1in]{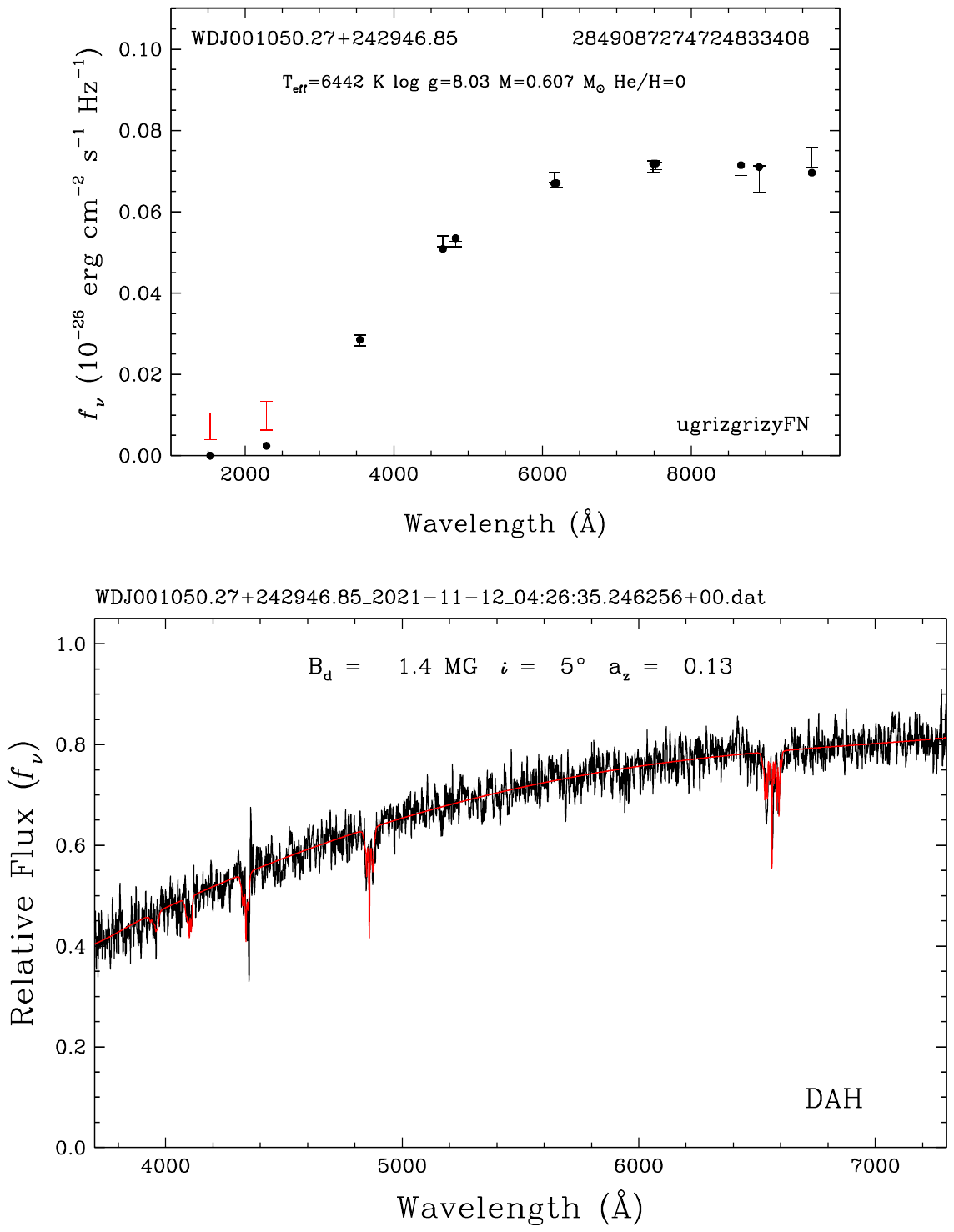}
\includegraphics[width=2.55in, clip=true, trim=0.4in 0.8in 0.2in 1.1in]{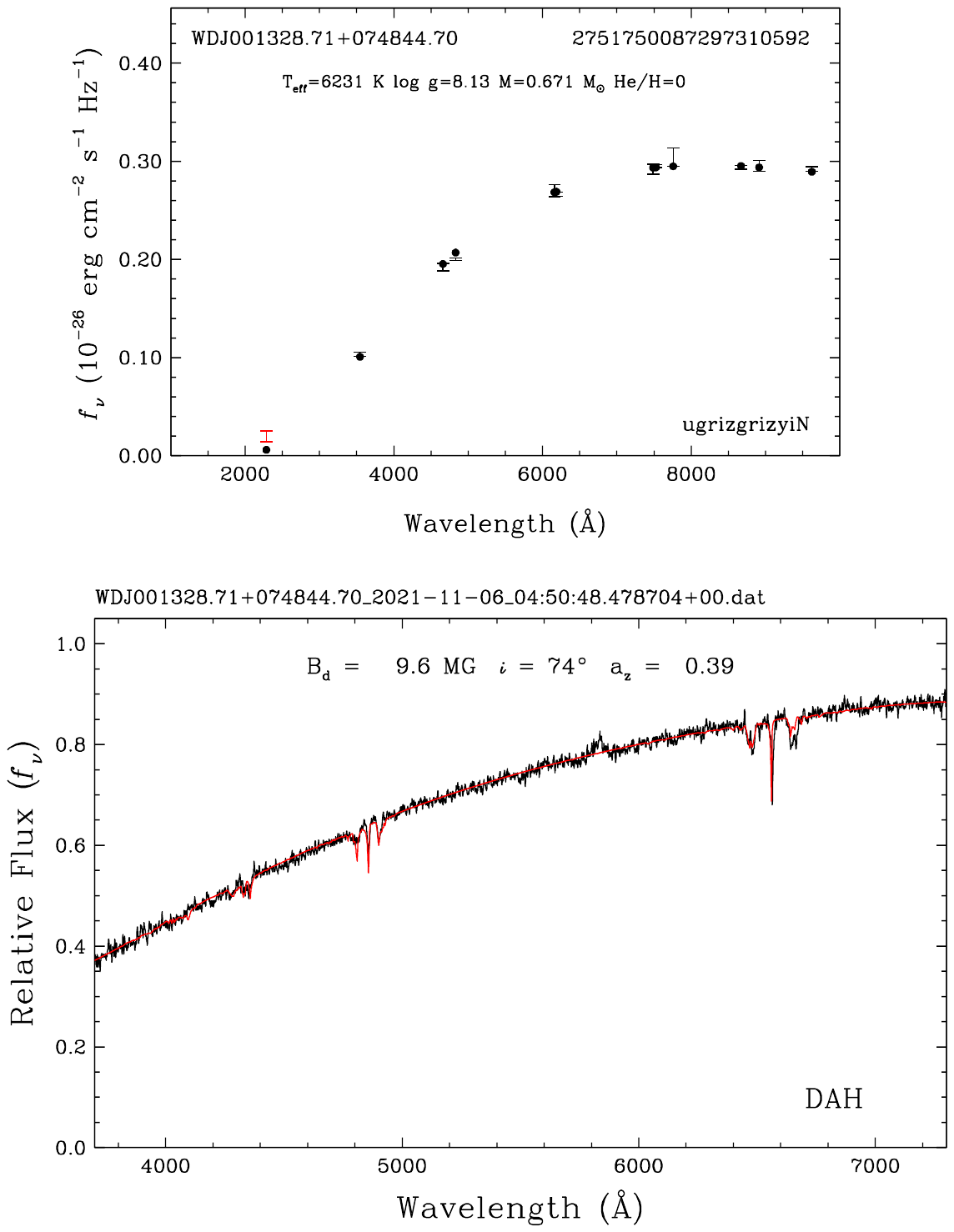}
\includegraphics[width=2.55in, clip=true, trim=0.4in 0.8in 0.2in 1.1in]{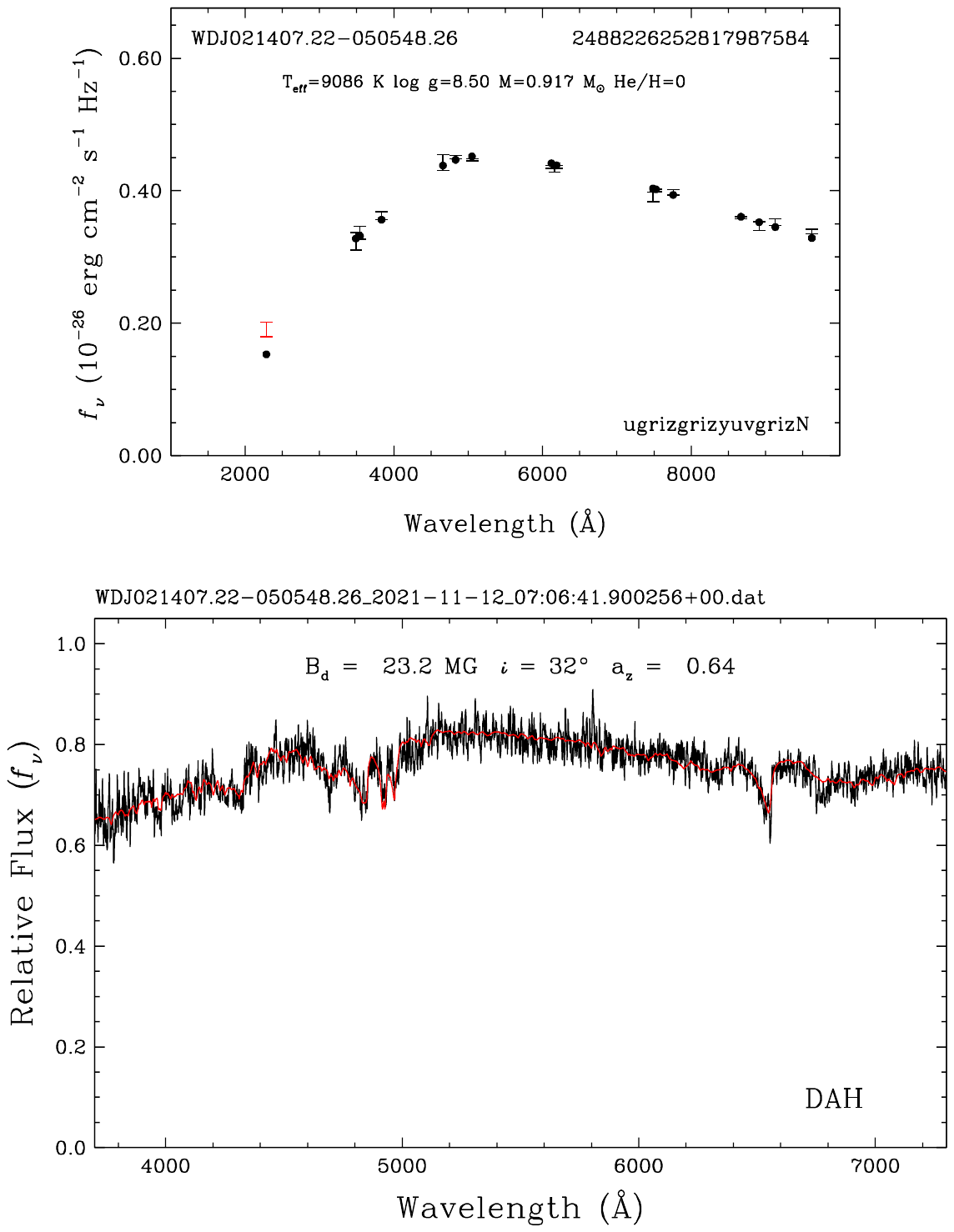}
\caption{Model fits to magnetic DA white dwarfs with field strengths ranging from 1 to 20 MG.}
\label{figdah}
\end{figure*}

Figure \ref{figdzh} shows the H/He ratios determined or constrained from the analysis of the metal-rich white dwarfs in our sample, excluding the DAZs with ${\rm He/H}=0$. This figure shows that (1) the H pollution increases as a function of decreasing $T_{\rm eff}$, with very high values observed at low temperatures (this is similar to the DC stars discussed below in Section \ref{secdc}), and (2) not all DZs are extremely H-rich, as DZs with $\log$ H/He $\leq-4$ exist essentially at all temperatures shown here. 

\subsection{Magnetic White Dwarfs}
\label{secmag}

We identify 535 magnetic white dwarfs in our sample, including 509 magnetic DAs, 10 DQs, and 16 DZs.
The magnetic DA sample also includes 36 DAHe objects that show emission features \citep[see][and references therein]{manser23}. 
We obtain a relatively good fit for one of these objects, WDJ111257.59+690255.98, but we simply use nonmagnetic DA models for the rest.
Since we do not have magnetic DQ or DZ models, we use nonmagnetic models to fit their spectra. For the magnetic DA, we use the photometric technique to constrain the effective temperature and surface gravity, and construct a grid of specific intensities at the surface by solving the radiative transfer equation for various field strengths. The details of our magnetic fits are presented in Paper I.
We use the PIKAIA genetic algorithm \citep{pikaia} to find the best-fitting magnetic model.

Figure \ref{figdah} shows example model fits to three magnetic DAs with fields ranging from $\sim$1 to 20 MG.
Thanks to DESI's medium resolution, it is possible to see Zeeman splitting in cool white dwarfs with relatively
low field strengths. Our model fits to all of the magnetic white dwarfs are included in the \href{https://doi.org/10.5281/zenodo.20775938}{Zenodo archive}. However, a more refined analysis is required to constrain the field geometry for low-field objects and patchy atmosphere (see below) white dwarfs. 
Our normal method of fitting the entire spectrum does not work well for low-field white dwarfs, as the $\chi^2$ is dominated by
the continuum. Better characterization of these low-field objects requires a tailored analysis of the H$\alpha$ region using a higher wavelength
resolution. 

\begin{figure}
\includegraphics[width=3in, clip=true, trim=0.3in 3.8in 0.4in 2.7in]{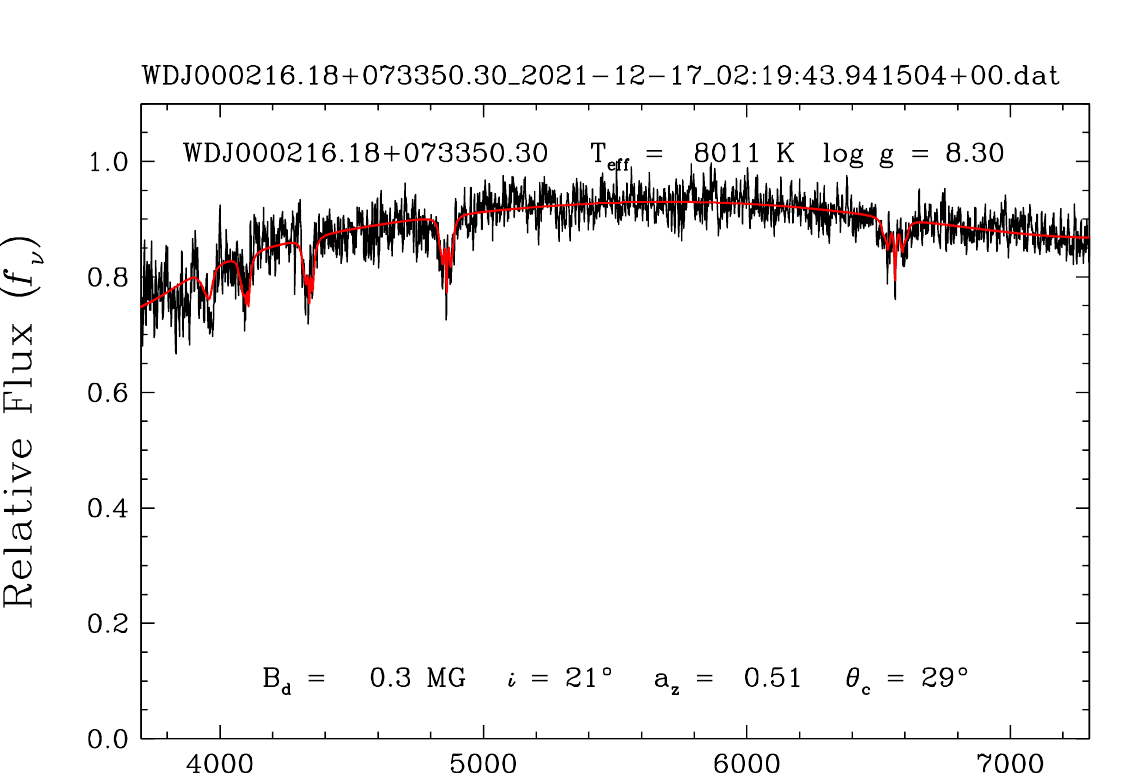}
\includegraphics[width=3in, clip=true, trim=0.3in 3.2in 0.4in 3.3in]{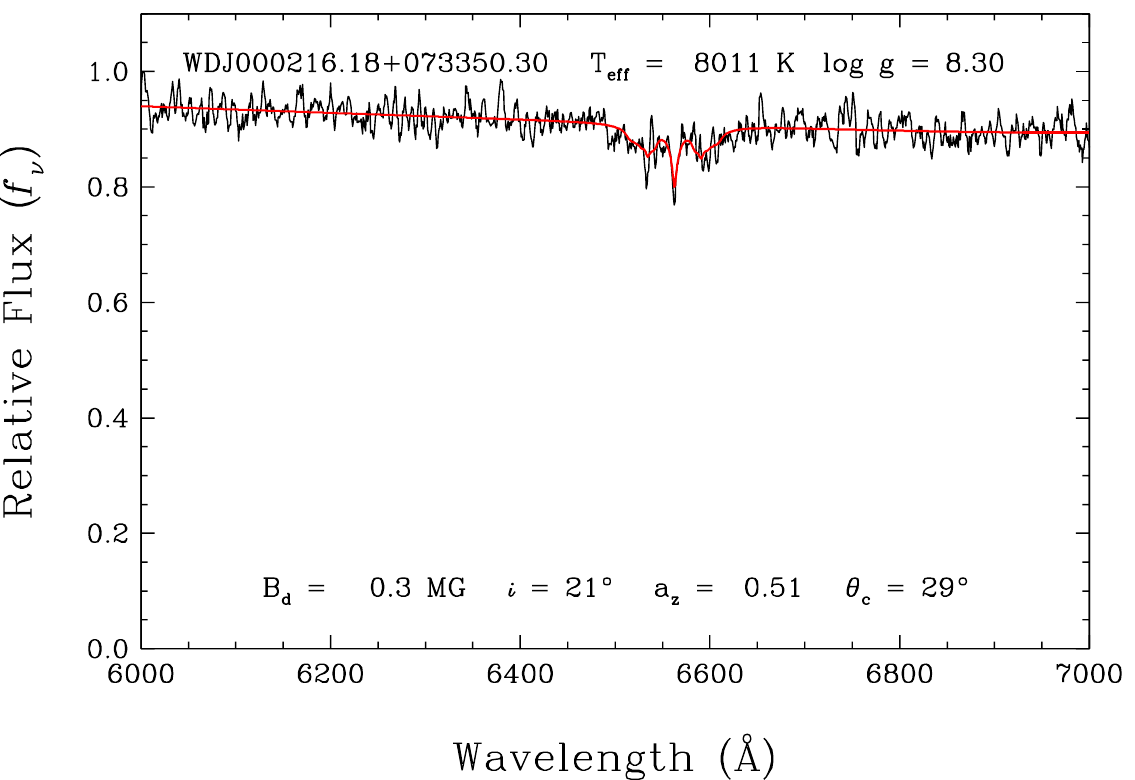}
\caption{Model fits to a magnetic white dwarf with relatively weak and sharp lines. The only way to match the observed Zeeman split lines is through
a patchy atmosphere with H caps and a He belt.}
\label{figpatchy}
\end{figure}

Similarly, some of the magnetic white dwarfs display Zeeman split lines that are much weaker and/or sharper than predicted based on homogeneous atmosphere models. For example, our model fits to WDJ000216.18+073350.30 under the assumption of a homogeneous atmosphere fail to match the H line profiles (see the fit shown on the \href{https://doi.org/10.5281/zenodo.20775938}{Zenodo archive}). The observed lines are significantly weaker than expected for a $T_{\rm eff}\approx8000$ K pure H atmosphere white dwarf. 
Figure \ref{figpatchy} shows our fits to the same object using a patchy atmosphere with H caps and an equatorial He belt \citep{moss24,moss25}. 
A patchy atmosphere model with a 0.3 MG field and H-caps extending $29^\circ$ from the poles provides a good fit to the sharp Zeeman split components.
A patchy atmosphere ensures that only a portion of the surface produces H lines, which helps match both the depth and the sharpness of the H lines. WDJ215408.18+195454.38 is another patchy atmosphere object with a similar temperature ($T_{\rm eff}=8872$ K), and there are others.
Given the sheer number of magnetic white dwarfs in DESI DR1, we delay a refined analysis of low-field and/or patchy atmosphere white dwarfs to a follow-up paper.

\subsection{DC White Dwarfs}
\label{secdc}

We left the discussion of the analysis of the DC white dwarfs last, because their analysis turned out to be the most complicated. Even though
DC white dwarfs have the simplest spectra with a featureless continuum, the H/He ratio in the atmosphere significantly impacts the derived masses
(like the DZ white dwarfs discussed above), and it is difficult to hide H$\alpha$ and also obtain a DC mass distribution that is consistent with the mean mass for white dwarfs in the solar neighborhood, $M\sim0.6~M_\odot$. 

\citet{bergeron19} demonstrated that the location of the DC white dwarfs in the Gaia H-R diagram requires them to have trace amounts of H, C, or
other electron donors in the atmosphere. C likely plays a significant role \citep{blouin23a,blouin23b,camisassa23}, though \citet{kilic25a} demonstrated
that electron donors from H are required below $T_{\rm eff}\sim9000$ K. 
\citet{caron23} and \citet{kilic20,kilic25a} adopted models where they adjusted the H/He ratio as a function of temperature for objects hotter than
6500 K based on the predictions of the convective mixing scenario from \citet{rolland18}. They also assumed a pure H composition below 5200 K, and adopted the pure He or mixed H/He solutions based on a $\chi^2$ analysis  for the 5200-6500 K temperature range. 

In order to come up with a
more uniform approach for the analysis of DC white dwarfs, we explored various model fits. To do so, we rely on an updated version of model grids from \citet{blouin24}, including a non-ideal equation of state \citep{becker14}, a refined treatment of helium ionization equilibrium following \citet{Kowalski2007}, a correction to the H$_3^+$ partition function (see Section \ref{seccmd}), the inclusion of the HeH$^+$ ion \citep{harris04}, the H$_2$–He collision-induced absorption opacity from \citet{abel12}, and the high-density correction to the He$^-$ free-free absorption coefficient \citep{iglesias02,bergeron22}.
Our model grids cover $T_{\rm eff}$ ranging from 3500 to 12,000 K, $\log{g}$ from 7.0 to 9.5, and $\log$ H/He from $-6$ to 0, as well as pure He models. We also extended the $\log {\rm H/He}= -6$ grid up to 15,000 K.

\begin{figure}
\center
\includegraphics[width=3in, clip=true, trim=0.55in 1.46in 0.7in 1.1in]{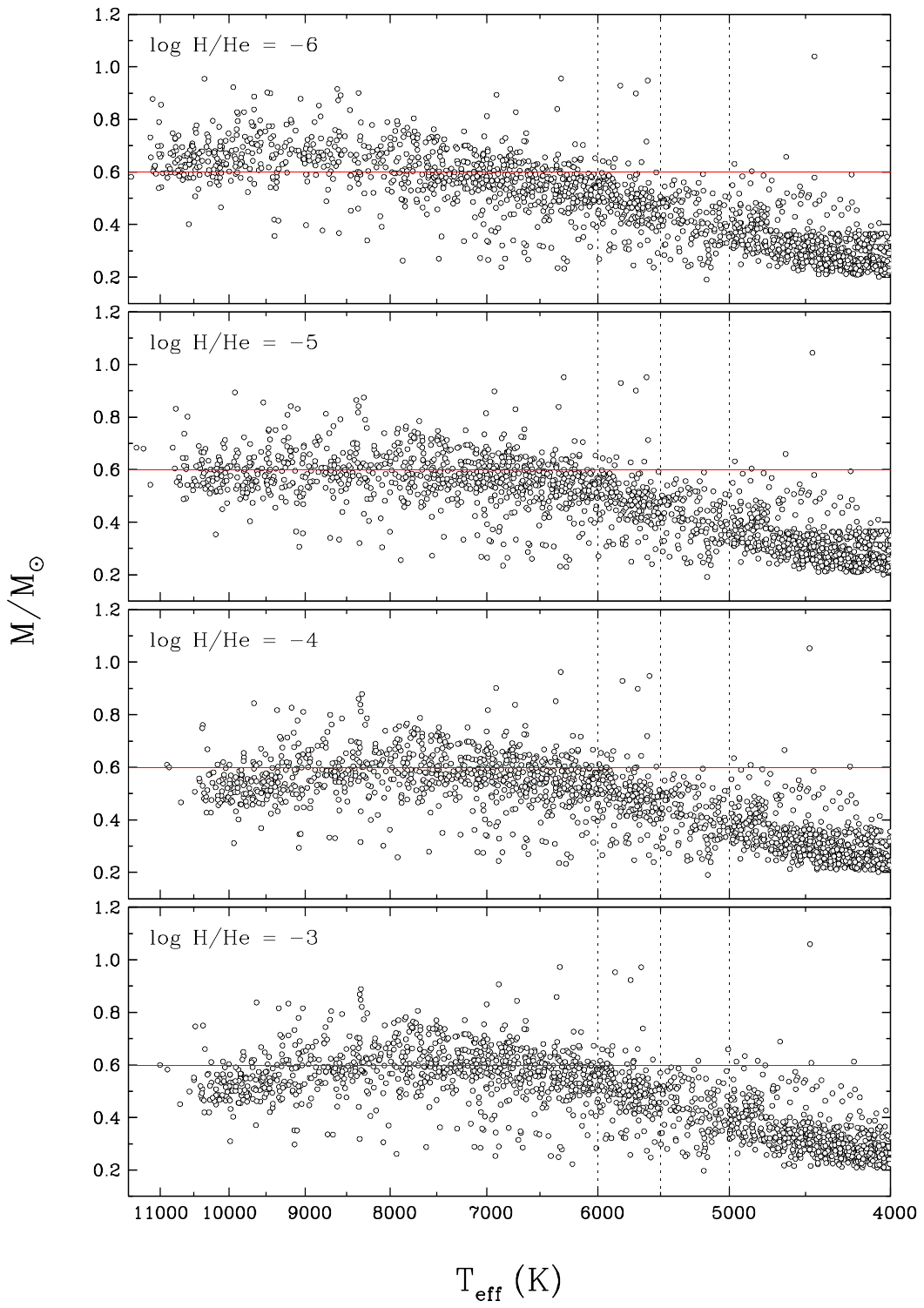}
\includegraphics[width=3in, clip=true, trim=0.55in 0.6in 0.7in 1.15in]{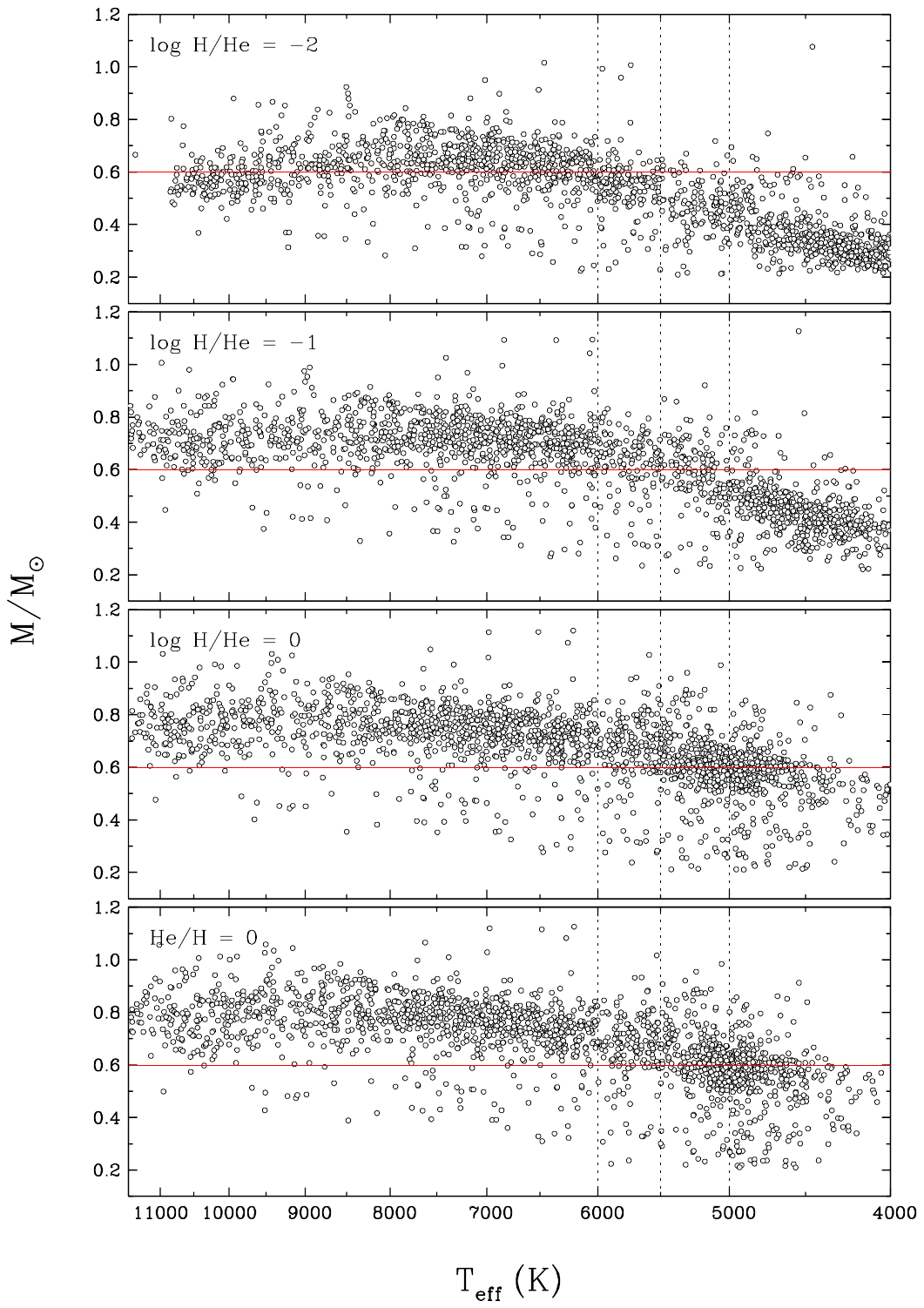}
\caption{Mass vs $T_{\rm eff}$ diagrams for DC stars with $<5$\% distance uncertainty for various assumed compositions. The dotted lines mark 5000, 5500, and 6000 K.}
\label{figdc} 
\end{figure}

Figure \ref{figdc} shows mass versus effective temperature diagrams for DC stars with $<5$\% distance uncertainty for
mixed atmosphere model fits with various H/He ratios. The bottom panel shows the same fits assuming pure H atmospheres. On the hot end,
\citet{bergeron19} showed that the masses of DC white dwarfs are too high for a pure He composition, and that these masses are lowered
when a small trace of H ($\log$ H/He $=-5$) is included. The resulting masses are in much better agreement with the average white dwarf mass
of $0.6~M_\odot$ (red line in each panel). However, inspecting the spectroscopic fits to the DESI data, we find that $\log$ H/He $=-5$ models have too much hydrogen at these temperatures because H$\alpha$ becomes visible. Hence, we need to account for small traces of H at the hot end of the DC sequence, but also make sure that H$\alpha$ remains invisible. C likely plays a role as well, but here we use hydrogen as a proxy for all electron donors in the atmosphere.

Figure \ref{figdc} demonstrates that the masses for DCs below 6000 K are not affected until we reach $\log$ H/He $= -2$, and the H/He ratio has
to be as high as $\log$ H/He $=-1$ between 5000 K and 5500 K, otherwise the masses are too small. This is the most difficult range of $T_{\rm eff}$ to have (1) the correct mean mass of $0.6~M_\odot$ and (2) no H$\alpha$ visible. Furthermore, there appears to be a population of massive ($M\geq0.8~M_\odot$) DC stars in that range of temperature (5000-6000 K) for $\log$ H/He $=-1$ or more H-rich models. These stars would have normal masses under the assumption of $\log$ H/He = $-3$ or $-2$. Hence, it is inherently difficult to constrain the masses of individual DC white dwarfs, but we can use the population characteristics to determine their overall atmospheric properties.

The bottom two panels in Figure \ref{figdc} show that below 5000 K, DC white dwarfs have to be extremely H-rich to have normal masses. Note that this is only possible because of the correction to the H$_3^+$ partition function.
Both H/He = 1 and pure H atmosphere models yield similar results. Not all DC stars need to have pure hydrogen atmospheres below 5000 K,
though we expect a significant fraction to evolve from DA white dwarfs, so they likely retain pure H atmospheres. In reality, the DC population below
5000 K is most likely composed of objects with both pure H atmospheres and mixed H/He atmospheres with H/He ratio of order unity, and probably some with H/He $\ll 1$.

\begin{figure}
\center
\includegraphics[width=2.7in, clip=true, trim=1in 2.7in 1.1in 3.7in]{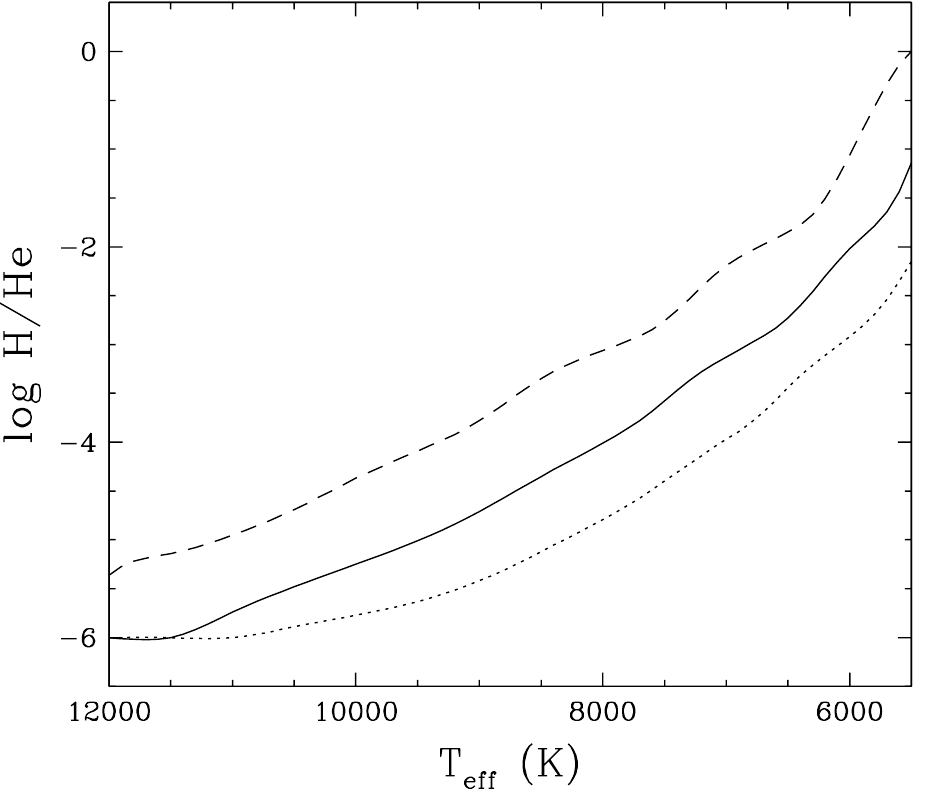}
\caption{H/He limits used in the analysis of the DC stars with $\log{g}$ = 7 (dotted), 8 (solid), and 9 (dashed line).}
\label{fighhe} 
\end{figure}

After carefully considering these model fits under various H/He ratios, we assume H/He ratios at the H$\alpha$ visibility limit for the analysis of the
DC stars. This limit is $\approx2$ \AA\ equivalent-width in typical (noisy) DESI spectra. 
Figure \ref{fighhe} shows the H/He limits used in the analysis of DCs in our sample. H/He ratios range from $\log$ H/He $\sim-6$
at the hot end to $-1$ near 5500 K. We adopt pure H atmosphere compositions below 5000 K. The resulting mass distribution is consistent with 
the mean white dwarf mass (see below), and the resulting H/He vs $T_{\rm eff}$ relation is also consistent with the convective mixing of DA stars with thicker H layers at cooler temperatures. Indeed, looking at the predictions of the evolutionary models for convectively mixed DA white dwarfs presented in Figure 19 of \citet[][see also \citealt{rolland18}]{bergeron22}, a typical $0.6~M_\odot$ DA white dwarf with a relatively thin H layer of $\log{M_{\rm H}/M_\star}=-9.5$ is expected
to mix at 9500 K and end up with $\log$ H/He $\sim-4$, whereas a similar star with a thicker H layer of  $\log{M_{\rm H}/M_\star}=-7.5$
would mix near 6000 K and end up with  $\log$ H/He $\sim-1$ in its atmosphere. These predicted H/He abundance ratios are exactly what we need
(and used) to obtain a reasonable mass distribution of DC white dwarfs in our sample. They are also consistent with the H/He ratios measured for DZ stars (see Figure \ref{figdzh}). 

We close this section by highlighting a rare class of DC white dwarfs, IR-faint white dwarfs. Collision induced absorption (CIA) due to molecular
hydrogen becomes the dominant opacity source at the low temperatures and high densities of cool white dwarf atmospheres \citep{hansen98,borysow01}. CIA is expected to dominate below 4000 K in pure H atmospheres, but it can be significant at hotter temperatures
in mixed H/He atmospheres. Since the CIA dominates in the near-infrared \citep{blouin24}, it can shift the peak of the spectral energy distribution
to the blue-optical region for these cool white dwarfs \citep{harris99,gates04}. \citet{kilic20} identified an IR-faint white dwarf sequence in Gaia color-magnitude diagrams, and \citet{bergeron22} took advantage of Gaia and Pan-STARRS photometry to significantly expand the IR-faint sample within 100 pc from 35 to about 105 stars.

We identify 1 IR-faint DZ and 28 IR-faint DC white dwarfs in the DESI DR1 cool white dwarf sample, including 5 stars that were not included
in \citet{bergeron22}. One of these objects, WDJ095106.36+645400.46, was highlighted in the DESI Early Data Release white dwarf catalog paper by
 \citet{manser24}. The remaining four objects, WDJ023807.51+181530.52, WDJ140435.27$-$020634.66, WDJ151515.50$-$011959.54, and  WDJ225817.72+012811.62 are new discoveries in DESI DR1. We include near-infrared photometry (if available) in our model fits for IR-faint white dwarfs, as this is the only way to get
meaningful parameters for many of these targets \citep{bergeron22}.

\begin{figure}
\center
\includegraphics[width=3in, clip=true, trim=0.4in 0.8in 0.3in 1.1in]{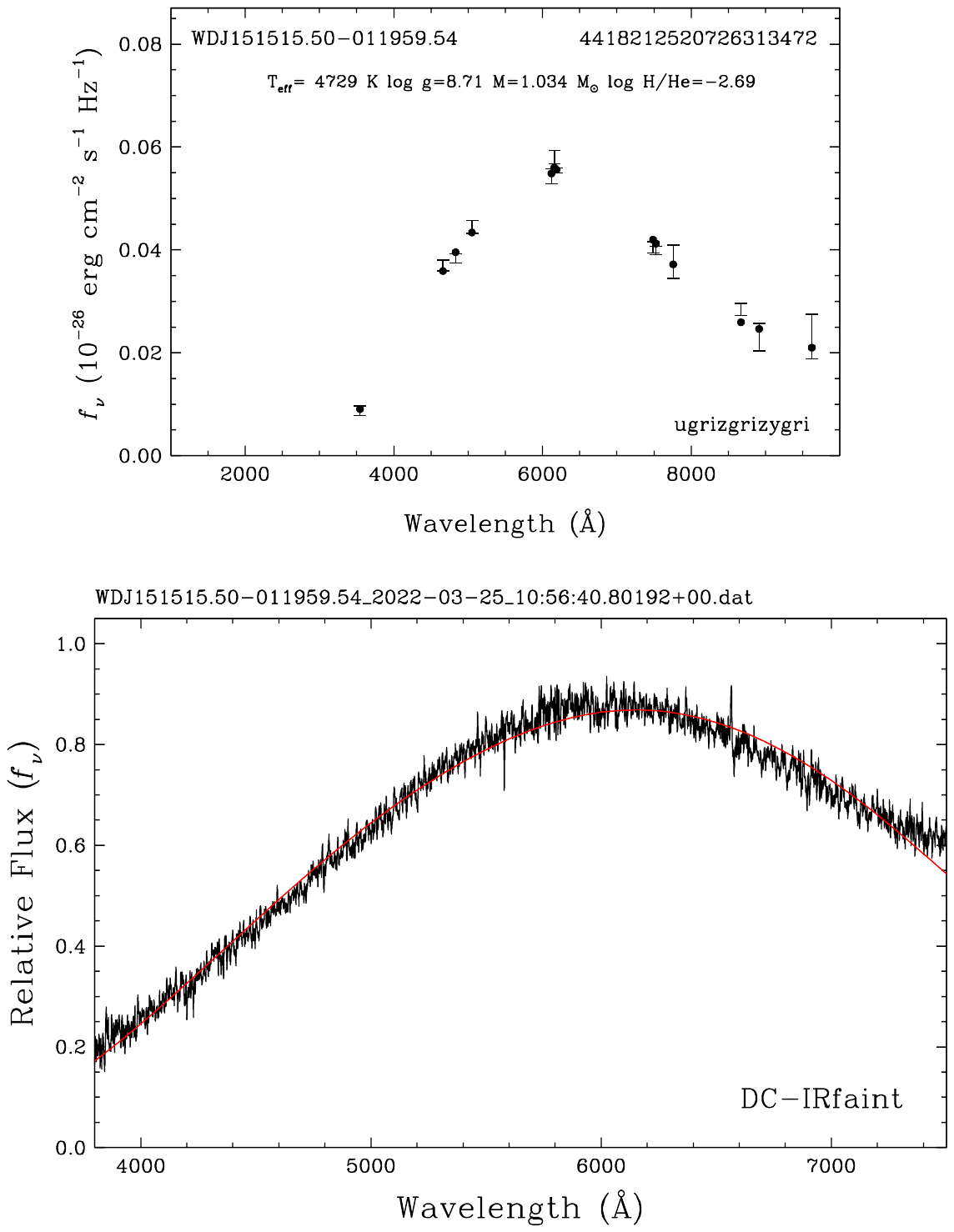}
\caption{Example model fits to a new IR-faint DC white dwarf identified in DESI DR1.}
\label{figirfaint}
\end{figure}

\begin{figure*}
\center
\includegraphics[width=6in, clip=true, trim=0.7in 1.4in 1.5in 1.6in]{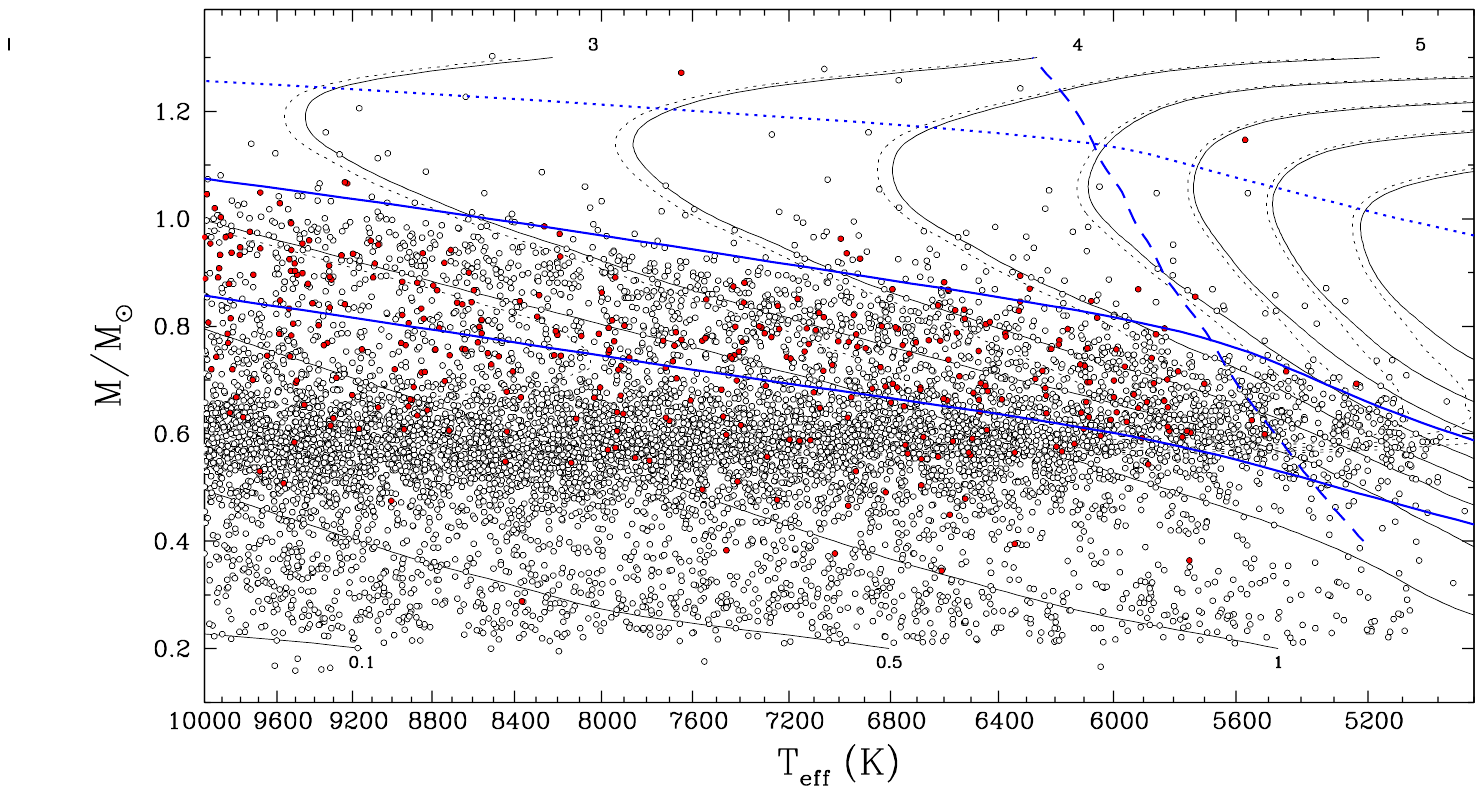}
\caption{Stellar masses as a function of effective temperature for cool DA (white) and DAH (red points) white dwarfs in DESI DR1. 
We restrict the sample to DAs with $<10$\% distance errors, but include all of the magnetic DAs in this figure.
The solid black curves display theoretical isochrones, labeled in units of Gyr, for C/O core white dwarfs with $q({\rm He}) = 10^{-2}$ and $q$(H) = $10^{-4}$, and the dotted curves show the same isochrones with the main-sequence progenitor lifetimes included. The lower blue solid curve marks the onset of crystallization at the center of evolving models, while the upper curve marks the locations where 80\% of the total mass has solidified. The dashed curve indicates the onset of convective coupling, while the blue dotted curve corresponds to the transition between the classical to the quantum (Debye cooling) regime in the ionic plasma.}
\label{figmt}
\end{figure*}

IR-faint white dwarfs are challenging to model. H$_2$-He CIA calculations from \citet{abel12} predict significantly different absorption spectra from the older calculations by \citet{jorgensen00}.
However, \citet{bergeron22} demonstrated that the models with \citet{jorgensen00} opacities provide
superior fits to the optical and near-infrared spectral energy distributions of IR-faint DCs. James Webb
Space Telescope observations of 3 IR-faint white dwarfs further complicated this picture, as \citet{blouin24}
discovered unexpected flux bumps at $2.4~\mu$m in two of these objects and interpreted them as evidence of
temperature inversions above the photosphere. Such temperature inversions could be common among the IR-faint
population. Given these complications and the problems with matching the spectral energy distributions of
IR-faint DCs using the \citet{abel12} opacities, here we rely on the \citet{bergeron22} model grids to fit
the IR-faint DC sample. These are the models used in Figure \ref{figcmd} for the $\log$ H/He = $-5$ sequence.
 
Figure \ref{figirfaint} shows our best-fitting model to the spectral energy distribution of  WDJ151515.50$-$011959.54. Incidentally, this object does not have any near-infrared photometry available. However, the impact of CIA is clearly visible in the available optical data. Even though this is a relatively cool white dwarf, the DESI spectrum peaks at around 6000 \AA. Our best-fitting model
with $T_{\rm eff}= 4729\pm74$ K, $M=1.034\pm0.021~M_\odot$, and $\log$ H/He = $-2.69$ provides an excellent match to the observed photometry of this star. Note that the IR-faint white dwarf sample in DESI DR1 is likely incomplete, as proper identification of IR-faint white dwarfs requires near-infrared photometry, where the CIA dominates. 

\subsection{Miscellaneous Objects}

Our sample also includes several DO white dwarfs with M dwarf companions, cataclysmic variables (CV) and AM CVn binaries. We use the same approach as in Paper I for the DO white dwarfs, and present photometric fits under the assumption of pure H and pure He atmospheres for CV and AM CVn. Given the various emission lines observed in CV and AM CVn, our model fits are simply included for completeness, and they are not meant to provide any meaningful constraints on the physical parameters of those objects. 

\section{Discussion}
\label{secdis}

\subsection{The DESI Cool White Dwarf Sample}

Previous spectroscopic surveys targeting cool white dwarfs either suffered from small numbers or severe spectroscopic incompleteness below a
certain temperature threshold \citep[e.g.][]{eisenstein06,kleinman13,kepler16,kilic10b,kilic20,kilic25a}. \citet{obrien24} presented spectroscopic observations of the volume-complete 40 pc sample of white dwarfs, though that sample includes only 1076 spectroscopically confirmed stars.
With spectroscopy of more than 25,000 stars, DESI DR1 cool white dwarf sample provides an unprecedented opportunity to study the properties
of this sample and identify trends.

Figure \ref{figmt} shows the photometric masses as a function of temperature for cool DA white dwarfs in DESI DR1. Here we restricted the sample to DAs with $<10$\% distance errors, but we include all of the magnetic DAs in the sample. We include the theoretical isochrones for C/O core white dwarfs with thick envelopes, $q({\rm He}) = 10^{-2}$ and $q$(H) = $10^{-4}$ for comparison. The dotted lines show the same isochrones with the progenitor lifetimes taken into account \citep[see][for details]{kilic25a}. The solid blue curves mark the onset of crystallization in the core (lower curve) and where 80\% of the star has solidified (upper curve), while the dashed curve indicates the onset of convective coupling \citep{fontaine01}. The blue dotted curve marks the transition from the classical to the Debye cooling regime. 

The DA population extends down to 5000 K, below which they turn into DC white dwarfs as H$\alpha$ disappears. Hence, we can expect many of the DC white dwarfs below 5000 K to be the descendants of DAs with pure H atmospheres. The main DA population is concentrated around $0.6~M_\odot$ with a significant number
of low-mass and high-mass DAs also visible. Our color selection for the sample (see Figure \ref{fighr}) leads to a large number of  over-luminous white dwarfs to be included. Over-luminosity can be due to low-mass, as
inferred from our model fits assuming single star evolution. However, it is more likely due to a binary system \citep{marsh95}. Hence, the inferred masses and temperatures for these `low-mass' white dwarfs are likely incorrect.  

\begin{figure}
\hspace{-0.2in}
\includegraphics[width=3.4in, clip=true, trim=0.7in 0.7in 0.4in 1.4in]{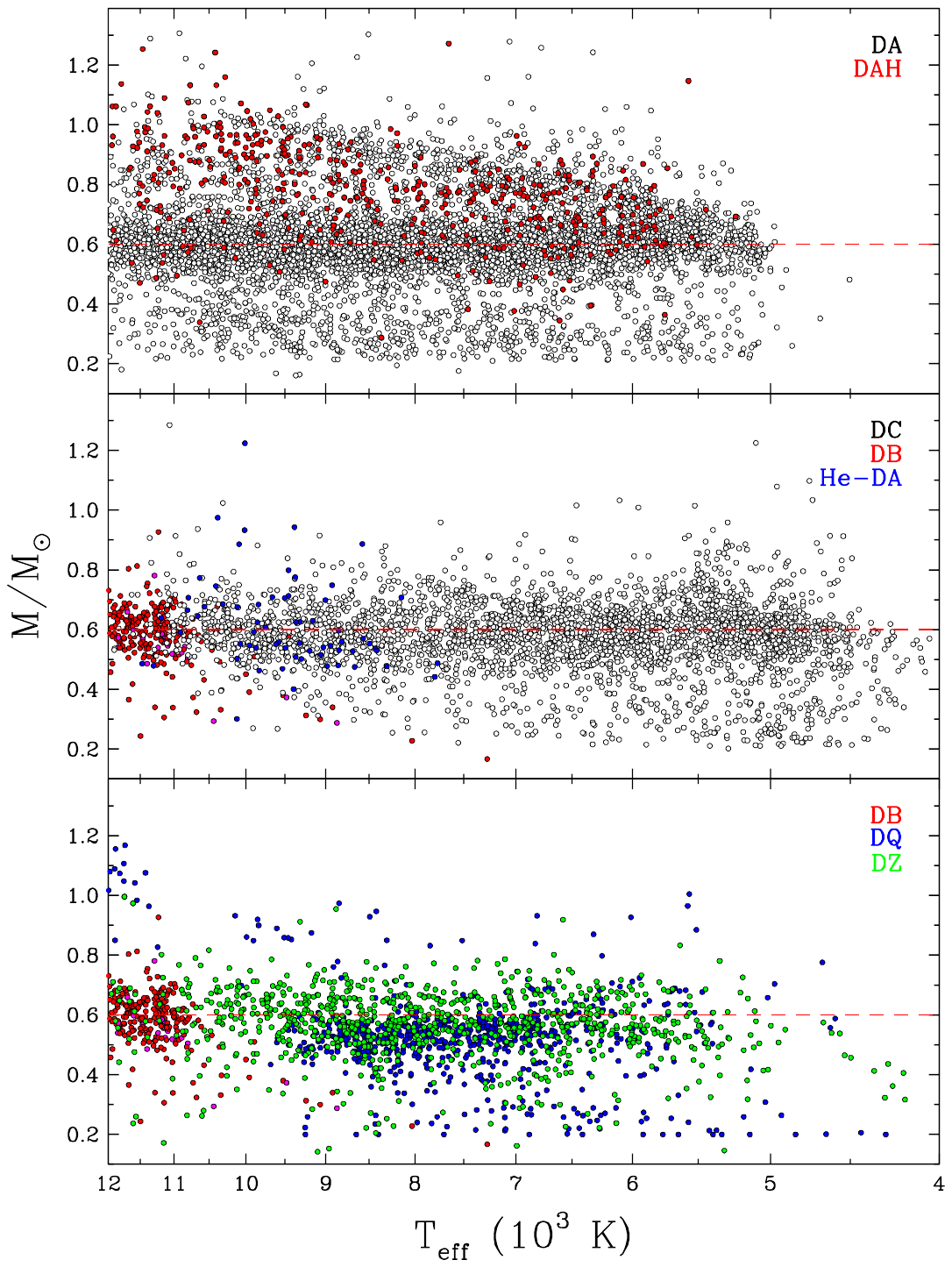}
\caption{Mass vs temperature diagrams for various types of white dwarfs in our sample. DB samples shown in the middle and bottom panels include DB(A)Z white dwarfs (magenta symbols). The dotted line marks $0.6~M_\odot$.}
\label{figsp} 
\end{figure}

More interestingly, a concentration of massive white dwarfs on the crystallization sequence is clearly visible, and this
sequence merges with the $0.6~M_\odot$ DA sequence below 6000 K, indicating that the majority of the cooler DAs observed are going through
crystallization. \citet{kilic25a} found a dearth of ultramassive DA white dwarfs with $M\geq1.1~M_\odot$ below 6000 K in the 100 pc SDSS sample,
and attributed this to the ultramassive white dwarfs rapidly fading away due to Debye cooling. We also make a similar observation here:
out of the more than 1000 DAs cooler than 6000 K, there is only one object with $M\geq1.1~M_\odot$, WDJ071057.12+430622.1, which is magnetic.
Hence, ultramassive DAs are unusually rare below 6000 K. 

Another interesting observation among the DA population is that magnetic DAs are found everywhere, and not just on the crystallization sequence.
The origin of magnetic fields in white dwarfs has been a puzzle for decades \citep{ferrario20}. Mergers clearly contribute to the population of young
and massive magnetic white dwarfs, but the prevalence of magnetism among cool white dwarfs suggest that crystallization-induced dynamos
or convective dynamos from earlier evolutionary phases are likely in play for older magnetic white dwarfs \citep[see e.g.][]{isern17,ginzburg22,bagnulo22,camisassa24,moss25,einramhof26}. Since many of the magnetic DAs in our sample are not on the crystallization sequence,
this observation would suggest that the fields in these stars are not induced by core crystallization, and instead likely due to the delayed emergence of fossil fields on the surface of a white dwarf \citep{camisassa24,einramhof26}.

Figure \ref{figsp} shows the stellar masses as a function of temperature for non-DA white dwarfs. Here, we also
include the DA sequence in the top panel for comparison. The dashed line in each panel marks $0.6~M_\odot$. We restrict this figure to DA and DB/DC stars with distance uncertainty of $<5$\% and $<10$\%, respectively, but we include all DAH, He-DA, DQ, and DZ stars. This is one of the most important figures presented in this paper; the population characteristics and the white dwarf spectral evolution are imprinted on the mass distributions of the various spectral types shown here.

The middle panel in Figure \ref{figsp} shows the mass distributions of DCs along with the DB and He-DA stars for comparison. On the hot end, the only difference between He-DA and DC white dwarfs is that He-DA stars are more H-rich, which leads to visible Balmer absorption features. When
we do see those lines, we can constrain the H/He abundance ratio in the atmosphere, and the resulting masses are consistent with the average
white dwarf mass of $0.6~M_\odot$. This is one of the main motivations for our prescription of varying H/He ratios as a function of effective temperature for DC white dwarfs. Trace amounts of H just below the visibility limit enables us to obtain masses close to $0.6~M_\odot$ for
the entire DC sequence.  With our new models including the H$_3^+$ correction, the mass distribution below 5000 K is also close to $0.6~M_\odot$, as opposed to previous studies.

The bottom panel in Figure \ref{figsp} shows the mass distributions of DB, DQ, and DZ white dwarfs. DB(A)Z stars are shown with a different color (magenta symbols). This figure clearly shows the transformation of DBs into DC and DZ white dwarfs below about 11,000 K.
Just like the DB sample here and in the 100 pc SDSS sample \citep{jewett24,kilic25a}, we do not see a large number of DZs above $0.8~M_\odot$. The DQ mass distribution is different; the hot white dwarf sample in DESI \citep{kilic26} includes warm and ultramassive DQs with $M\gtrsim1~M_\odot$, whereas the cool DESI sample shown here includes mostly lower-mass DQs with
$M\sim0.55~M_\odot$; see the next section for more details. There are also a number of low-mass DQs with $M\sim0.2~M_\odot$ visible in this figure.
The mass estimates and the resulting C/He abundances are likely wrong for those targets since the spectra are likely contaminated by a companion.

\subsection{The DQ Sequence}

\begin{figure}
\hspace{-0.2in}
\includegraphics[width=3.4in, clip=true, trim=0.3in 3.9in 0.4in 2.4in]{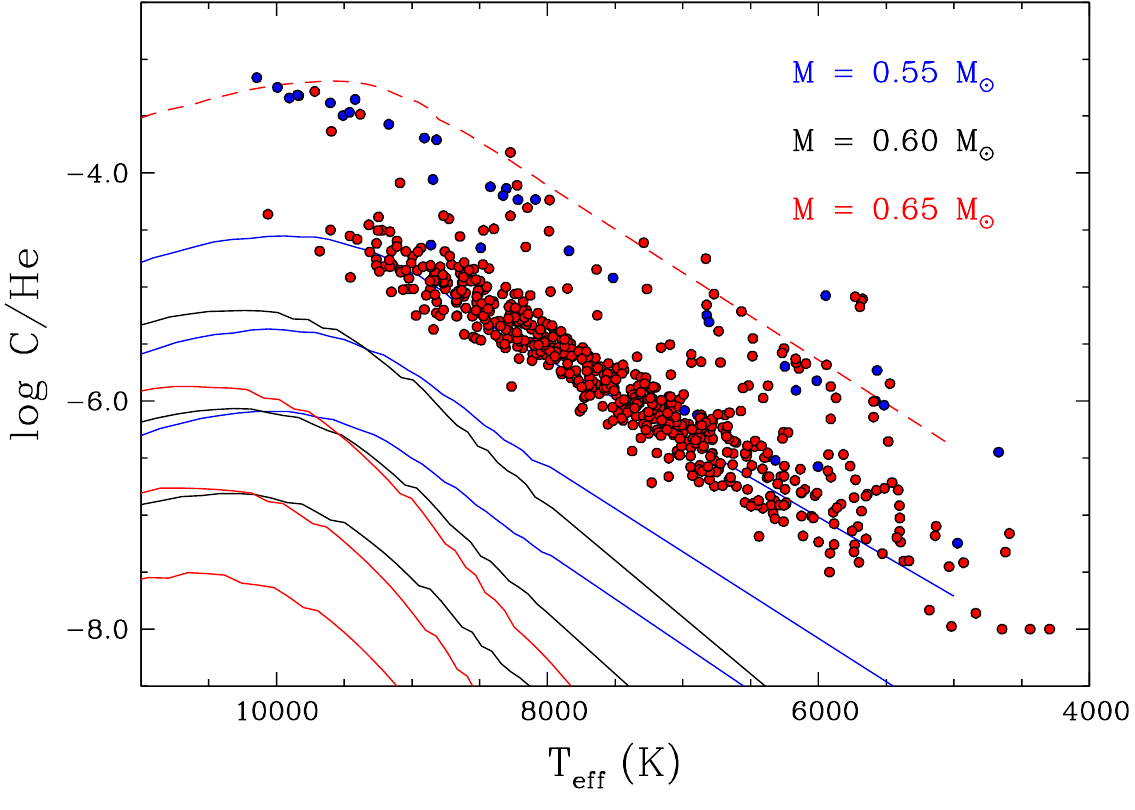}
\caption{Atmospheric carbon abundance as a function of temperature for the classical DQ white dwarfs in our sample. Blue points mark stars with $M > 0.7~M_\odot$. The solid lines show the theoretical predictions for the convective dredge-up of carbon for three different masses, and with initial carbon mass fractions of $X_{\rm C}=0.2$, 0.4, and 0.6 (from bottom to top), and a standard envelope with $M_{\rm env} = 10^{-2}~M_\star$. The red dashed line shows the predictions for a $0.65~M_{\odot}$ white dwarf with a thin envelope, $M_{\rm env} = 10^{-3.5}~M_\star$. The theoretical predictions are linearly extrapolated below $T_{\rm eff}=8000$ K.}
\label{figdqseq} 
\end{figure}

\citet{coutu19} presented an analysis of 319 DQ white dwarfs with trigonometric parallax measurements from Gaia DR2. 
Their sample and the model atmosphere grid are limited to DQs hotter than $T_{\rm eff}>6000$ K. We identify 701 DQs
in the cool DESI white dwarf sample, significantly enlarging the number of DQs known in the solar neighborhood.

Figure \ref{figdqseq} shows the carbon abundances for the classical DQs in our sample along with the theoretical predictions for convective dredge up
of carbon \citep{dufour05,bedard22b} for three different masses (0.55, 0.60, and $0.65~M_{\odot}$). 
The average mass for DQs in the solar neighborhood is lower than $0.6~M_\odot$ (see Figure \ref{figsp}).
A likely explanation for the relatively low-mass DQ population involves the mass dependence of the convective dredge-up process.
Carbon is likely dredged up in most cool He-atmosphere white dwarfs \citep{pelletier86}, but only the lower mass white dwarfs are enriched enough to
show visible C features in the optical data \citep{bedard22b}. 

For each mass track shown in Figure \ref{figdqseq}, we show three different predictions for a standard envelope mass of $M_{\rm env} = 10^{-2}~M_\star$ and initial carbon mass fractions of $X_{\rm C}=0.2$, 0.4, and 0.6 (from bottom to top).
The red dashed line shows an evolutionary sequence for a $0.65~M_{\odot}$ white dwarf with a thin envelope, $M_{\rm env} = 10^{-3.5}~M_\star$, and an initial carbon mass fraction of $X_{\rm C}=0.6$. Blue and red points mark objects with mass above and below $0.7~M_\odot$, respectively.

Because warm DQs in DESI DR1 were discussed in detail in Paper I, here we concentrate on the classical DQs. 
A comparison between this figure and Figure 12 of \citet{coutu19} shows that we have a much larger sample of DQs in DESI, and that
the cool DQ sequence in DESI extends below 6000 K, though the number of DQs is clearly much smaller below that temperature. 
As also seen in the \citet{coutu19} sample and the 100 pc SDSS sample \citep{kilic25a}, the classical DQs in DESI DR1 start to appear around
$T_{\rm eff}=10,000$ K, and they form a tight sequence in C/He versus temperature. A comparison with the predictions of the convective dredge up
scenario shows that the observed decline in C/He is exactly what is expected from element transport in white dwarfs \citep{bedard24b}.
The majority of the cool DQs are consistent with $0.55~M_\odot$ white dwarfs with relatively thick ($M_{\rm env} = 10^{-2}~M_\star$)
envelopes and large initial carbon mass fractions with $X_{\rm C}=0.6$. The observed scatter in the C/He relation can be explained by differences
in stellar and envelope masses. The bottom of the sequence corresponds to the optical detection limit for carbon in DESI DR1 spectra.

Evolutionary models predict that to display Swan bands, a He-atmosphere white dwarf needs (1) a low mass and (2) a high carbon abundance in the envelope of its PG 1159 progenitor.
Therefore, low-mass non-DAs that have a low initial carbon abundance should not appear as DQs, but instead appear as DC or DZ white dwarfs (if they accreted metals). 
Hence, a significant fraction of the $\sim$$0.55~M_\odot$ DC and DZs observed in Figure \ref{figsp} may be the descendants of low $X_{\rm C}$ PG 1159 stars. 

One of the most striking features of the DQ sequence is the disappearance of the DQs at cooler temperatures. There are $\sim$200 DQs with
$T_{\rm eff}$ between 8000-9000 K and 7000-8000 K. However, the number decreases to 154 for 6000-7000 K, 57 for 5000-6000 K, and only
9 below 5000 K. This is not an observational bias \citep[also see][]{coutu19,blouin19,bedard24b}. 

Coolest DQs with $\log$ C/He $\sim-7$ are clearly
detectable in DESI DR1 (see Figure \ref{figdq} above). Hence, there is a real decrease in the number of DQs observed in the solar neighborhood below 6000 K. \citet{kilic25a} argue that this could be due to the relatively low-masses of DQ white dwarfs, which leads to long progenitor
main-sequence lifetimes. Hence, DQs may simply not have enough time to evolve down to 4000 K within a Hubble time. 

\subsection{Metal-Rich White Dwarfs}

\begin{figure}
\center
\includegraphics[width=3.4in, clip=true, trim=0.3in 2.7in 0.7in 4.3in]{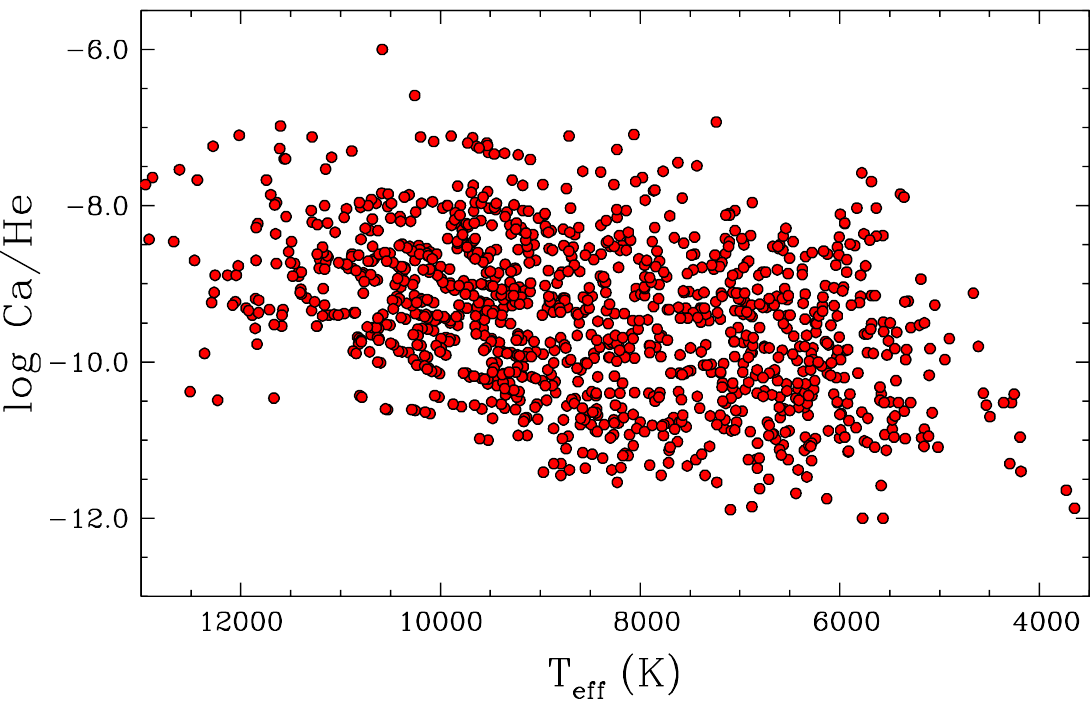}
\caption{Ca/He ratio versus photometric effective temperature for the metal-rich white dwarfs in our sample.}
\label{figdzseq} 
\end{figure}

We identify 1002 metal-rich white dwarfs in our sample, including DAZ, DBZ, DBAZ, DZ, and DZA white dwarfs, and excluding DQZ. Figure \ref{figdzseq}
shows the Ca/He abundance ratio versus temperature for these stars. The absence of objects in the bottom left portion of this diagram (at high temperatures
and low calcium abundances) is due to the detection limit of \ion{Ca}{2} H and K lines in DESI DR1 spectra. 

The DZ sequence extends down to about 4000 K, however just like the DQs, the number of DZs decreases significantly below 5500 K. The number goes from 73
with $T_{\rm eff}=$ 5500-6000 K down to 30 with 5000-5500 K, and 13 with 4500-5000 K. Hence, DZs appear to be rare below 5000 K.
The decrease in frequency of metal-pollution could be related to the disappearance of cool DQs, or it could simply be due to a decrease in the number of tidal disruption events that occur around
white dwarfs. Theoretical simulations predict that the majority of tidal disruption events occur within the first few hundred Myr after the star
becomes a white dwarf, and after a peak at $\sim$30 Myr, the frequency of metal pollution is predicted to decrease with time \citep[see for example,][]{debes12}. Observationally, the picture is more complicated. \citet{hollands18b} found a decrease of three orders of magnitude in the accretion rate of material within 6.5 Gyr of the formation of a white dwarf. On the other hand, \citet{blouinxu22} found the mass accretion rates decrease by less than
a factor of 10 between 1 and 8 Gyr. 

The observed Ca/He abundance ratios range from $\log$ Ca/He = $-12$, which is also the lower boundary of our model grid, to $-6.4$. 
\citet{dufour12} measured $\log$ Ca/He = $-6.23 \pm 0.15$ for the heavily polluted DBZ white dwarf J0738+1835, which is one of the most metal-polluted
white dwarfs known. The most metal-rich white dwarfs identified here have similar Ca/He ratios. These objects are prime targets for detailed characterization
of their abundances through high-resolution spectroscopy, which would help constrain the composition of the accreted minor body \citep[e.g,][]{zuckerman07,xu14,xu17}.

\subsection{Spectral Evolution}

\begin{figure}
\center
\includegraphics[width=3.4in]{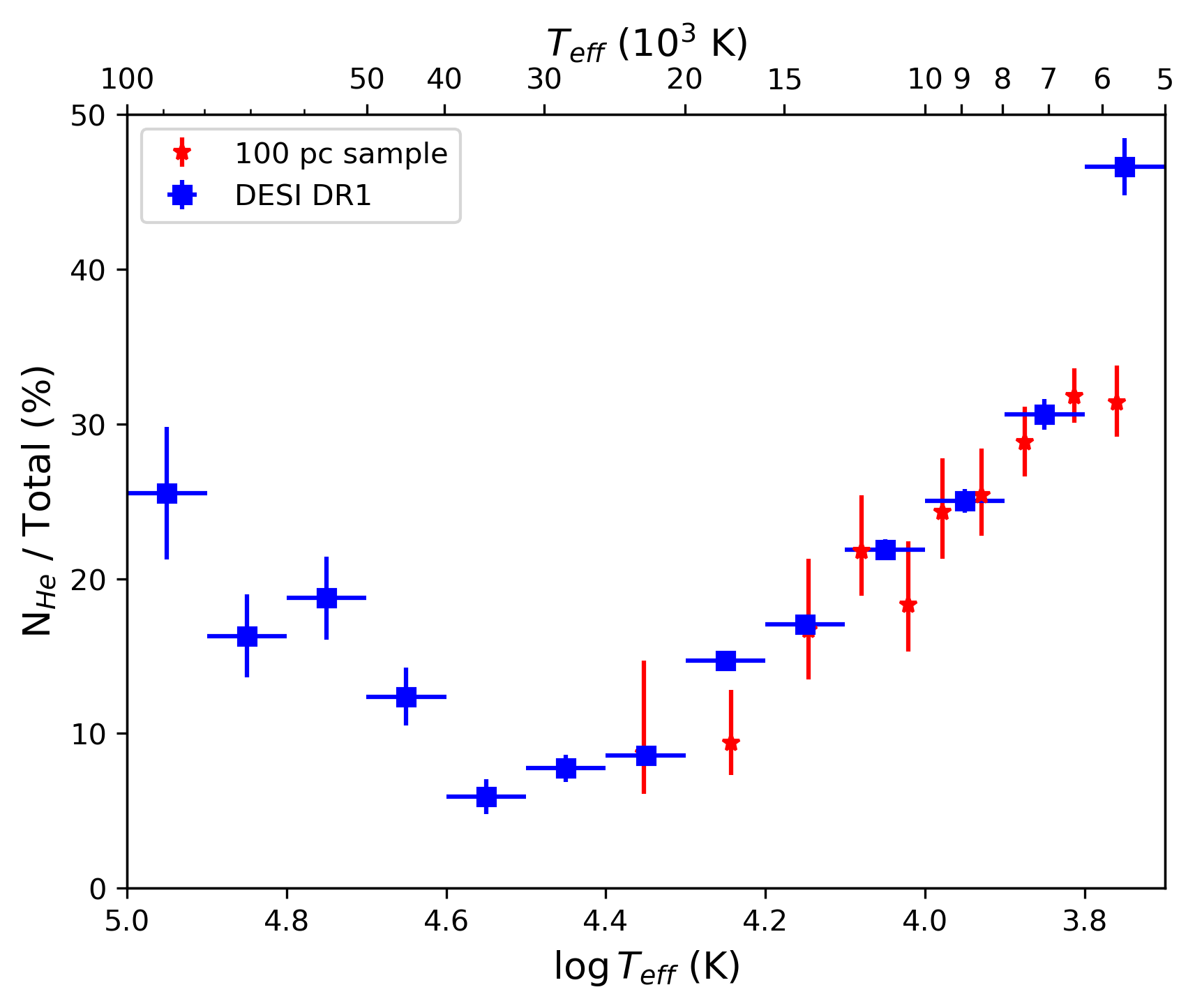}
\caption{Fraction of He-atmosphere white dwarfs as a function of effective temperature from DESI DR1 (blue points) and the 100 pc white dwarf sample in the SDSS footprint \citep{kilic25a}.}
\label{fighe} 
\end{figure}

The surface composition of a white dwarf can change due to gravitational settling, radiative levitation, winds, convection, external accretion, and magnetism (which can inhibit
convection). These processes become important in various temperature ranges, and lead to significant changes
in the H- to He-atmosphere ratio as a function of effective temperature \citep[see][for a review]{bedard24b}.
The overall trend is that the He-atmosphere fraction starts around 25\% for hot white dwarfs, but then decreases down to about 10\% around 30,000 K, before rebounding at cooler temperatures 
\citep[e.g.,][]{genest19,ourique19,cunningham20,bedard20,lopez22,jimenez23,torres23,obrien24}.

H-atmosphere white dwarfs develop a convection zone at the surface at about 18,000 K \citep{cunningham19}. The convection zone deepens with decreasing temperature and expands to include
a large portion of the envelope below $T_{\rm eff}=12,000$ K. Depending on the thickness of the surface H layer, the
convection zone may be deep enough to convectively mix the surface layers, and convert a DA white dwarf
into a He-DA or a non-DA. For example, \citet{rolland18} estimate that a DA white dwarf with a surface H layer
mass of $10^{-8}~M_\odot$ will transition to a He-atmosphere at $T_{\rm eff}\approx7000$ K.
Observationally, there is considerable uncertainty in the He-atmosphere white dwarf fraction below 10,000 K
\citep[see Figure 2 in][]{bedard24b}. 

\begin{deluxetable}{cccc}
\tablecolumns{4} \tablewidth{0pt}
\tablefontsize{\small}
\tablecaption{Fraction of He-atmosphere white dwarfs in DESI DR1 as a function of effective temperature.\label{tabhe}}
\tablehead{\colhead{$\log{T_{\rm eff}}$} & \colhead{He-fraction} & \colhead{$\log{T_{\rm eff}}$} & \colhead{He-fraction}\\
(K) & (\%) & (K) & (\%)}
\startdata
4.95 & 25.5 $\pm$ 4.3 & 4.25 & 14.7 $\pm$ 0.6 \\
4.85 & 16.3 $\pm$ 2.7 & 4.15 & 17.1 $\pm$ 0.6 \\
4.75 & 18.8 $\pm$ 2.7 & 4.05 & 21.9 $\pm$ 0.7 \\
4.65 & 12.4 $\pm$ 1.9 & 3.95 & 25.0 $\pm$ 0.8 \\
4.55 & 5.9 $\pm$ 1.1 & 3.85 & 30.6 $\pm$ 1.0 \\
4.45 & 7.8 $\pm$ 0.9 & 3.75 & 46.6 $\pm$ 1.8 \\
4.35 & 8.6 $\pm$ 0.6 \\
\enddata
\end{deluxetable}

With a Gaia-based target selection and a relatively large sample size, DESI DR1 provides an excellent
opportunity to revisit this issue. In Paper I, we presented the non-DA fraction in the $10^5-10^4$ K
temperature range. With the cool white dwarf sample presented here, we can now extend this analysis to cool
white dwarfs, which make up the majority of the white dwarfs in the solar neighborhood.

Figure \ref{fighe} shows the fraction of He-atmosphere white dwarfs as a function of effective temperature in DESI DR1 (blue points) and
the 100 pc white dwarf sample in the SDSS footprint \citep[red points,][]{kilic25a}. Table \ref{tabhe} lists the He-fractions in the DESI DR1 sample. This is the first time we are able to use a single survey to study the He-fraction over the entire temperature range from 100,000 K
down to 5000 K. This figure shows that the He-fraction starts at about 25\% for the hottest white dwarfs, but decreases to $\sim$8\% near 30,000 K, and then gradually increases to $\approx22$\% near 11,000 K. This gradual increase leads to a He-fraction of 31\% at 7000 K.

The DESI results are consistent with the He-fractions obtained using the 100 pc SDSS sample within the errors. The only discrepancy is observed at the coolest temperature bin, which corresponds to the 5500-6000 K range in the 100 pc sample and the 5000-6300 K range in DESI. The He-fraction for the 5500-6000 K range in DESI is $41.2\pm2.4$\%, still significantly higher than estimated from the 100 pc sample.   
However, our assumptions for the analysis of the cool DC sample between 6500 and 5200 K are significantly different from the assumptions used in the 100 pc sample (see Section \ref{secdc}). 
This is also one of the most problematic ranges of temperature in our analysis; given the relatively noisy DESI spectra, it is easy to miss weak H$\alpha$ features, which could significantly impact the He-fractions obtained in this study. Furthermore, the distinction between H and He atmosphere white dwarfs become unclear at these temperatures, because we require H/He ratios of order unity (see Figure \ref{fighhe}). Hence, the coolest DESI data point in this figure should be used with caution. 

The overall agreement between the DESI results and the 100 pc SDSS sample demonstrate that the fraction of helium-rich white dwarfs increases with decreasing
temperature due to convective mixing. Given the white dwarf luminosity function, this also means that more white dwarfs undergo mixing at low temperatures than at high temperatures.

\subsection{Unusual Binaries}

Our target selection based on Gaia colors ($G_{\rm BP} - G_{\rm RP}>0$) leads to a bias for the inclusion of a large number of white dwarf + main-sequence binaries, as well as double-degenerates. Here we discuss three unusual classes of binary systems, extremely low-mass (ELM, $M \lesssim0.3~M_\odot$) white dwarfs and double-lined spectroscopic binaries involving DA+DB and DA+DQ white dwarfs. Even though our sample also includes many likely DA+DA binaries, there are significant degeneracies in the DA+DA atmosphere model fits. Without radial velocity constraints on the mass ratios, a change in the surface gravity for one star can be compensated by a similar change in the surface gravity of the companion to match
the observed spectral energy distribution. A detailed model atmosphere analysis of the DA+DA candidates is beyond the scope of this paper.

\subsubsection{ELM White Dwarfs}

ELM white dwarfs are some of the strongest sources of gravitational wave sources known in the mHz frequency range \citep{brown11,hermes12,kupfer24}. Previous works identified and characterized the orbits for $\sim$150 ELM binaries in the Galaxy \citep{brown10,brown22,kilic10a,burdge20a,burdge20b,kosakowski20,kosakowski23,kosakowski25,chickles25,barrientos25}. 

A large sample of ELM white dwarf candidates can be identified through Gaia photometry and astrometry \citep{pelisoli19}. However, metal-poor subdwarf A-type (sdA) stars can be a major source of confusion with cooler $\sim$8000~K white dwarfs at the spectral resolution of surveys like SDSS and DESI. 
\citet{brown17} demonstrated that pure H atmosphere model fits to sdA stars result in surface gravity estimates that are systematically higher by $\sim$1 dex. These model fits lead to an overdensity of sdA stars with $T_{\rm eff}\sim8000$ K and $\log{g}\sim5$-6 in the SDSS \citep{brown17,pelisoli18}. We also see the same over-density in DESI DR1 based on our pure H atmosphere model fits. 

\begin{figure}
\includegraphics[width=3.3in]{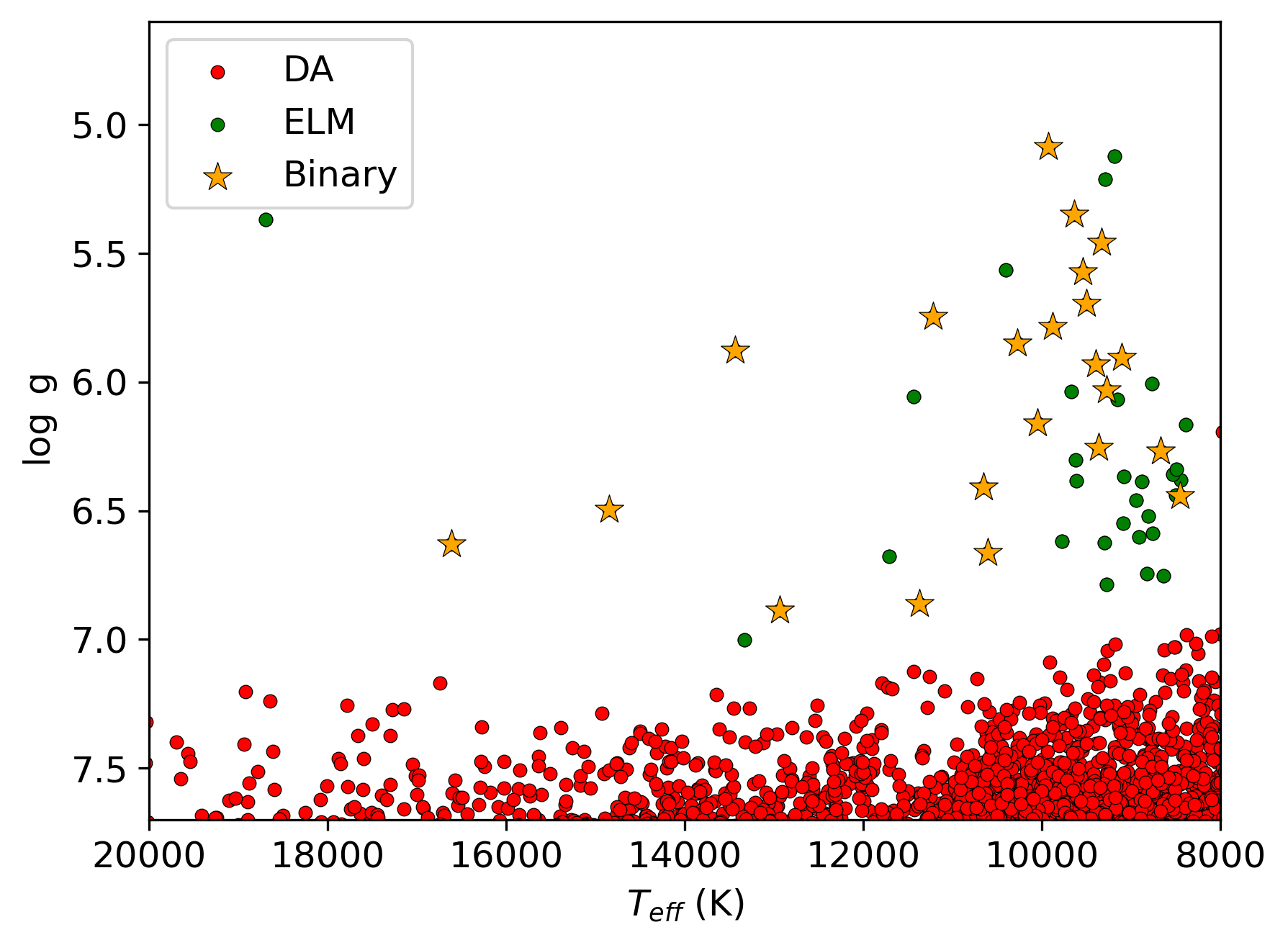} 
\caption{Spectroscopic parameters of our DA white dwarf sample in the ELM region. Stars mark the previously known ELM white dwarf binaries, and green dots mark the 29 ELM white dwarf candidates that are prime targets for follow-up radial velocity observations.}
\label{figelm} 
\end{figure}

\begin{figure}
\includegraphics[width=3.3in]{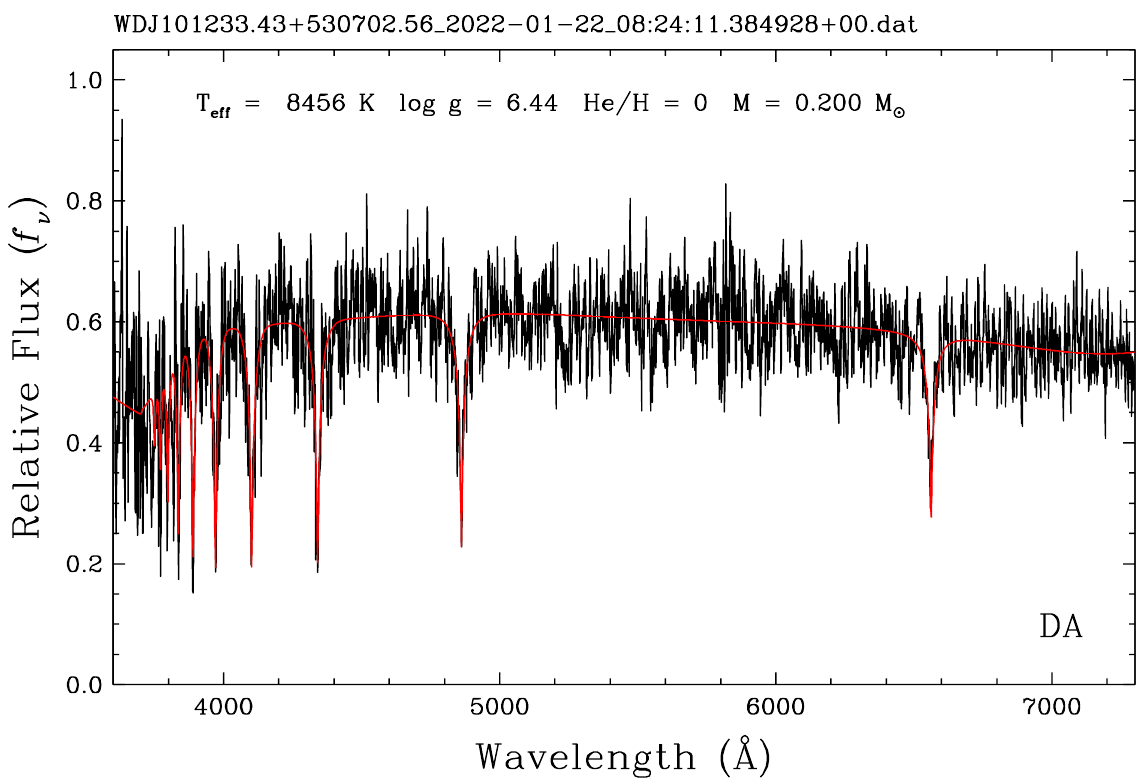} 
\caption{Spectroscopic model fits to the DESI DR1 spectrum of the white dwarf companion to PSR J1012+5307.}
\label{figpsr} 
\end{figure}

\begin{deluxetable*}{lrrcrc}
\tabletypesize{\tiny}
\tablecolumns{8} \tablewidth{0pt}
\tablecaption{New ELM white dwarf candidates.\label{tabelm}}
\tablehead{\colhead{Name} & \colhead{SourceID} & \colhead{$T_{\rm eff,phot}$} & \colhead{$\log{g,{\rm phot}}$} & \colhead{$T_{\rm eff,spec}$} & \colhead{$\log{g,{\rm spec}}$}\\
 &  & (K) & (cm s$^{-2}$) & (K) & (cm s$^{-2}$) }
\startdata
WDJ011213.23+090938.85 & 2579536673415380480 & 8912 $\pm$ 106 & 6.105 $\pm$ 0.230 & 9281 & 6.787 \\
WDJ012751.27+032117.37 & 2560031577656622208 & 9044 $\pm$ 112 & 5.613 $\pm$ 0.418 & 8811 & 6.520 \\
WDJ013900.23+240258.42 & 291000256395937408 & 8193 $\pm$ 62 & 6.627 $\pm$ 0.051 & 8758 & 6.587 \\
WDJ022621.34$-$014139.97 & 2499604377099621248 & 8859 $\pm$ 122 & 6.398 $\pm$ 0.370 & 8950 & 6.458 \\
WDJ025734.19+045047.74 & 5266626337620992 & 15901 $\pm$ 470 & 6.409 $\pm$ 0.320 & 18696 & 5.370 \\
WDJ042237.74+004714.24 & 3254939109848050304 & 9527 $\pm$ 96 & 6.903 $\pm$ 0.129 & 9618 & 6.383 \\
WDJ065252.72+562434.10 & 1000249779605696896 & 9079 $\pm$ 108 & 5.189 $\pm$ 0.634 & 9296 & 5.213 \\
WDJ070330.75+694243.75 & 1109446814844698496 & 9601 $\pm$ 196 & 5.795 $\pm$ 0.394 & 9671 & 6.037 \\
WDJ073801.97+513536.94 & 982734353215948672 & 10134 $\pm$ 196 & 6.301 $\pm$ 0.176 & 9192 & 5.124 \\
WDJ082331.28+045721.87 & 3093100069211196160 & 8093 $\pm$ 100 & 5.723 $\pm$ 0.268 & 8445 & 6.381 \\
WDJ090008.70$-$070113.11 & 5757292884492107136 & 8272 $\pm$ 72 & 5.754 $\pm$ 0.224 & 8392 & 6.165 \\
WDJ105436.50$-$051626.68 & 3764920880875585024 & 10570 $\pm$ 234 & 6.263 $\pm$ 0.429 & 10407 & 5.566 \\
WDJ142907.79$-$010038.14 & 3652672391630956032 & 8824 $\pm$ 101 & 5.690 $\pm$ 0.386 & 9089 & 6.550 \\
WDJ143147.21$-$003234.30 & 3653088183120009472 & 8890 $\pm$ 115 & 6.062 $\pm$ 0.503 & 9300 & 6.625 \\
WDJ144337.97+335754.86 & 1286885314140963072 & 8256 $\pm$ 194 & 6.246 $\pm$ 0.410 & 8539 & 6.358 \\
WDJ154045.22+024010.21 & 4423906955870722816 & 11370 $\pm$ 303 & 7.005 $\pm$ 0.222 & 13333 & 7.001 \\
WDJ154946.01+093923.91 & 4455519186478809856 & 8439 $\pm$ 120 & 5.643 $\pm$ 0.441 & 8502 & 6.440 \\
WDJ163209.15+260554.01 & 1304020687463771648 & 8470 $\pm$ 100 & 5.324 $\pm$ 0.546 & 8771 & 6.007 \\
WDJ170008.13+131754.51 & 4544499638859040000 & 9043 $\pm$ 152 & 6.279 $\pm$ 0.303 & 8829 & 6.744 \\
WDJ171436.52+182640.53 & 4548513199897544960 & 8978 $\pm$ 70 & 5.868 $\pm$ 0.355 & 9159 & 6.069 \\
WDJ171801.07+151048.36 & 4546525523392712064 & 9673 $\pm$ 123 & 5.448 $\pm$ 0.473 & 9778 & 6.618 \\
WDJ172243.94+354859.74 & 1336671441165227264 & 8512 $\pm$ 69 & 6.656 $\pm$ 0.101 & 8492 & 6.338 \\
WDJ175828.43+720756.96 & 1639303701652898432 & 10819 $\pm$ 156 & 6.276 $\pm$ 0.108 & 11441 & 6.057 \\
WDJ204339.83$-$013229.55 & 4225477577010044416 & 8848 $\pm$ 86 & 5.633 $\pm$ 0.651 & 9627 & 6.304 \\
WDJ220636.30+163743.55 & 1774968627474349184 & 14095 $\pm$ 544 & 7.077 $\pm$ 0.282 & 11714 & 6.677 \\
WDJ223716.61+052228.50 & 2706009304070920192 & 8645 $\pm$ 100 & 6.288 $\pm$ 0.255 & 8886 & 6.387 \\
WDJ224010.20+055459.64 & 2706430279585237120 & 8451 $\pm$ 164 & 6.301 $\pm$ 0.409 & 8915 & 6.603 \\
WDJ232309.45$-$111922.05 & 2436733305790413824 & 8248 $\pm$ 105 & 6.283 $\pm$ 0.180 & 8637 & 6.754 \\
WDJ234200.53+272007.53 & 2865184197972062336 & 8748 $\pm$ 114 & 6.127 $\pm$ 0.232 & 9086 & 6.367 \\
\enddata
\end{deluxetable*}

Figure \ref{figelm} shows the spectroscopic parameters of the DA white dwarfs in our sample in the ELM region. After excluding the sdA population with $T_{\rm eff}<9000$ K and $\log{g}<6$ based on pure H atmosphere model fits, we identify 51 potential targets of interest with $T_{\rm eff}>9000$ K and $\log{g}=5$-7, or $T_{\rm eff}>8000$ K and $\log{g}=6$-7. Out of the 51 selected targets, 22 are confirmed as ELM white dwarf binaries in the literature, including the companion to PSR J1012+5307. Figure \ref{figpsr} shows our model fits to the DESI spectrum of this object. \citet{vankerkwijk96} obtained $T_{\rm eff} = 8550 \pm 25$ K and $\log{g}= 6.75 \pm 0.07$ for the companion to PSR J1012+5307 based on 1D model atmospheres. Including the 3D corrections from \citet{tremblay15} brings down these values to $T_{\rm eff} = 8440$ K and $\log{g} = 6.43$, which are almost identical to the parameters that we obtained using the DESI spectrum. Hence, the remaining 29 candidates with similar parameters are also likely to be
ELM white dwarfs. Table \ref{tabelm} presents the physical parameters of these 29 systems; follow-up radial velocity observations are needed to confirm their binary nature and constrain their orbital parameters.

\subsubsection{Double-lined Spectroscopic Binaries: DA+DB and DA+DQ}
\label{secdqda}

\begin{figure*}
\center
\includegraphics[width=2.8in, clip=true, trim=0.4in 0.8in 0.1in 1.1in]{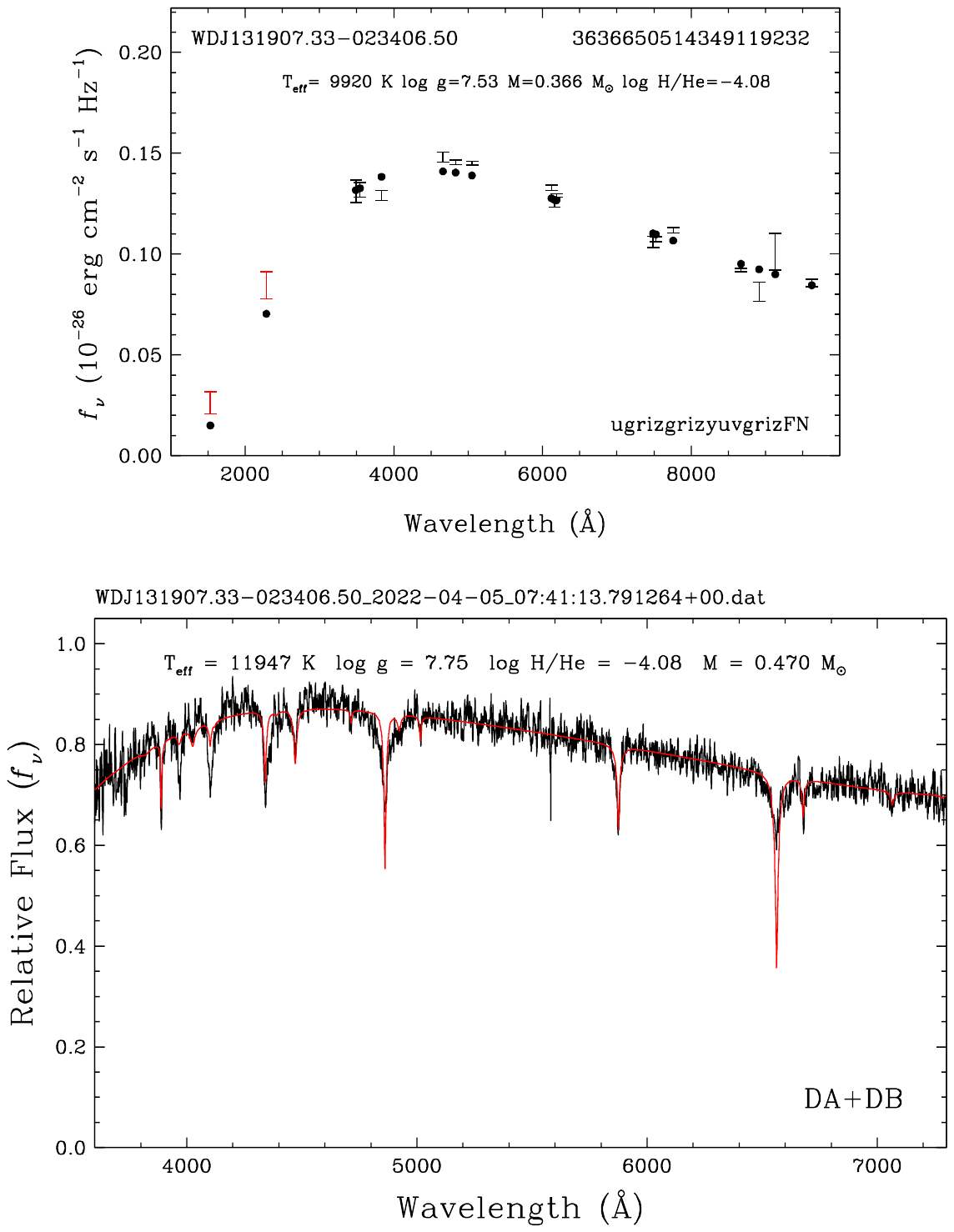}
\includegraphics[width=2.8in, clip=true, trim=0.4in 0.8in 0.1in 1.1in]{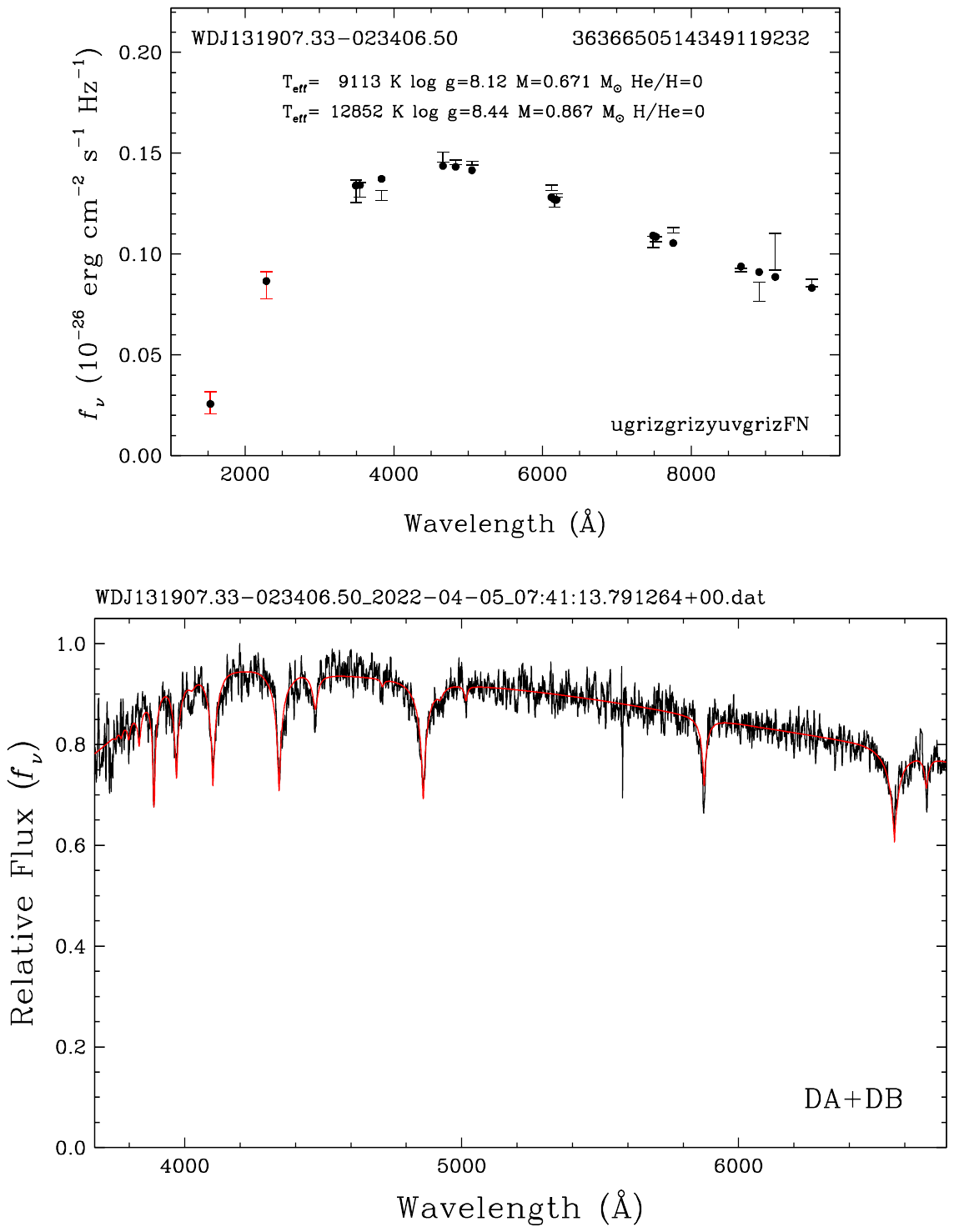}
\caption{Model fits to WDJ131907.33$-$023406.50 under the assumption of a single star (left panels) and a double degenerate binary containing a DA+DB system (right panels, the adopted photometric and spectroscopic parameters are given in the upper panel). The top and bottom panels show the photometric and spectroscopic fits, respectively. Note that there are significant degeneracies in the joint DA+DB model fits.}
\label{figdadb} 
\end{figure*}

We identify two types of double-lined spectroscopic binaries in DESI DR1 white
dwarf sample: DA+DB and DA+DQ systems. In Paper I, we identified 23 DA+DB binaries. Here, we identify
14 additional systems, where the model fits under the assumption of a single DBA star fail. 
Figure \ref{figdadb} shows our model fits for one of the newly identified targets, WDJ131907.33$-$023406.50. 
Assuming a single star, the photometric fit (using the same H/He ratio as the spectroscopic fit) indicates a mass of only $0.37~M_\odot$, whereas the spectroscopic fit results in a mass of $0.47~M_\odot$. In addition,
the spectroscopic fit fails to reproduce the depth of the H lines in the DESI spectrum. 

The right panel in Figure \ref{figdadb} shows our model fit under the assumption of a binary. We use the
PIKAIA algorithm \citep{pikaia} with $T_{\rm eff}$ and $\log{g}$ for the DA and DB components as free parameters, where the $\chi^2$ used is a combination of the photometric and the spectroscopic fit $\chi^2$.
The joint fit indicates a $0.67~M_\odot$ DA white dwarf with $T_{\rm eff}=9113$ K with a $0.87~M_\odot$ DB
white dwarf with $T_{\rm eff}=12,852$ K. However, these fits are based on a single DESI spectrum, and there are degeneracies in the model fits. Follow-up radial velocity observations would be needed to precisely constrain the physical and orbital parameters of this system. Table \ref{tabdadb} provides the list of 14 DA+DB binaries identified in this work. We provide both sets of fits, the fits assuming a single object and the deconvolved fits,
to all 14 systems in the \href{https://doi.org/10.5281/zenodo.20775938}{Zenodo archive}. 

\begin{deluxetable}{lr}
\tablecolumns{8} \tablewidth{0pt}
\tablefontsize{\tiny}
\tablecaption{DA+DB binary white dwarfs.\label{tabdadb}}
\tablehead{\colhead{Object} & \colhead{Gaia SourceID}}
\startdata
WDJ010835.18$-$112229.19 & 2469718276666730880  \\
WDJ020851.09+312531.08 & 301366761260727808 \\
WDJ022228.39+283007.72 & 130960218644587136  \\
WDJ041656.73$-$085514.81 & 3192584465407223424  \\
WDJ065845.28+591502.72 & 1002481586347664256  \\
WDJ074419.82+302203.33 & 879166817510277632  \\
WDJ074508.80+311659.57 & 880073193049049216  \\
WDJ092659.86+001457.65 & 3840889059815758336  \\
WDJ113917.53+203258.26 & 3979143786491250304  \\
WDJ131907.33$-$023406.50 & 3636650514349119232  \\
WDJ151032.82+034434.44 & 1155613113367500032  \\
WDJ170604.93+423822.00 & 1355079434773397120  \\
WDJ173213.07+170242.66 & 4550185935039440384  \\
WDJ224348.27+255120.94 & 1877613196183230080 \\
\enddata
\end{deluxetable}

DA+DQ binaries are rare; \citet{adamane26} report only 4 such systems in the literature. However, \citet{coutu19} list six other cool DQ white dwarfs that also show H$\alpha$, three of which are low-mass, and therefore they are most probably unresolved DA+DQ binaries. Hence, including the \citet{coutu19} sample brings the number of DA+DQ binaries in the literature to 7. Only two of the previously known DA+DQ systems have orbital constraints based on radial velocity measurements. Interestingly, both involve 0.5 $M_\odot$ DQ white dwarfs orbiting a DA companion in a $\approx30$ hour orbit \citep{vennes25,adamane26}.

Through an inspection of the DQ white dwarfs in DESI DR1, we identify 11 DA+DQ systems, six of which are new discoveries. This is the beauty of the current and the next generation multi-plexed spectroscopic surveys like DESI; with a target selection based on Gaia, they are destined to find rare systems like these. Table \ref{tabdqda} presents the names and Gaia Source IDs for these 11 systems. 

\begin{figure}
\centering
\includegraphics[width=3in]{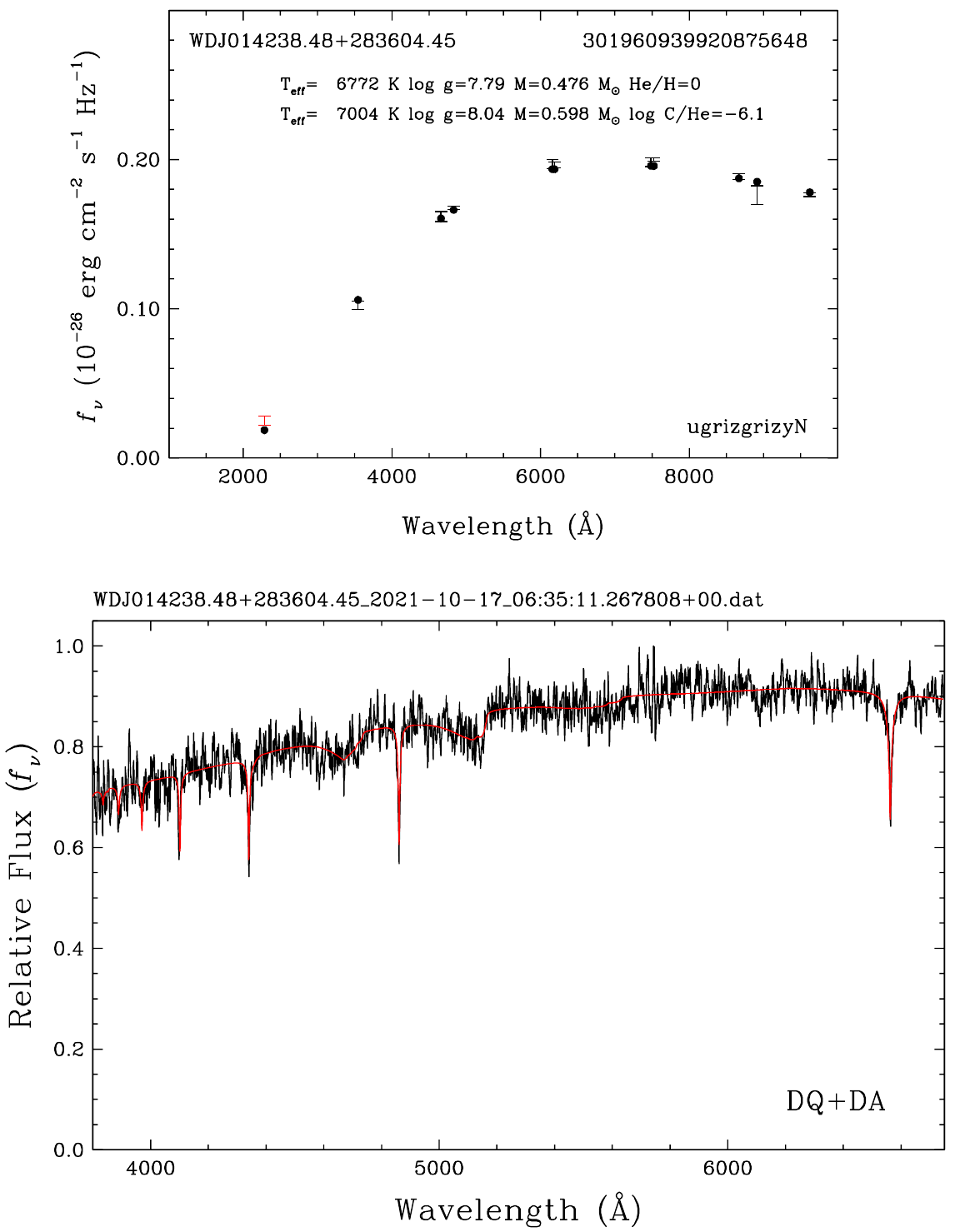}
\caption{Joint fits to the photometric and spectroscopic spectral energy distribution of the newly identified DA+DQ binary WDJ014238.48+283604.45. The data can be explained by a binary system consisting of a 6772 K and $M=0.476~M_\odot$ DA white dwarf and a 7004 K and $M=0.598~M_\odot$ DQ white dwarf with $\log$ C/He = $-6.1$. Note that there are significant degeneracies in the DA+DQ joint model fits.}
\label{figdadq}
\end{figure}

Figure \ref{figdadq} shows our joint model fits to the spectral energy distribution of one of the newly identified DA+DQ systems. 
We use a similar approach to our analysis of the DA+DB systems discussed above, and employ the PIKAIA algorithm \citep{pikaia} with $T_{\rm eff}$ and $\log{g}$ for each component (and $\log$ C/He for the DQ white dwarf) as free parameters. Our best-fit model indicates a binary consisting of a 6772 K and $M=0.476~M_\odot$ DA white dwarf with a 7004 K and $M=0.598~M_\odot$ DQ white dwarf companion. However, there are many degeneracies in the model fits.
Our best-fitting solutions represent only one of the many possible solutions; the best-fitting model to the DESI spectrum of WDJ090618.44+022311.66 requires masses that are significantly different from the masses inferred through radial velocity observations of the same system by \citet{adamane26}. However, both solutions provide excellent matches to the DESI spectrum. Hence, follow-up radial velocity observations of this system, as well as the other DA+DQ binaries are essential for constraining the parameters of these double-lined spectroscopic binaries and obtain dynamical mass measurements for the DQ white dwarfs in these systems.

\begin{deluxetable}{lrl}
\tabletypesize{\tiny}
\tablecolumns{8} \tablewidth{0pt}
\tablecaption{DA+DQ White Dwarfs in DESI DR1.\label{tabdqda}}
\tablehead{\colhead{Name} & \colhead{SourceID} & \colhead{Reference}}
\startdata
WDJ014238.48+283604.45 &  301960939920875648 & \\
WDJ090618.44+022311.66 &  577257520277310848 & \citet{adamane26} \\
WDJ092053.05+692645.36 & 1117655837377358976 & \citet{manser24} \\
WDJ092317.43+050309.64 &  585478637437191424 &  \\
WDJ092825.16+263856.56 &  694176429973819520 & \citet{coutu19} \\
WDJ094948.20+001000.02 & 3834035425883229056 &  \\
WDJ095753.41+104635.42 & 3879512807355721600 &  \\
WDJ102433.90+695300.60 & 1073394550124605184 &  \\
WDJ140641.02+340200.70 & 1481854770426696704 & \citet{coutu19} \\
WDJ153210.05+135616.11 & 1193974077628437760 & \citet{giammichele12} \\
WDJ190342.60+611736.29 & 2252082731770989696 &  \\
\enddata
\end{deluxetable}

\subsection{The Future: DESI, SDSS-V, 4MOST, and WEAVE}

There are 359,073 high-confidence white dwarfs in the Gaia white dwarf catalog of \citet{gentile21}. DESI DR1 has provided usable spectra for about 45,000 of these objects. DESI DR1 targeted Gaia white dwarfs with $G < 20$ and declination $\delta\gtrsim-20^\circ$, achieving 46\% completeness within its footprint (see Figure 2 in \citealt{kilic26}). We expect that future DESI releases will complete observations of the majority of the Gaia white dwarfs in its footprint. However, DESI avoids the Galactic plane, and its observations are limited to Galactic latitudes $|b|>20^\circ$. Roughly 40\% of the high-confidence white dwarfs in Gaia are within $20^\circ$ of the plane. Hence, there is a gap in sky coverage for the northern white dwarfs near the plane. 

WEAVE will target white dwarfs as flux calibrators, and it plans to observe more than 50,000 white dwarfs as part of its targeted white dwarf survey \citep{jin24}. It will also observe around 1200 deg$^2$ of the northern Galactic Plane. However, most of the northern Galactic plane will not be covered by either DESI or WEAVE. 4MOST is planning to cover the southern Galactic plane \citep{dejong19}. Hence, a combination of these
multi-plexed spectroscopic surveys will provide spectroscopic data for the majority of the high-confidence white dwarfs in Gaia over the next 5-10 years, excluding the northern Galactic plane. The total number of spectroscopically confirmed Gaia white dwarfs will likely reach $\sim$$3\times10^5$, or a factor of $\sim$6 more than the DESI DR1 sample. This sample would provide an excellent opportunity to refine our understanding of the white dwarf population characteristics and evolution, and discover rare objects like warm DQs and unusual binaries including but not limited to DA+DQ, DA+DB, ELM white dwarfs, and AM CVn.

\section{Conclusions}
\label{seccon}

We present a detailed model atmosphere analysis of cool ($T_{\rm eff} \lesssim 10,000$ K), red white dwarf candidates with $G_{\rm BP}-G_{\rm RP}>0$
in DESI DR1. The cool white dwarf sample is dominated by DAs (73\%), followed by DC (16\%), DZ (3.6\%), DB (3.5\%), and DQ (2.7\%) white dwarfs.
Along with the hot DESI sample presented in Paper I, this paper completes a detailed model atmosphere analysis of nearly 45,000 unique
white dwarf candidates with DESI DR1 spectra. We identify 535 magnetic white dwarfs in the cool DESI DR1 sample. Along with the 298 magnetics identified in Paper I, this brings the total number of magnetic white dwarfs in DESI DR1 to 833, or 1.9\% of the sample. 

In Paper I, we found a systematic offset between the photometric and spectroscopic parameters of hot DA white dwarfs based on DESI DR1 spectra. 
This systematic problem is not seen in the cool DA sample analyzed here, indicating that DESI's calibration problems have a negligible effect for
cooler DAs with narrower Balmer lines. In addition to the dominant peak at $0.6~M_\odot$, we find relatively large numbers of low-mass DAs and white dwarf + M dwarf binaries among our cool white dwarf targets. Since DESI is a magnitude-limited survey, over-luminous systems, i.e. low-mass white dwarfs and binaries, are over-represented in the resulting population. We also find a relatively large number of DAs on the crystallization sequence, but we observe a dearth of ultramassive DAs with $M\geq1.1~M_\odot$ below $T_{\rm eff}=6000$ K. This is most likely because these ultramassive white dwarfs are
in the Debye cooling range, and therefore they disappear quickly from observational samples. 

Thanks to DESI's higher resolution compared to the SDSS, it is possible to identify magnetic DAs with relatively low fields, $B\sim1$ MG.
We find cool magnetic DAs throughout the white dwarf cooling sequence, and not just on the crystallization sequence. Hence, a crystallization induced
dynamo cannot solely explain the emergence of magnetism in cool white dwarfs. Instead, a mechanism that can generate fields in a large number of white dwarfs, like a convective dynamo that forms during earlier evolutionary phases \citep{camisassa24,einramhof26} and finally emerges during the white dwarf sequence, seems more plausible to explain our observations. 

We use the DESI DR1 sample to constrain the fraction of He-atmosphere
white dwarfs as a function of temperature over the range $100,000$ to 5000 K, and demonstrate that the He-fraction increases significantly below 10,000 K due to convective mixing. 

One of the main challenges that we faced during the analysis of the cool white dwarf sample in DESI DR1 was the treatment of DC white dwarfs.
Previous work on the mass distribution of DC white dwarfs in Gaia demonstrated that they must have C and/or H as electron donors in their atmospheres
\citep{bergeron19}. However, there was no uniform approach presented to keep the DC masses at $0.6~M_\odot$ for the entire temperature range from
11,000 K down to 4000 K, and also have H$\alpha$ remain invisible. 

Based on a detailed analysis of the DESI DC sample, we find that the H/He abundance ratio in He-atmosphere white dwarfs increases at lower temperatures, reaching H/He ratios of order unity at $T_{\rm eff}=5000$ K. This statement is, of course, model dependent. It is possible that there is something wrong with the currently available He-atmosphere models at low temperatures, as we do not yet have a consistent set of He-atmosphere models that can simultaneously fit the cool DC stars and the IR-faint white dwarfs \citep{blouin24}. 
The current models are likely inadequate or incomplete for the extreme conditions and high atmospheric pressures that characterize He-rich atmospheres \citep[e.g.,][]{bergeron22}. Therefore, we cannot rule out a scenario where we require more and more H for cooler DCs because the currently available He-atmosphere models become progressively worse at low temperatures, forcing us to chose the H-rich models. Further theoretical and observational work to understand the compositions and physical parameters of cool He-rich white dwarfs would be essential to resolve these issues.

On the other hand, contrary to DC stars, the metallic lines in DZ stars provide a proxy to estimate the H/He ratio.  
There is considerable uncertainty in the precise values of H/He in DZs, but there are clear trends emerging below $T_{\rm eff}=6000$ K. Just like in DC stars, we find that the H pollution increases as a function of decreasing $T_{\rm eff}$. There are indeed some cool DZs that require H/He $\sim1$, as is the case for the DZA stars. However, not all DZs are extremely H-rich, as DZs with $\log$ H/He $\leq-4$ exist essentially at all temperatures between 11,000 and 4000 K. Given the assumed and measured H/He ratios in DC and DZ white dwarfs, the majority of cool He-atmosphere white dwarfs likely result from convective mixing at low $T_{\rm eff}$, and therefore contain relatively large amounts of H. However, a small fraction must be the cooled off versions of DB/DBA stars or DA stars that mixed between $\sim$10,000 and 7000 K, with much less H. These proportions must be reflected in the H/He ratios of the DC and DZ populations. 

Finally, DESI DR1 also includes a number of rare binary systems, including extremely low-mass and double-lined spectroscopic binaries involving DA+DB and DA+DQ white dwarfs. With the current
and upcoming multi-plexed surveys like DESI, SDSS-V, 4MOST, and WEAVE, we will be able to observe the majority of the high-confidence white dwarfs in Gaia. Such a large sample size will enable us to find diamonds in the rough, including large samples of warm DQs that are going through crystallization and distillation, exotic binaries, and will lead to unexpected discoveries. However, there remains a gap in (DESI's) sky coverage in the northern Galactic plane. When the main DESI survey is done, we hope that DESI can expand its survey to the northern Galactic plane.

\begin{acknowledgements}

This work is supported in part by the NSF under grant  AST-2508429, the NASA under grants 80NSSC22K0479, 80NSSC24K0380, and 80NSSC24K0436, the NSERC Canada, the Fund FRQ-NT (Qu\'ebec), and the Smithsonian Institution.

This research used data obtained with the Dark Energy Spectroscopic Instrument (DESI). DESI construction and operations is managed by the Lawrence Berkeley National Laboratory. This material is based upon work supported by the U.S. Department of Energy, Office of Science, Office of High-Energy Physics, under Contract No. DE–AC02–05CH11231, and by the National Energy Research Scientific Computing Center, a DOE Office of Science User Facility under the same contract. Additional support for DESI was provided by the U.S. National Science Foundation (NSF), Division of Astronomical Sciences under Contract No. AST-0950945 to the NSF’s National Optical-Infrared Astronomy Research Laboratory; the Science and Technology Facilities Council of the United Kingdom; the Gordon and Betty Moore Foundation; the Heising-Simons Foundation; the French Alternative Energies and Atomic Energy Commission (CEA); the National Council of Humanities, Science and Technology of Mexico (CONAHCYT); the Ministry of Science and Innovation of Spain (MICINN), and by the DESI Member Institutions: www.desi.lbl.gov/collaborating-institutions. The DESI collaboration is honored to be permitted to conduct scientific research on I’oligam Du’ag (Kitt Peak), a mountain with particular significance to the Tohono O’odham Nation. Any opinions, findings, and conclusions or recommendations expressed in this material are those of the author(s) and do not necessarily reflect the views of the U.S. National Science Foundation, the U.S. Department of Energy, or any of the listed funding agencies.

\end{acknowledgements}

\facilities{Mayall (DESI)}


\end{document}